\documentclass{ustthesis}
\usepackage{xspace}
\usepackage{xcolor}
\usepackage{pifont}
\usepackage{mathpazo,amsmath,amssymb,epsfig,enumerate,bbm,calc,color,ifthen,capt-of} 
\usepackage[margin=25mm,textheight=247mm,textwidth=145mm]{geometry}

\usepackage{url}
\usepackage{amsfonts}
\usepackage{mathrsfs}
\usepackage{bookmark}

\usepackage[TABBOTCAP]{subfigure}
\usepackage{float}

\usepackage{tabularx}
\newcolumntype{Y}{>{\centering\arraybackslash}X}

\usepackage{algorithmic}
\usepackage[linesnumbered,ruled,vlined]{algorithm2e}
\let\oldnl\nl 
\newcommand{\nonl}{\renewcommand{\nl}{\let\nl\oldnl}} 

\usepackage{tikz}
\usetikzlibrary{fit,calc}
\newcommand*{\tikzmk}[1]{\tikz[remember picture,overlay] \node (#1) {};\ignorespaces}
\newcommand{\boxit}[1]{\tikz[remember picture,overlay]{\node[yshift=5pt,fill=#1,opacity=.25,fit={($(A)+(0.1em,0em)$)($(B)+(.94\linewidth,0.6\baselineskip)$)}] {};}\ignorespaces}
\newcommand{\boxend}[1]{\tikz[remember picture,overlay]{\node[yshift=5pt,fill=#1,opacity=.25,fit={($(A)+(0.1em,0em)$)($(B)+(.94\linewidth,-0.2\baselineskip)$)}] {};}\ignorespaces}
\colorlet{mygrey}{black!40}
\colorlet{myblue}{cyan!60}
\colorlet{mypink}{red!40}

\usepackage{hyperref}
\hypersetup{
  colorlinks=true,      
  linkcolor=blue,       
  citecolor=magenta,    
  filecolor=cyan,       
  urlcolor=red          
}

\newenvironment{icompact}{
  \begin{list}{$\bullet$}{
    \parsep 1pt plus 1pt
    \partopsep 1pt plus 1pt
    \topsep 1pt plus 2pt minus 1pt
    \itemsep 1.5pt plus 1pt
    \parskip 0pt plus 2pt
    \leftmargin 0.15in}
       }
  {\normalsize\end{list}}

\newcommand{\paraspace}{\vspace{0.05in}}
\newcommand{\parab}[1]{\paraspace\noindent{\bf #1} }
\newcommand*\circled[1]{\raisebox{0.5pt}{\textcircled{\raisebox{-0.9pt} {#1}}}}

\usepackage{cleveref}
\crefformat{chapter}{\S#2#1#3} 
\crefformat{section}{\S#2#1#3} 
\crefformat{subsection}{\S#2#1#3}
\crefformat{subsubsection}{\S#2#1#3}
\crefname{section}{§}{§§}

\def\ie{\textit{i.e.}}
\def\eg{\textit{e.g.}}

\newcommand{\sys}{{GEMINI}\xspace}
\newcommand{\sysA}{{GEMINI}\xspace}
\newcommand{\sysB}{{FlashPass}\xspace}

\title{Congestion Control Mechanisms for Inter-Datacenter Networks}  
\author{Gaoxiong~Zeng}     
\degree{\PhD}              
\subject{Computer Science and Engineering}      
\department{Computer Science and Engineering}       
\advisor{Prof.~Kai~Chen}     
\depthead{Prof.~Dit-Yan~YEUNG}    
\defencedate{2022}{1}{10}      

\begin{document}
\maketitle

\pdfbookmark[section]{\contentsname}{toc}
\newpage

\thispagestyle{empty}
\null\skip0.2in
\begin{center}

\vspace{75mm}
\emph{To my family and friends}
\end{center}

\acknowledgments
The last six years at HKUST were an incredible journey in my life. There were dilemmas and breakthroughs, depressions and endeavors, failures and successes along the journey. After all, I did not only become stronger in ability, but also got to know many brilliant people.
First of all, I am gratitude to my supervisor, Prof. Kai Chen. He provides me all kinds of help throughout my PhD studies, and encourages me to do great research for life.

I have been very fortunate to collaborate with many outstanding researchers from both academia and industry. I am grateful to Prof. Dongsu Han from Korea Advanced Institute of Science and Technology, Dr. Kun Tan and Dr. Lei Cui from Huawei, Dr. Yibo Zhu from then Microsoft Research and now ByteDance, Dr. Hongqiang Liu and Dr. Yifei Yuan from Alibaba, etc. for the help and guidance on my research.

I am thankful to Prof. Yongshun Cai, Prof. Qiong Luo, Prof. Brahim Bensaou, Prof. Jiang Xu, and Prof. Hong Xu for serving on my thesis defense committee. I would also like to thank my thesis proposal committee and qualifying examination committee members: Prof. Bo Li, Prof. Shueng-Han Gary Chan, and Prof. Wei Wang, etc.

I am grateful to my labmates of the HKUST SING Group. Many thanks to Wei Bai and Shuihai Hu for their tremendous help on my research and life. I enjoy chatting with Hong Zhang and Li Chen, who can always come up with inspiring ideas. I am fortunate to work with many amazing colleagues: Hengky Susanto, Yuchao Zhang, Wenxin Li, Jinbin Hu,  Weiyan Wang, Junxue Zhang, Ge Chen, Yiding Wang, Han Tian, Ziyang Li, Hao Wang (from SJTU), Jiacheng Xia, Qinghe Jing, Bairen Yi, Zhaoxiong Yang, Justinas Lingys, Jinzhen Bao, Zhouwang Fu, Zhuotao Liu, Xuya Jia, Baochen Qiao, Jingrong Chen, Xinchen Wan, Chaoliang Zeng, Yiqing Ma, Liu Yang, Zhaorong Liu, Duowen Liu, Zhuoyi Peng, Cengguang Zhang, Ding Tang, Hao Wang (from PKU), Zilong Wang, Di Chai, Yilun Jin, Hao Xu, Xudong Liao, Kaiqiang Xu, Yuxuan Qin, Zhenghang Ren, Tianjian Chen, Wenxue Li, Xiaodian Cheng, Decun Sun, Xinyang Huang, Shuowei Cai, Jianxin Qiu, Xiaoyu Hu, etc.
Also, I am very grateful to Connie Lau, Isaac Ma, and Lily Chan for their administration work; and staff from CS system that helps to maintain the lab facilities.

I would like to thank my HKUST friends outside SING Group for their company and help: Mingfei Sun, Peng Xu, Shiheng Wang, Kaiyi Wu, Chen Chen, Lili Wei, Jieru Zhao, etc. I am also thankful to my friends outside HKUST who enriched my life outside work: Hao Wei, Jinghua Zhong, Hongwei Yuan, Jin Xiao, Shuguang Hu, etc.

Last but not least, I would like to thank my family for their continuous love and support in the past 30 years. Without them, this thesis would have not been possible.

\endacknowledgments

\tableofcontents

\listoffigures

\listoftables

\listofalgorithms
\addcontentsline{toc}{chapter}{List of Algorithms}
\newpage

\begin{abstract} \label{sec-abstract}

Applications running in geographically distributed setting are becoming prevalent. Large-scale online services often share or replicate their data into multiple data centers (DCs) in different geographic regions. Driven by the data communication need of these applications, inter-datacenter network (IDN) is getting increasingly important. 

However, we find congestion control for inter-datacenter networks quite challenging. Firstly, the inter-datacenter communication involves both data center networks (DCNs) and wide-area networks (WANs) connecting each data center. Such a network environment presents quite heterogeneous characteristics (\eg, buffer depths, RTTs). Existing congestion control mechanisms consider either DCN or WAN congestion, while not simultaneously capturing the degree of congestion for both. 

Secondly, to reduce evolution cost and improve flexibility, large enterprises have been building and deploying their wide-area routers based on shallow-buffered switching chips. However, with legacy congestion control mechanisms (\eg, TCP Cubic), shallow buffer can easily get overwhelmed by large BDP (bandwidth-delay product) wide-area traffic, leading to high packet losses and degraded throughput.

This thesis describes my research efforts on optimizing congestion control mechanisms for the inter-datacenter networks. First, we design \sysA~--- a reactive congestion control mechanism that simultaneously handles congestions both in DCN and WAN. 
Second, we present \sysB~--- a proactive congestion control mechanism that achieves near zero loss without degrading throughput under the shallow-buffered WAN. Extensive evaluation shows their superior performance over existing congestion control mechanisms.

\end{abstract}

\chapter{Introduction}\label{sec-introduction}

Applications running in geographically distributed setting are becoming prevalent~\cite{laoutaris2011inter, wu2013spanstore, pu2015low, hung2015scheduling, jin2016optimizing, gaia}. Large-scale online services often share or replicate their data into multiple DCs in different geographic regions. For example, a retailer website runs a database of in-stock items replicated in each regional data center for fast serving local customers. These regional databases synchronize with each other periodically for the latest data. Other examples include image sharing on online social networks, video storage and streaming, geo-distributed data analytics, etc.

With the prevalence of the geo-distributed applications and services, inter-datacenter network (IDN) is becoming an increasingly important cloud infrastructure~\cite{b4, swan, bwe, b4after}. For example, Google~\cite{b4after} reveals that its inter-datacenter wide-area traffic has been growing exponentially with a doubling of every 9 months in recent 5 years. This pushes the IDN facility to evolve much faster than the rest of its infrastructure components. 

However, we find congestion control for inter-datacenter networks quite challenging. Firstly, the inter-datacenter communication involves both data center networks (DCNs) and wide-area networks (WANs) connecting each data center. Such a network environment presents quite heterogeneous characteristics (\eg, buffer depths, RTTs). Existing congestion control mechanisms consider either DCN or WAN congestion, while not simultaneously capturing the location and degree of congestion for both network segments. Secondly, to reduce evolution cost and improve flexibility, large enterprises have been building and deploying their wide-area routers based on shallow-buffered switching chips.
However, with legacy congestion control mechanisms (\eg, TCP Cubic), shallow buffer can easily get overwhelmed by large BDP (bandwidth-delay product) wide-area traffic, leading to high packet losses and degraded throughput.

This thesis describes my research efforts on optimizing congestion control mechanisms for the inter-datacenter networks. First, we design \sysA~\cite{gemini}~--- a reactive congestion control mechanism that simultaneously handles congestions both in DCN and WAN. The key idea is to strategically integrate ECN and delay signal for congestion control. 
Second, we present \sysB~\cite{flashpass}~--- a proactive congestion control mechanism that achieves near zero loss without degrading throughput under the shallow-buffered WAN. A sender-driven emulation mechanism is adopted to achieve accurate bandwidth allocation. Finally, we conduct extensive experiments for evaluation, and results validate their superior performance over existing congestion control mechanisms.

\section{Contributions}
This thesis focuses on the congestion control mechanisms for the inter-datacenter networks (IDNs). In the following subsections, we overview our two key contributions.

\subsection{Congestion Control under Network Heterogeneity}
Geographically distributed applications hosted on cloud are becoming prevalent~\cite{laoutaris2011inter, pu2015low, jin2016optimizing, gaia}. They run on \textit{cross-datacenter network} (cross-DCN) that consists of multiple data center networks (DCNs) connected by a wide area network (WAN). Such a cross-DC network poses significant challenges in transport design because the DCN and WAN segments have vastly distinct characteristics (\eg, buffer depths, RTTs). 

In this work, we find that existing DCN or WAN transport reacting to ECN or delay alone do not (and cannot be extended to) work well for such an environment. The key reason is that neither of the signals, by itself only, can simultaneously capture the location and degree of congestion, mainly due to the discrepancies between DCN and WAN. 

Motivated by this, we present the design and implementation of \sys~\cite{gemini} that strategically integrates both ECN and delay signals for cross-DC congestion control. To achieve low latency, \sys bounds the inter-DC latency with delay signal and prevents the intra-DC packet loss with ECN. To maintain high throughput, \sys modulates the window dynamics and maintains low buffer occupancy utilizing both congestion  signals. 

\sys is supported by rigorous theoretical analysis, implemented in Linux kernel 4.9.25, and evaluated by extensive testbed experiments. Results show that \sys achieves up to 53\%, 31\%, 76\% and 2\% reduction of small flow completion times (FCTs) on average, and up to 34\%, 39\%, 9\% and 58\% reduction of large flow average completion times compared to Cubic~\cite{cubic}, DCTCP~\cite{dctcp}, BBR~\cite{bbr}, and Vegas~\cite{vegas}.
Furthermore, \sys requires no customized hardware support and can be readily deployed in practice.

\subsection{Congestion Control under Shallow-buffered WAN}
To reduce evolution cost and improve flexibility, large enterprises (\eg, Google~\cite{google-dc}, Alibaba~\cite{alibaba-dc}, etc.) have been building and deploying their wide-area routers based on shallow-buffered switching chips.
However, with legacy reactive transport (\eg, TCP Cubic~\cite{cubic}), shallow buffer can easily get overwhelmed by large BDP wide-area traffic, leading to high packet losses and degraded throughput. 
To address it, current practice seeks help from traffic engineering, rate limiting, or multi-service traffic scheduling.

Instead, we ask: can we design a transport to simultaneously achieve high throughput and low loss for shallow-buffered WAN? 
We answer this question affirmatively by employing proactive congestion control (PCC).
However, two issues exist for existing PCC to work on WAN. Firstly, wide-area traffics have diverse RTTs. The interleaved credits can still trigger data crush due to RTT difference. Secondly, there is one RTT delay for credits to trigger data sending, which can degrade network performance.

Therefore, we propose a novel PCC design --- \sysB~\cite{flashpass}.
To address the first issue, \sysB adopts sender-driven emulation process with send time calibration to avoid the data packet crush.
To address the second issue, \sysB enables early data transmission in the starting phase, and incorporates an over-provisioning with selective dropping mechanism for efficient credit allocation in the finishing phase.

Our evaluation with production workload demonstrates that \sysB reduces the overall flow completion times of Cubic~\cite{cubic} and ExpressPass~\cite{expresspass} by up to 32\% and 11.4\%, and the 99-th tail completion times of small flows by up to 49.5\% and 38\%, respectively.

\section{Organization}
The remainder of the dissertation is organized as follows.
Chapter~\ref{sec-all:background} briefly goes through some background regarding network optimization mechanisms for data centers (DCs), including network architecture design, transport-layer congestion control, etc.
Chapter~\ref{sec:gemini} introduces \sysA, where we present the detailed design, theoretical analysis, implementation and evaluation of a reactive congestion control mechanism under the heterogeneous cross-datacenter network (cross-DCN). 
Chapter~\ref{sec:flashpass} describes \sysB, where we demonstrate the design challenges we observe and design choices we made to the proactive congestion control mechanisms under the shallow-buffered inter-datacenter network (IDN).
Finally, Chapter~\ref{sec-conclusion} concludes the thesis work and presents the future directions.

\newpage

\chapter{Background: Network Optimization for Data Centers}\label{sec-all:background}
In this chapter, we present a general picture of intra-datacenter and inter-datacenter networks\footnote{We often refer to the intra-datacenter network as DCN, and the inter-datacenter network as IDN.} and the related optimization efforts. In \S\ref{intro-dc}, we give an introduction to data centers (DCs). We show some characteristics and requirements of the applications hosted in DCs. In \S\ref{dc-optimization}, we introduce some research directions that try to optimize intra-DCN or inter-DCN, including network architecture designs and congestion control mechanisms.

\section{Introduction to Data Centers}\label{intro-dc}

Driven by the need of fast computation and large data storage of various web applications and services, large clusters of commodity PCs or servers have been built around the globe at a large scale rapidly~\cite{alibaba-dc, facebook-dc, google-dc, microsoft-dc}. Such kind of commodity server clusters together with the associated components (\eg, telecommunication, storage and power systems) is called a \textbf{data center (DC)}. Figure~\ref{fig:googledc} shows some typical views of data centers~\cite{google-dc}. Each data center often consists of 10s of thousands of servers or more. Dozens of DCs are connected with each other via a wide-area network (WAN) globally.


As shown by work~\cite{fattree, dctcp, swan, b4}, one of the principle performance bottlenecks of large-scale data centers lies in the inter-node network communication (for both intra-DCN and inter-DCN). 
Industry measurements from \cite{vl2, dctcp, benson2010network, inside-facebook} imply that data centers host a variety of applications. These applications have distinct network requirements. Some desire low latency for small messages (mice flows), while others prefer large long-term throughput for bulk transfers (elephant flows). Typically, as shown in \cite{vl2}, mice flows are numerous (over 99\% are smaller than 100 MB). However, more than 90\% of bytes come from elephant flows that are larger than 100MB. These application characteristics impose great pressure on the network architectures and mechanisms.

\section{Optimizing Intra- and Inter-Datacenter Networks}
\label{dc-optimization}
Driven by the stringent requirements of various web applications and services, network research community proposes numbers of designs in various aspects to deliver high network performance. We now discuss some key directions in the following subsections:

\begin{figure*}
\centering
\subfigure[Inside View of Datacenter]
{
\centering
\includegraphics[width=0.47\textwidth]{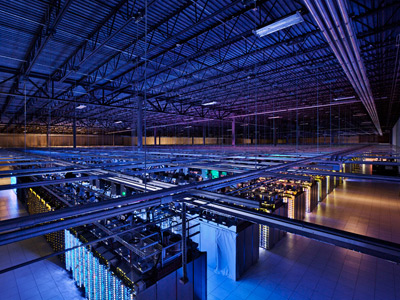}
\label{fig:inside-dc}
}
\subfigure[Outside View of Datacenter]
{
\centering
\includegraphics[width=0.47\textwidth]{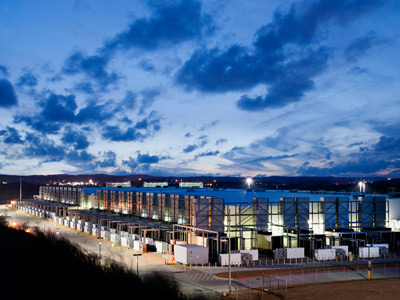}
\label{fig:outside-dc}
}
\subfigure[Global View of Inter-Datacenter Network (IDN)]
{
\centering
\includegraphics[width=\textwidth]{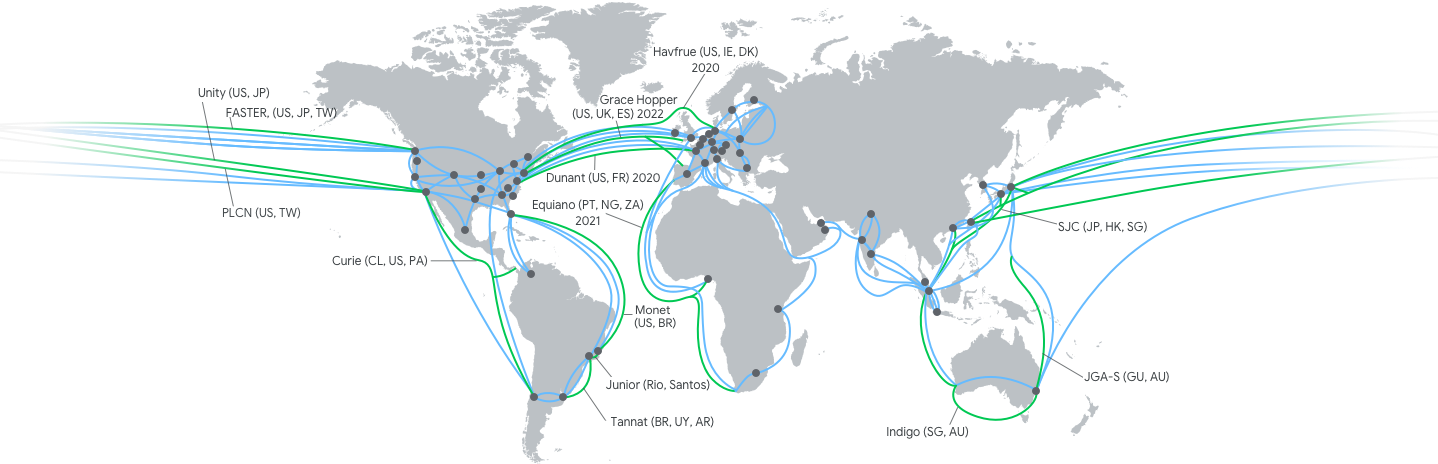}
\label{fig:global-dc}
}
\caption{Data Centers (DCs) in Google~\cite{google-dc}.}
\label{fig:googledc}
\end{figure*}

\subsection{Network Architectures}
Network architecture refers to the layout of the network, consisting of the hardware, software, connectivity, etc. One dividing crest of DCN architectures is the underlying switching technology: (1)~Electric packet switching (EPS); (2)~Optical circuit switching (OCS). Currently, the EPS network is dominant, while OCS is under exploration~\cite{cthrough, helios, osa, megaswitch}. 


Network topology design has also drawn great interests from both academia and industry~\cite{fattree, f10, dcell, bcube, vl2, jellyfish, jupiter}. Traditionally, DCNs are built with multi-level, multi-rooted trees of switches. The leaves of the tree are the so-called top-of-rack (TOR) switches, each connecting to dozens of servers downward with the access links and the network aggregates and cores upward with the core links. The access links vary from 1-10 Gbps while the core links are 40-100 Gbps typically. One obvious downside of these networks is that they do not scale. Building a non-blocking network requires large port counts and high internal backplane bandwidth of core switches, which won't be cheap even available. 

Fattree~\cite{fattree} is one of the seminal DCN topology optimization solutions. It is essentially a special instance of the Clos topology~\cite{fattree, dc-computer}. It also uses a multi-level, multi-rooted tree structure. However, based on the well-designed wiring scheme, it manages to deliver scalable bandwidth for non-blocking network communication with commodity Ethernet switches. F10~\cite{f10} is a variant of Fattree topology that are resistant to switch or link failures. Other works include DCell~\cite{dcell}, BCube~\cite{bcube}, VL2~\cite{vl2}, Jellyfish~\cite{jellyfish}, Jupiter Rising~\cite{jupiter}, etc.

\subsection{Transport Protocols and Congestion Control}
Network protocols characterize or even standardize the behavior of network communications. In particular, transport protocol (layer 4 in OSI network model~\cite{osi}) is of great importance. It provides the communication service for upper application layers, such as connection-oriented data streaming, reliability and congestion control. The well-known transport protocols of the Internet include the connection-oriented Transmission Control Protocol (TCP)~\cite{tcp} and the connectionless User Datagram Protocol (UDP)~\cite{udp}. 

Congestion control (CC) is one of the fundamental building blocks of the TCP transport protocol, the goal of which is to allocate network bandwidth among hosts efficiently, \eg, avoiding congestion collapse. At a high level, TCP congestion control is an end-to-end protocol built upon the principle of ``conservation of packets''~\cite{tcp1988}. That is, a flow ``in equilibrium'' should run stably with a full window of data in transit. There should be no new packet injection into the network until old ones leave. Following this principle, the implementation of TCP often employs the ``ACK clocking'' to trigger new packets into the network. 

TCP Tahoe~\cite{tcp1988} is the seminal TCP congestion control algorithm. It consists of two stages: (1) Slow Start; (2) Congestion Avoidance. For slow start, the sender begins with an initial congestion window (CWND). Then for each acknowledgement (ACK) received from receiver, it increases the CWND by the same ACKed size, resulting in roughly window doubling in each round-trip time (RTT). Thus, it can probe for the available bandwidth at an exponential speed. Congestoin avoidance follows the additive increase multiplicative decrease (AIMD) principle. It increments CWND linearly until it encounters a packet loss, which indicates possible network congestion. It then saves the half of the current window as a threshold value, resets CWND to one, and restarts from slow start.

TCP Tahoe~\cite{tcp1988} adopts ``Fast Retransmission'' for fast loss detection. Specifically, it takes 3 duplicate ACKs as a sign for packet loss and thus avoid waiting for long timeout. 
TCP Reno~\cite{reno} adds the ``Fast Recovery'' mechanism. Fast recovery suggests to reset the CWND to its half instead of one in case of 3 duplicate ACKs. This helps TCP to handle single packet loss, but not for consecutive losses as it may cut CWND multiple times. 
TCP NewReno~\cite{newreno} remedies the problem by keeping fast recovery state until all outstanding data gets ACKed. However, both TCP Reno and NewReno can retransmit at most one packet per RTT. 
TCP with ``Selective Acknowledgment'' (TCP SACK)~\cite{sack} enables multiple retransmissions by ACKing received data selectively instead of cumulatively. 


In the early years of DCN, the legacy TCP congestion control~\cite{tcp1988, reno, newreno, sack, cubic} has been adopted directly from the Internet for DCN data communication. These protocols typically leverage packet loss signal for congestion feedback. There are often two reasons for packet losses: packet corruption in transit, or the network congestion with insufficient buffer capacity. On most network paths, loss corruption is extremely rare. If packet loss is (almost) always due to congestion and if a timeout is (almost) always due to a lost packet, we can take it as a good indicator for the ``network is congested'' signal.

Although loss-based congestion control gets widely adopted due to its simplicity, it tends to fill switch buffer and cause excessive packet losses, thus failing to meet the harsh low latency requirements in DCN. Motivated by this observation, the seminal DCN transport design, DCTCP~\cite{dctcp}, is proposed in 2010.  DCTCP detects the network congestion with the Explicit Congestion Notification (ECN)~\cite{ecn} signal and reacts to the extent of congestion based on the ACK fraction with ECN marks. 

Since then, many congestion control mechanisms~\cite{dcn-transport, dctcp, hull, d2tcp, pfabric, l2dct, pase, dcqcn, timely, mlt, dlcp} have been proposed for high performance data communication in DCN. For example, HULL~\cite{hull} trades off some network bandwidth to achieve near zero queueing with a phantom queue. D2TCP~\cite{d2tcp} and L2DCT~\cite{l2dct} modulate the window adjustment function of DCTCP to meet deadlines and minimize FCT, respectively. DCQCN~\cite{dcqcn} is built on the top of DCTCP and QCN~\cite{qcn}. It enables the realistic deployment of Remote Direct Memory Access (RDMA) in large-scale ethernet DCNs.
The other line of work leverages delay signal with microsecond-level accuracy for congestion feedback, which is enabled by recent advances~\cite{softnic, timely, dx} in NIC technology. For example, TIMELY~\cite{timely} uses RTT signal for congestion control in RDMA networks.

There are also congestion control mechanisms aiming to optimize data communication for inter-datacenter networks. For example, BBR~\cite{bbr} is proposed primarily for the enterprise WAN. The core idea is to work at the theoretically optimal point~\cite{kleinrock1979power} with the aid of sophisticated network sensing (\eg, precise bandwidth and RTT estimation). Copa~\cite{copa} adjusts sending rate towards $1/(\delta d_q)$, where $d_q$ is the queueing delay, by additive-increase additive-decrease (AIAD). It detects buffer-fillers by observing the delay evolution and switches between delay-sensitive and TCP-competitive mode. These wide-area transport protocols usually assume little help (\eg, no ECN support) from the network switches so as to work across the complex wide-area network (WAN) environment.

\subsection{Emerging Technologies}
\textbf{Remote Direct Memory Access (RDMA):}
Datacenter has been upgrading its link bandwidth from 10Gbps to 40Gbps and more to meet its rising application need. Traditional TCP/IP stacks fall short to run at such speed due to the high CPU overhead. Moreover, some applications require ultra-low latency message transfers (a few microseconds). Traditional TCP/IP stacks have much higher latency. 
To address it, Remote Direct Memory Access (RDMA) is adopted from the high performance computing (HPC)  community to the DCN scenario~\cite{dcqcn, rdma-scale, irn}. With RDMA, network interface cards (NICs) directly transfer data in and out of pre-registered memory buffers, only involving host CPUs during the initialization step. This reduces CPU consumption and network latency to a great extent.

\textbf{Programmable Networks:}
Software-Defined Networking (SDN) provides administrators flexible control over the network control planes. Unlike conventional switches, SDN separates the network control plane from the data plane, and leverages a central controller to manage multiple switch data planes. However, it targets at fixed-function switches that support a predetermined set of header fields and actions. 
Data plane programmability~\cite{rmt, p4} is one step towards more flexible switches whose data plane can be changed. For example, P4~\cite{p4} allows the programmers to control the packet processing of the forwarding plane without worrying about the underlying realization. This enables lots of optimization mechanisms for data centers. 

While these emerging technologies (\eg, RDMA, programmable networks, etc.) seem promising in theory, there are still plenty of practical challenges to be addressed as well as research opportunities for realistic deployment at scale \cite{dcqcn, rdma-scale, irn, understand-pcie, hpcc, gfc, pcn, p4com}.

\chapter{\sysA: Reactive Congestion Control for Inter-Datacenter Networks} 
\label{sec:gemini}


Applications running in geographically distributed setting are becoming prevalent~\cite{laoutaris2011inter, b4, swan, wu2013spanstore, pu2015low, hung2015scheduling, jin2016optimizing, gaia}. Large-scale online services often share or replicate their data into multiple DCs in different geographic regions. For example, a retailer website runs a database of in-stock items replicated in each regional data center for fast serving local customers. These regional databases synchronize with each other periodically for the latest data. Other examples include image sharing on online social networks, video storage and streaming, geo-distributed data analytics, etc.

\begin{figure}[!t]
\centering
\includegraphics[width=0.8\textwidth]{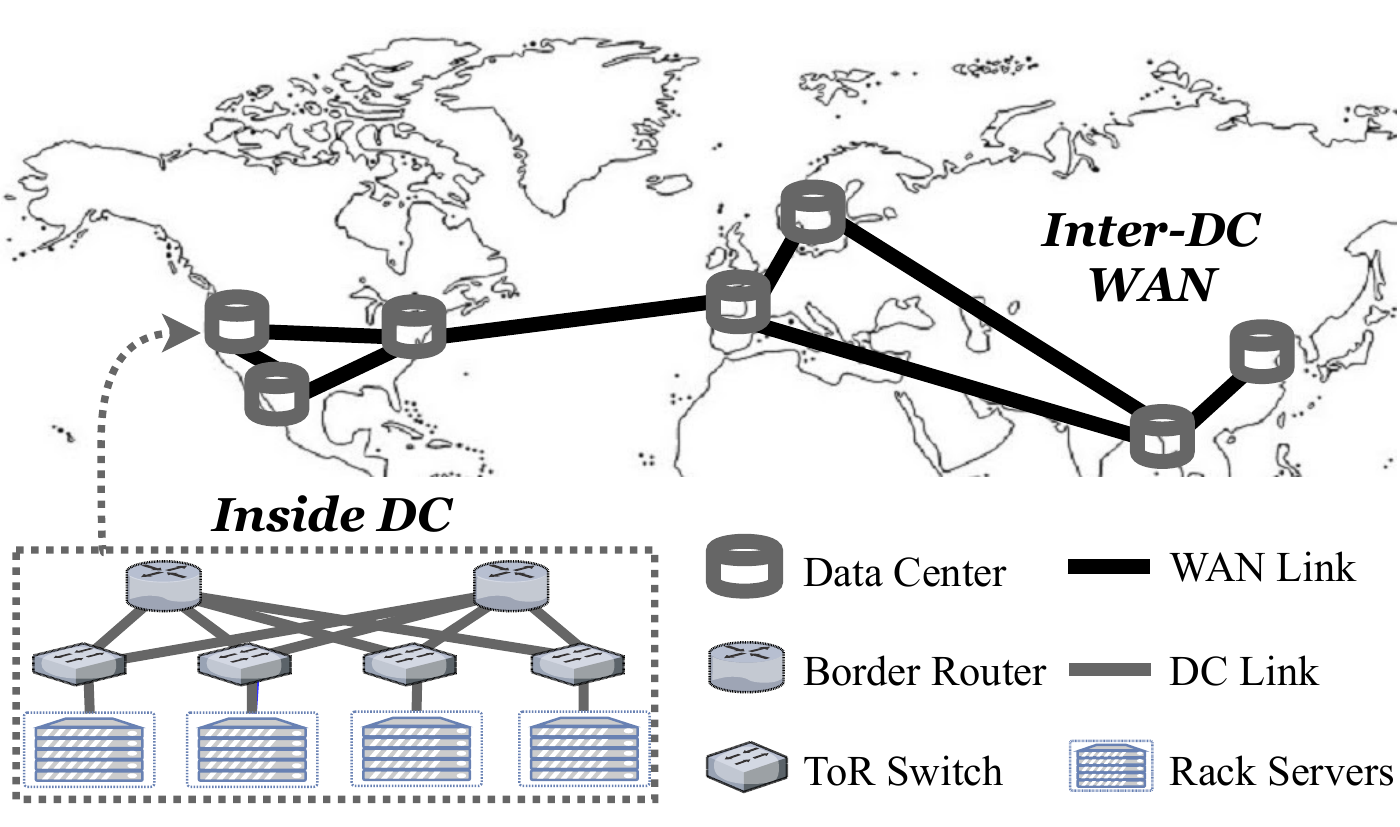}
\caption{Cross-Datacenter Network (Cross-DCN).}
\label{fig:crossdcnet}
\end{figure}

These applications run on \textit{cross-datacenter (DC) network} (Figure~\ref{fig:crossdcnet}) that consists of multiple data center networks (DCNs) connected by a wide area network (WAN). The wide area and intra-DC networks have vastly distinct characteristics (\S\ref{sec:background}). For WAN, achieving high network utilization is a focus and switches have deep buffers. In contrast, latency is critical in DCN and switches have shallow buffers. While there are numerous transport protocols designed for either DCN or WAN individually, to the best of our knowledge, little work has considered a heterogeneous cross-DC environment consisting of both parts.

To handle congestion control in either DCN or WAN, existing solutions have leveraged either ECN (\eg, DCTCP~\cite{dctcp} and DCQCN~\cite{dcqcn}) or delay (\eg, Vegas~\cite{vegas} and TIMELY~\cite{timely}) as the congestion signal, and successfully delivered compelling performance in terms of high-throughput and low-latency~\cite{dctcp, dcqcn, vegas, timely, bbr, copa}. Unfortunately, due to the discrepancies between DCN and WAN, none of existing solutions designed for DCN or WAN works well for a cross-DC network (\S\ref{sec:problem}). Even worse, it is unlikely, if not impossible, that they can be easily extended to work well.

The fundamental reason is that these solutions only exploit one of the signals (either ECN or delay), which suffices for a relatively homogeneous environment. However, by their nature, ECN or delay alone cannot handle heterogeneity. First, ECN is difficult to configure to meet requirements of mixed flows. The inter-DC and intra-DC flows coexist in cross-DC network, with RTTs varying by up to 1000$\times$. Small RTT flows require lower ECN thresholds for low latency; while large RTT flows require larger ones for high throughput. In fact, tuning ECN threshold may not work, because DC switch shallow buffers can be easily overwhelmed by bursty large-BDP cross-DC traffic. For example, DCN can account for 4$-$20$\times$ more packet losses than WAN in experiments under realistic workload (see Table~\ref{tab:drop1}). Moreover, ECN may not be well supported in WAN.

Meanwhile, delay signal, by itself, is limiting in {\em simultaneously} detecting congestion in WAN and DCN. Cross-DC flows may congest either in WAN or DCN, while delay signal cannot distinguish them given its end-to-end nature. This leads to a dilemma of either under-utilizing WAN (deep-buffered) links with small delay thresholds or increasing DCN (shallow-buffered) packet losses with higher thresholds. For example, Vegas, when scaling its default parameters by 20, achieves 1.5$\times$ higher throughput at the cost of $>\,$30$\times$ more intra-DC packet losses. Furthermore, low delay thresholds impose harsh requirements on accurate delay measurement~\cite{timely}, for which hardware support is needed.

The above problems call for a new synergy that considers not just one of, but both ECN and delay signals in congestion control for cross-DC network communications. Specifically, the new solution must be able to handle the following key challenges (\S\ref{sec:design_rationale}) that have not been exposed to any of prior works: 
(1) How to achieve persistent low latency in the heterogeneous environment, even if DC switches (more likely to drop packet) and WAN routers (more likely to accumulate large buffering) have vastly different buffer depths. 
(2) How to maintain high throughput for inter-DC traffic with shallow-buffered DC switches, even if the propagation delay is in tens of milliseconds range, instead of $<\,$250$\,\mu$s assumed by DCN transport protocols such as DCTCP.
(3) How to achieve ideal RTT-fairness between intra-DC and inter-DC flows, even if the RTTs differ by several orders of magnitude (\eg, 100$\,$ms inter-DC vs. 100$\,\mu$s intra-DC).

Toward this end, we present \sys to organically integrate ECN and delay through the following three main ideas (\S\ref{sec:detailed_design}) to combat the above two challenges:
\begin{icompact}
\item
\textit{Integrating ECN and delay signals for congestion detection.}
Delay signal is leveraged to bound the total in-flight traffic over the entire network path including the WAN segment, while ECN signal is used to control the per-hop queue inside DCN. With bounded end-to-end latency and limited packet losses, persistent low latency is guaranteed.
\item
\textit{Modulating the ECN-triggered window reduction aggressiveness by the RTT of a flow.}
Unlike conventional TCPs that drain queues more for larger RTT flows, we make large RTT flows decrease rates more gently, resulting in smoother ``sawtooth'' window dynamics. This, in turn, prevents bandwidth under-utilization of inter-DC traffic, while sustaining low ECN threshold for intra-DC traffic.
\item
\textit{Adapting to RTT variation in window increase.}
We scale the additive window increase step in proportion to RTT, which better balances the convergence speed and system stability under mixed inter-DC and intra-DC traffic.
\end{icompact}

Finally, we show the superior performance of \sys with theoretical analysis (\S\ref{sec-gemini:analysis}) as well as extensive testbed experiments (\S\ref{sec-gemini:evaluation}). We implement \sys with Linux kernel 4.9.25 and commodity switches. We show that \sys achieves up to 49\% higher throughput compared to DCTCP under DCN congestion, and up to 87\% lower RTT compared to Cubic under WAN congestion; converges to bandwidth fair-sharing point in a quick and stable manner regardless of different RTTs; and delivers persistent low flow completion times (FCT)---up to 53\%, 31\%, 76\% and 2\% reduction of small flow average completion times, and up to 34\%, 39\%, 9\% and 58\% reduction of large flow average completion times compared to TCP Cubic, DCTCP, BBR, and TCP Vegas.
Furthermore, \sys requires no customized hardware support and can be readily deployed in practice.

\section{Background and Motivation} \label{sec-gemini:motivation}

We show heterogeneity of cross-DC networks in \S\ref{sec:background}, and demonstrate transport performance impairments in \S\ref{sec:problem}.

\subsection{Heterogeneity in Cross-Datacenter Networks} \label{sec:background}
The real-world cross-datacenter networks present heterogeneous characteristics in the following aspects:

\begin{table*}[!t]
\centering
\resizebox{\textwidth}{!}{%
\begin{tabular}{|c|c|c|c|c|c|}
\hline
                           & \multicolumn{3}{c|}{DCN}                                 \\ \hline
Switch$\,$/$\,$Router              & Arista 7010T & Arista 7050T & Arista 7050QX \\ \hline
Capacity (ports$\times$BW) & 48$\times$1Gbps & 48$\times$10Gbps  & 32$\times$40Gbps \\ \hline
Total buffer size          & 4$\,$MB            & 9$\,$MB             & 12$\,$MB              \\ \hline
Buffer over Capacity   & 85$\,$KB            & 19.2$\,$KB              & 9.6$\,$KB           \\ \hline
\end{tabular}%
}

\vspace{1em}

\resizebox{\textwidth}{!}{%
\begin{tabular}{|c|c|c|c|c|c|}
\hline
                           & \multicolumn{2}{c|}{WAN}                                                     \\ \hline
Switch$\,$/$\,$Router              & Arista 7504R                         & Arista 7516R                          \\ \hline
Capacity (ports$\times$BW) & 576$\times$10$\,$Gbps/144$\times$100Gbps & 2304$\times$10Gbps/576$\times$100Gbps \\ \hline
Total buffer size          & 96$\,$GB                                 & 384$\,$GB                                 \\ \hline
Buffer over Capacity   & 16.7$\,$/$\,$6.7$\,$MB                           & 16.7$\,$/$\,$6.7$\,$MB                            \\ \hline
\end{tabular}%
}
\caption{Buffer size for commodity DCN switches and WAN routers.}
\label{tab:devices}
\end{table*}

\parab{Heterogeneous networking devices.} A cross-DC network consists of heterogeneous networking devices (\eg, with distinct buffer depths) from intra-DC network (DCN) and inter-DC WAN.  Table~\ref{tab:devices} gives a survey of switches or routers~\cite{switch-arista} commonly used in DCN and WAN. DCN switches have shallow buffers, up to tens of kilobytes per port per Gbps. In contrast, WAN routers adopt deep buffers, up to tens of megabytes per port per Gbps.

\begin{figure}[!t]
\centering
\includegraphics[width=0.8\textwidth]{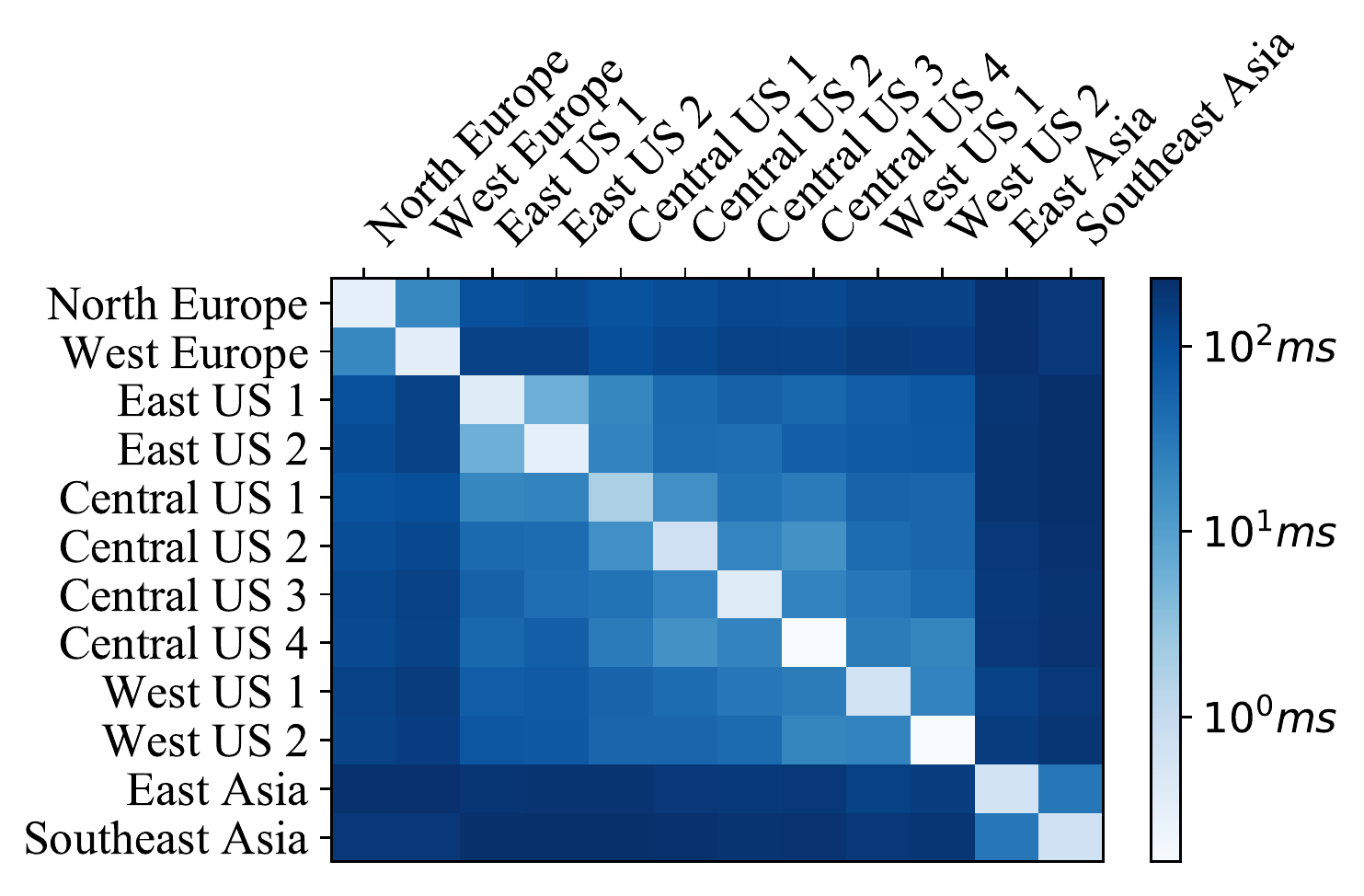}
\caption{RTT Heat Map in Cross-DC Network.}
\label{fig:rttmap}
\end{figure}

\parab{Mixed intra-DC and inter-DC traffic.}
Intra-DC and inter-DC traffic coexists in the cross-DC network~\cite{roy2015inside, jupiter}. They exhibit very different RTTs. To demonstrate this, we conduct RTT measurements on one of the major cloud platforms with 12 representative DCs across the globe. Figure~\ref{fig:rttmap} shows the result. The intra-DC RTTs are as small as hundreds of microseconds. In contrast, the inter-DC RTTs vary from several milliseconds to hundreds of milliseconds. 

\parab{Different administrative control.} Cloud operators have full control over DCN, but do not always control the WAN devices. This is because many cloud operators lease the network resource (\eg, guaranteed bandwidth) from Internet service providers (ISPs) and WAN gears are maintained by the ISPs.
As a result, some switch features, \eg, ECN, may not be well supported~\cite{bauer2011measuring, kuhlewind2014using} (either disabled or configured with undesirable marking thresholds) in WAN.



The heterogeneity imposes great challenges in transport design. Ideally, transport protocols should take congestion location (buffer depth), traffic type (RTT) and supported mechanism (\eg, ECN) into consideration. We show how prior designs are impacted without considering the heterogeneity in the following subsection (\S\ref{sec:problem}).

\subsection{Single Signal's Limitations under Heterogeneity}
\label{sec:problem}

Most of the existing transport protocols~\cite{dctcp, dcqcn, vegas, timely, bbr, copa} use either ECN or delay as the congestion signal. While they may work well in either DCN or WAN, we find that ECN or delay alone cannot handle heterogeneity. We conduct extensive experiments to study the performance impairments of leveraging ECN or delay signal alone in cross-DC networks.

\begin{figure}
\centering
\includegraphics[width=0.8\textwidth]{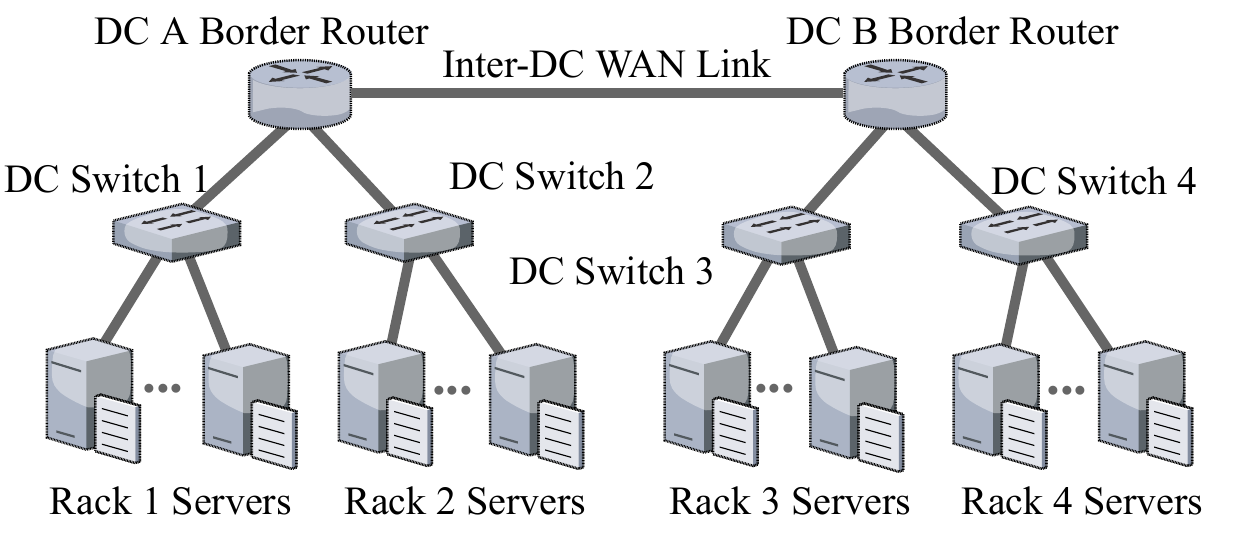}
\caption{Cross-Datacenter Network Testbed.}
\label{fig:testbed}
\end{figure}

\noindent \textbf{Testbed:}
We build a testbed (Figure~\ref{fig:testbed}) that emulates 2 DCs connected by an inter-DC WAN link. Each DC has 1 border router, 2 DC switches and 24 servers. All links have 1$\,$Gbps capacity. The intra-DC and inter-DC base RTTs (without queueing) are $\sim\,$200$\,\mu$s and $\sim\,$10$\,$ms \footnote{Our DC border routers are emulated by servers with multiple NICs, so that we can use \textsc{netem}~\cite{netem} to emulate inter-DC propagation delay.}, respectively. The maximum per-port buffer size of DC switch and border router are $\sim$450 and 10,000 1.5$\,$KB-MTU-sized packets, respectively. 

\noindent \textbf{Schemes Experimented:}
Instead of enumerating every transport protocol, we select several transport solutions that are representative for their own category based on the congestion signal and are readily deployable with solid Linux kernel implementation. Specifically, we experiment Cubic~\cite{cubic}, Vegas~\cite{vegas}, BBR~\cite{bbr} and DCTCP~\cite{dctcp}. Cubic is experimented with and without ECN. ECN threshold at DC switches is set to 300 packets\footnote{We have tuned the ECN threshold for both DCTCP and Cubic. The selected ECN marking threshold achieves the best throughput. A higher one leads to higher loss rate and thus lower throughput. A lower one also results in lower throughput due to frequent congestion notification.} to guarantee high throughput for inter-DC traffic (as suggested by Figure~\ref{fig:ecntradeoff_10ms}). ECN is not enabled in the WAN segment. Vegas uses two parameters $\alpha$ and $\beta$ to control the lower and upper bound of excessive packets in flight. We experiment the default setting (\(\alpha=2, \beta=4\)) and scaled by 10 settings (\(\alpha=20, \beta=40\)).

\begin{figure*}[h]
\centering
\subfigure[Small Flow - Average]
{
\centering
\includegraphics[width=0.45\textwidth]{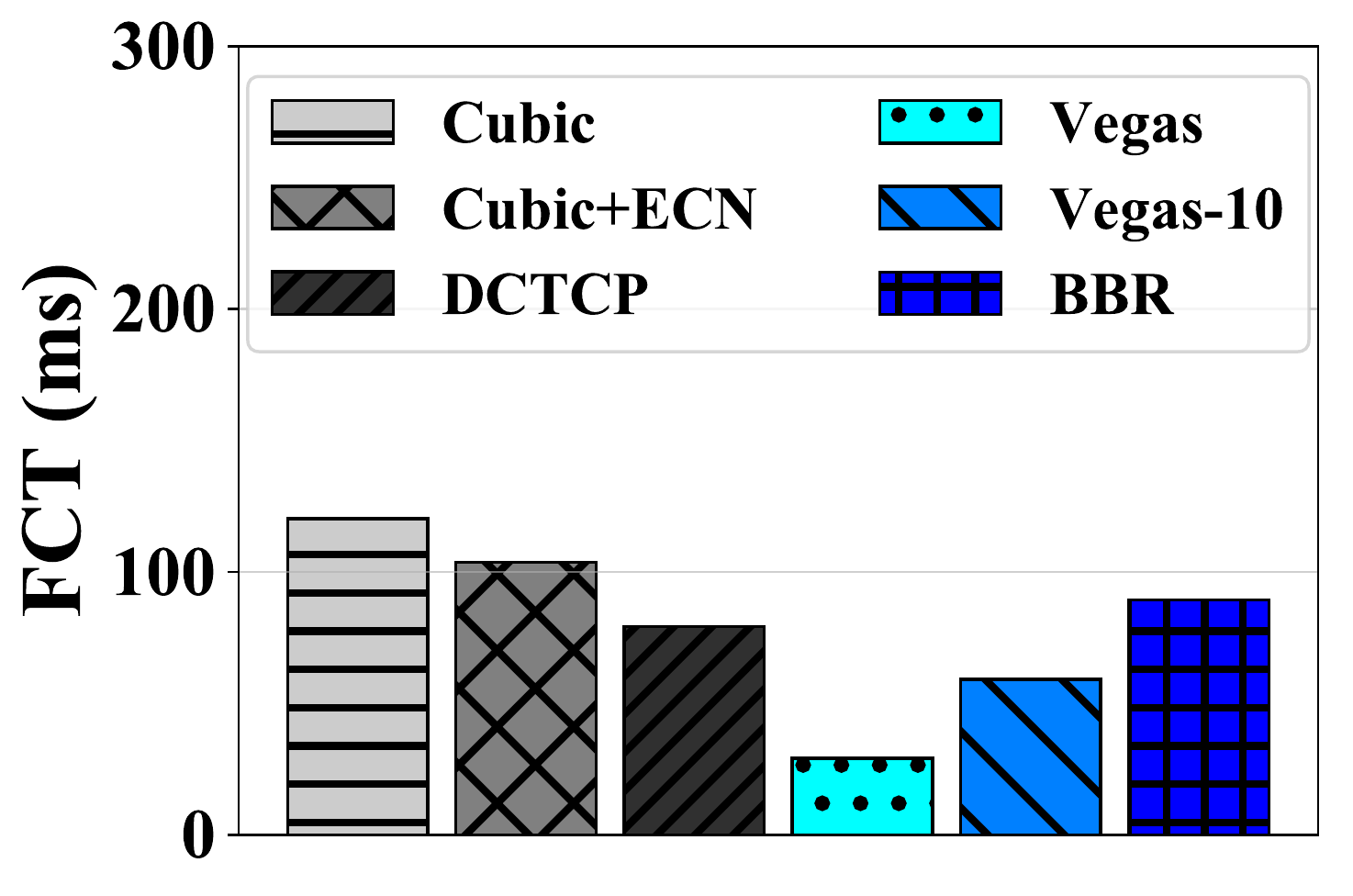}
\label{fig:interdcsmallave}
}
\subfigure[Small Flow - 99th Tail]
{
\centering
\includegraphics[width=0.45\textwidth]{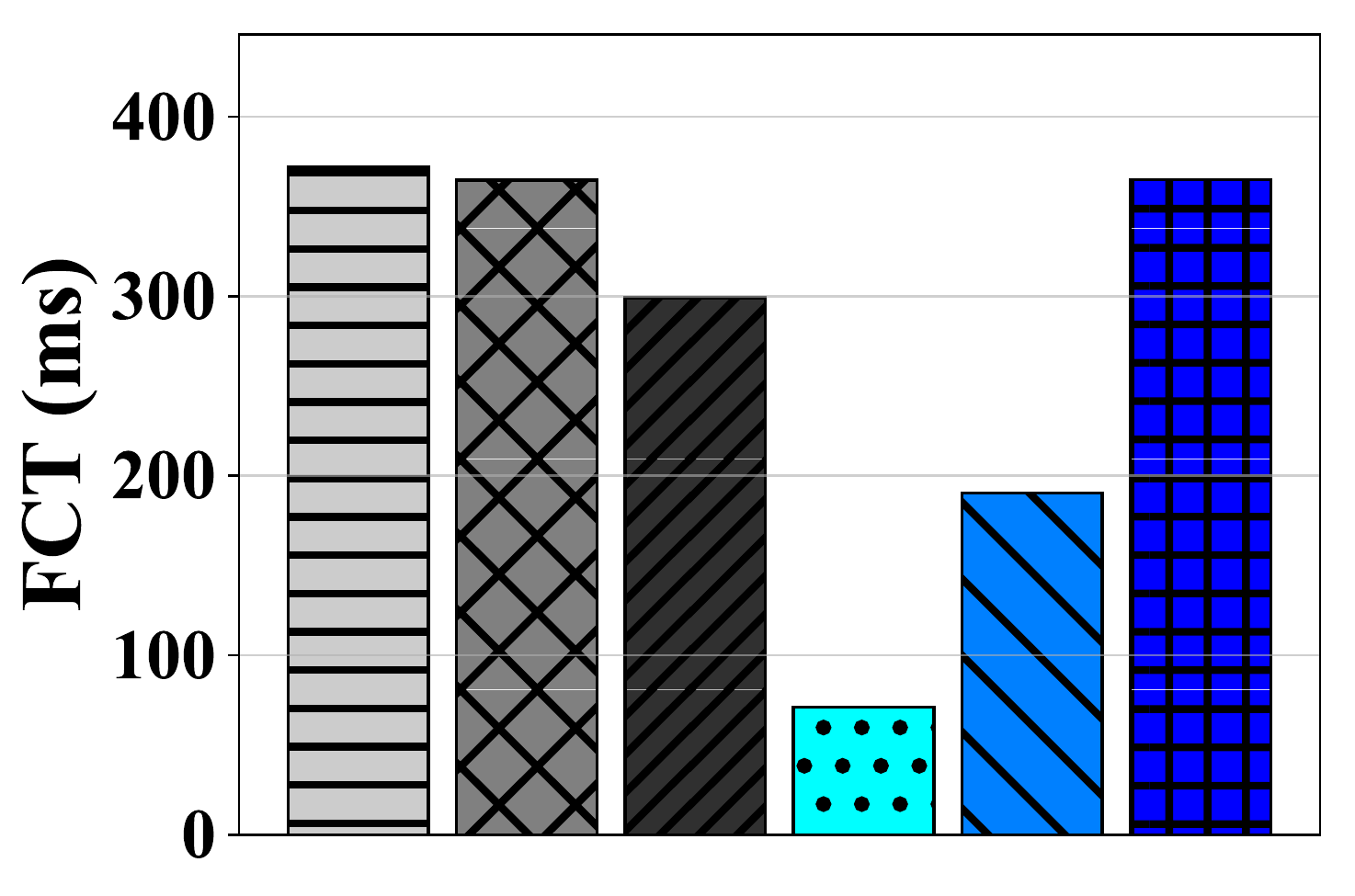}
\label{fig:interdcsmalltail}
}
\subfigure[Large Flow - Average]
{
\centering
\includegraphics[width=0.45\textwidth]{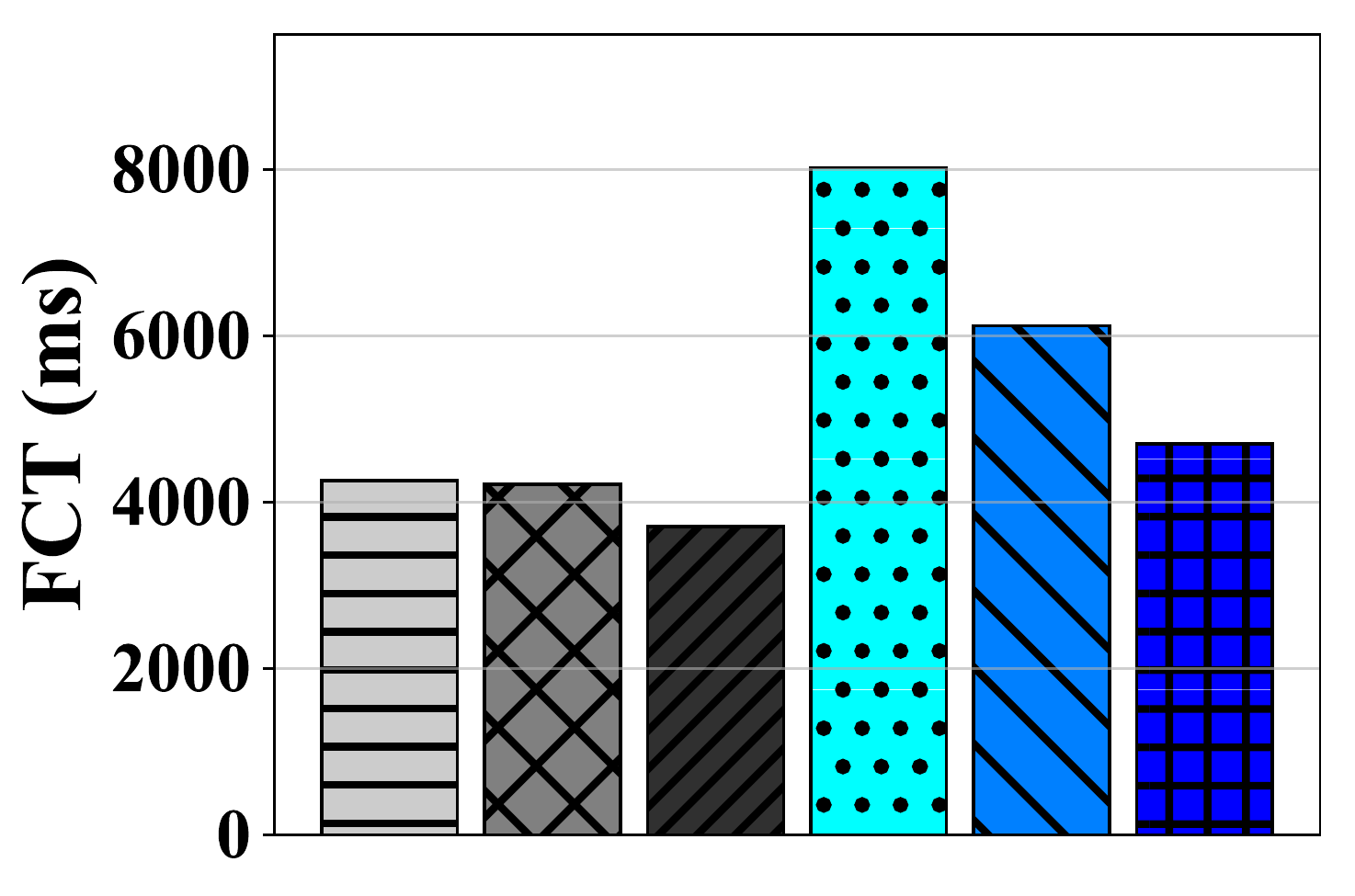}
\label{fig:interdclarge}
}
\subfigure[All Flow - Average]
{
\centering
\includegraphics[width=0.45\textwidth]{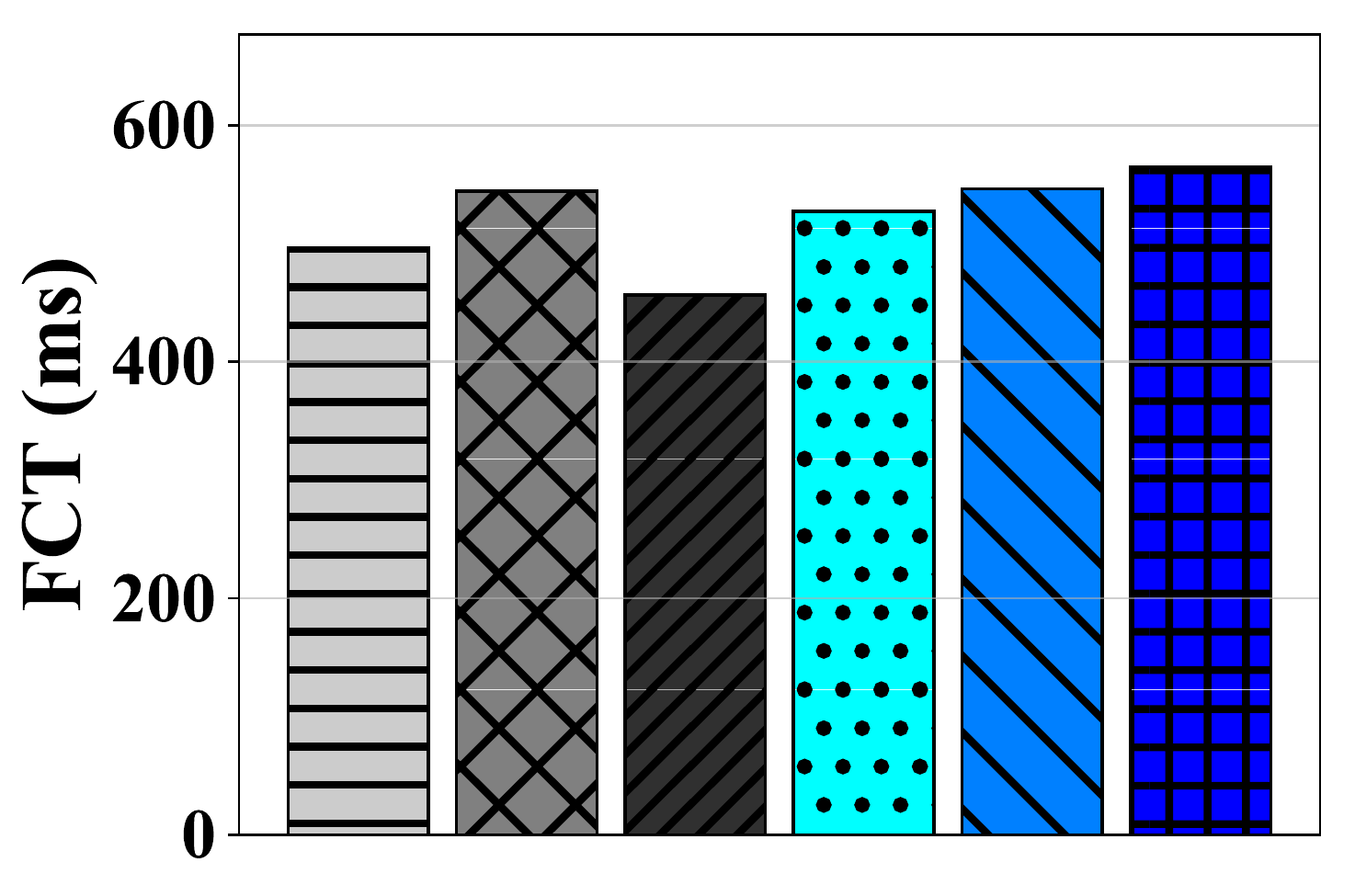}
\label{fig:interdcaverage}
}
\caption{Flow completion time (FCT) results. Small flow: Size $<\,$100$\,$KB. Large flow: Size $>\,$10$\,$MB.}
\label{fig:interdc}
\end{figure*}

\begin{figure}[t]
\subfigure[Intra-DC Flows: {\em RTT}=200$\,\mu$s]
{
\includegraphics[width=0.45\textwidth]{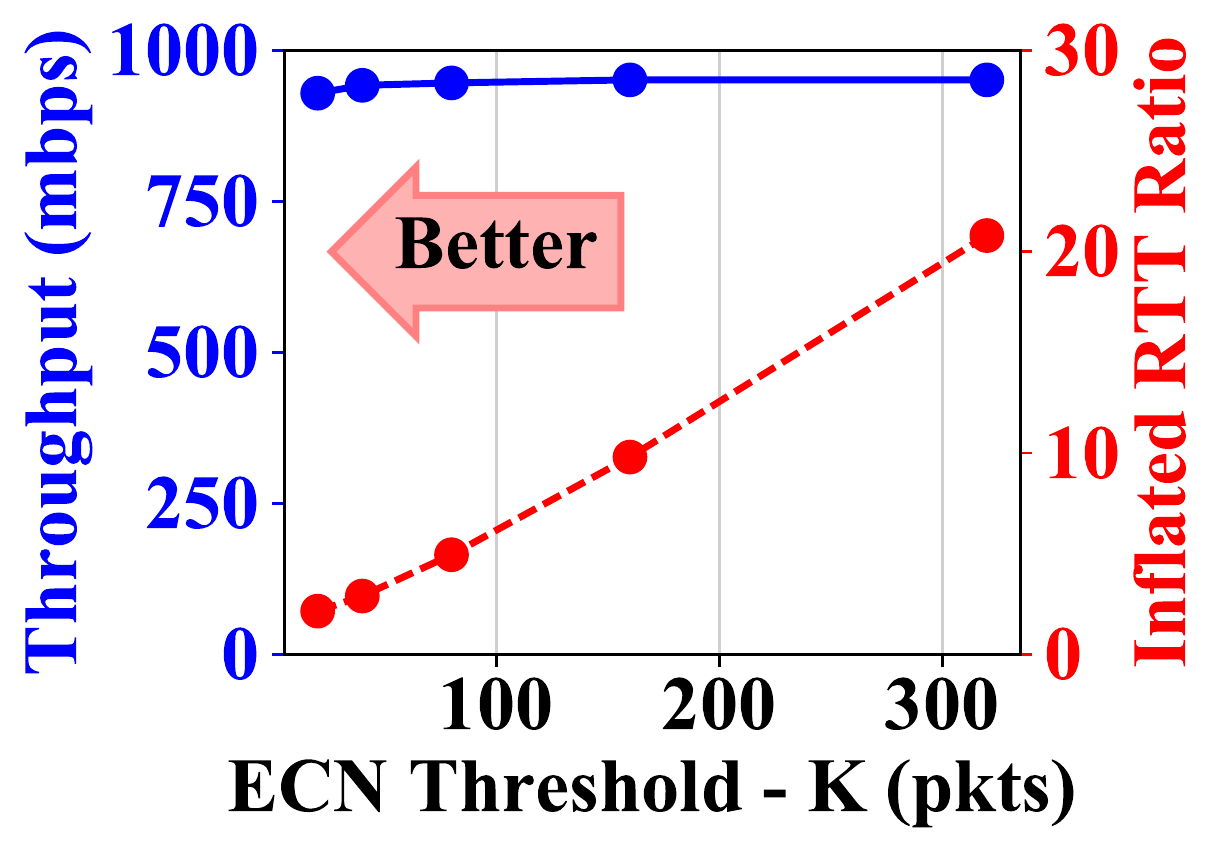}
\label{fig:ecntradeoff_200us}
}
\subfigure[Inter-DC Flows: {\em RTT}=10$\,$ms]
{
\includegraphics[width=0.45\textwidth]{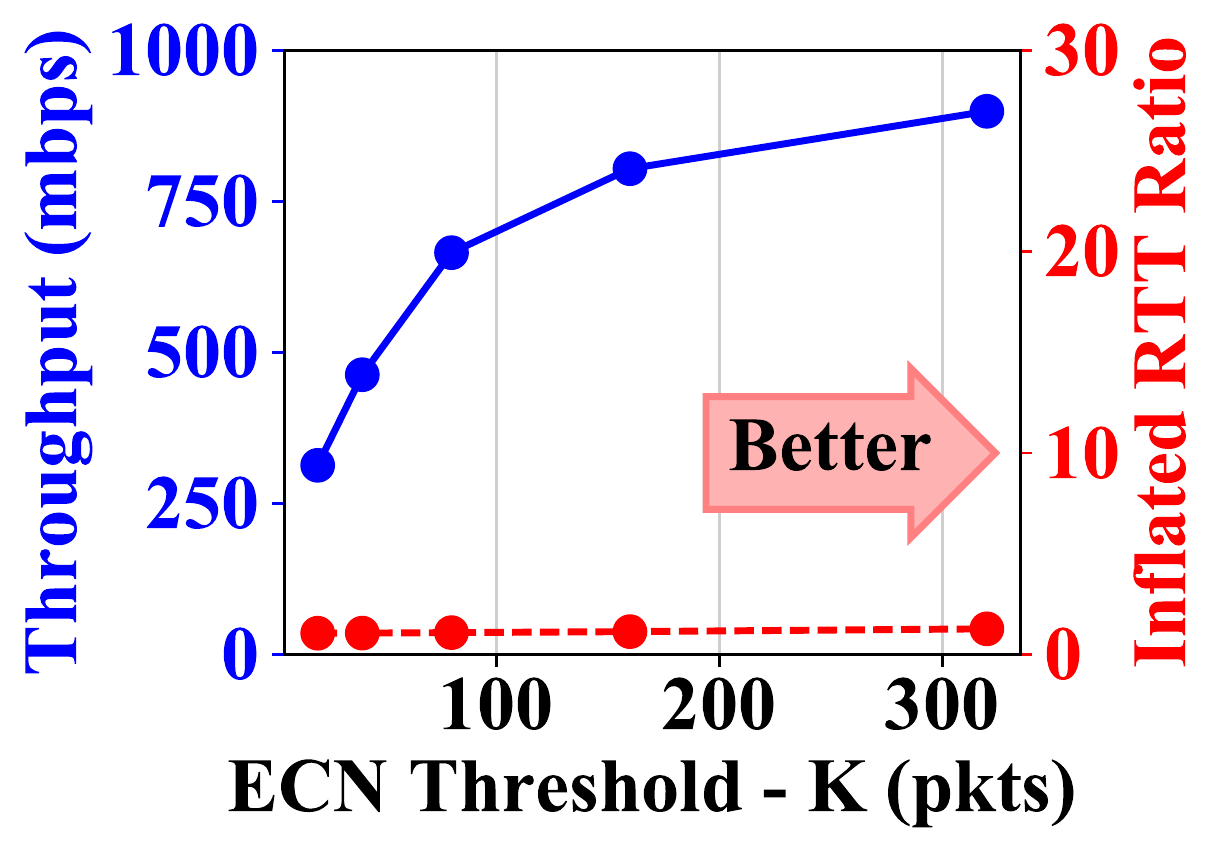}
\label{fig:ecntradeoff_10ms}
}
\centering
\caption{Conflicting ECN requirements in DCTCP. The right y-axis shows latency by the inflated RTT ratio --- the queueing-inflated RTT normalized by the base RTT (w/o queueing).}
\end{figure}

We run realistic workload based on a production trace of web search~\cite{alizadeh2010dctcp}. All flows cross the inter-DC WAN link. The average utilization of the inter-DC and intra-DC links are $\sim$90\% and $\sim$11.25--45\%.
The flow completion time (FCT) results are shown in Figure~\ref{fig:interdc}. We make the following observations and claims, and elaborate them later in the section:

\begin{icompact}
\item Transport protocols based on loss or ECN signal only (\eg, Cubic, Cubic$\,$+$\,$ECN and DCTCP) perform poorly in small flow FCTs (Figure~\ref{fig:interdcsmallave} and~\ref{fig:interdcsmalltail}). This is because they experience high packet losses in shallow-buffered DCN (Table~\ref{tab:drop1}) and large queueing delay without ECN in WAN. We further find that configuring ECN threshold is fundamentally difficult under \textit{mixed traffic}.

\item Transport protocols based on delay signal, when using small thresholds (\eg, Vegas), achieve good performance for small flows (Figure~\ref{fig:interdcsmallave} and~\ref{fig:interdcsmalltail}) at the cost of slowing down large flows (Figure~\ref{fig:interdclarge}). In contrast, when using large thresholds (\eg, Vegas with the scaled by 10 parameters), they greatly degrade the performance of small flows. We further demonstrate the dilemma on setting delay thresholds under \textit{distinct buffer depths}.

\item BBR suffers from high packet loss rates ($>\,$0.1\%), leading to poor small flow FCTs (Figure~\ref{fig:interdcsmallave} and~\ref{fig:interdcsmalltail}).
BBR requires precise estimates of available bandwidth and RTT, which is difficult to achieve under dynamic workload. For example, the bandwidth probing based on a multiple-phase cycle (8 RTTs by default) may not catch up with bursty traffic quickly enough to avoid buffer overflow. And it does not explicitly react to loss signal, leading to continuous losses until the mismatched estimation expires.
\end{icompact}

\parab{Problems of ECN-signal-only solutions.}
ECN-based transport uses the ECN signal~\cite{ramakrishnan1990binary} that often reflects the exceeding queue length at the congested network link.
For it to deliver high throughput, switch should not mark ECN until queue length reaches the bandwidth-delay product (BDP) of the network path\footnote{BDP is an attribute of a network path calculated by multiplying the bottleneck link bandwidth and the zero-queueing round-trip delay. Thus, BDP varies for flows of different network paths.} or a constant fraction of it~\cite{buffer-sizing, dctcp, alizadeh2011analysis}.
\textit{However, in a cross-DC setting, it is difficult to configure the marking parameters due to the large difference in RTT among different paths and divergent requirements imposed by intra-DC and inter-DC flows.} Intra-DC flows impose small buffer pressure but have stringent latency requirement (\eg, hundreds of microseconds).
In contrast, inter-DC flows have looser latency requirement given the large base latency of WAN, instead require large buffer space for high WAN utilization. 

To demonstrate the problem, we generate incast flows from hosts in the same rack to a remote server using DCTCP. We perform two experiments in this setting. In the first experiment, we choose a destination server in the same DC so there are intra-DC flows only. In the second experiment, we choose a destination server in a remote DC so there are inter-DC flows only. In both cases, the bottleneck link is at the source DC switch due to the incast traffic pattern. We vary the ECN marking threshold of the bottleneck switch between 20, 40, 80, 160, and 320 packets per port.

Figure~\ref{fig:ecntradeoff_200us} and~\ref{fig:ecntradeoff_10ms} show the throughput and latency results of intra-DC and inter-DC flows, respectively. From Figure~\ref{fig:ecntradeoff_200us}, we observe a small threshold is desirable to achieve low latency for intra-DC flows. In contrast, from Figure~\ref{fig:ecntradeoff_10ms}, we observe inter-DC flows require a high threshold for high throughput. Clearly, there is a conflict: one cannot achieve high throughput and low latency simultaneously for both inter-DC and intra-DC flows in the cross-DC network.

In fact, achieving high utilization over cross-DC is non-trivial because intra-DC switches have shallow buffers --- the shallow buffer is easily overwhelmed by bursty large-BDP cross-DC flows (we call it \textit{buffer mismatch}). We confirm that by measuring the packet loss rate (PLR) in previous dynamic workload experiments. Table~\ref{tab:drop1} shows the results. We find that packet losses happen within DCN mostly ($>\,$80\%), even though inter-DC WAN is more heavily loaded than intra-DC links. The high losses then lead to low throughput for loss-sensitive protocols. Large-BDP cross-DC traffic is a key factor of the problem. We repeat the same experiments with the inter-DC link delay set to 0. All traffic is now with low BDPs. We observe small PLRs ($<\,$10$\times10^{-5}$) within DCN for all ECN-based schemes this time. Further, we find that naively pacing packets like in BBR cannot completely resolve the problem. For example, Cubic with FQ/pacing~\cite{fq} has similar high PLR (66$\times10^{-5}$) in DCN compared to raw Cubic.

\begin{table}[h]
\centering
\begin{tabularx}{0.55\textwidth}{|Y|Y|Y|}
\hline
\textbf{Cubic} & \textbf{Cubic$\,$+$\,$ECN} & \textbf{DCTCP} \\ \hline
78 / 10         & 24 / 6             & 19 / $<\,$1         \\
\hline
\end{tabularx}
\caption{DCN / WAN Packet Loss Rate (\(10^{-5}\)).}
\label{tab:drop1}
\end{table}

In addition, ECN-based transport protocols require ECN marking support from all network switches. However, ECN marking may not be well supported. It is either disabled or configured with undesirable marking thresholds in WAN (discussed in \S\ref{sec:background}). 
As a result, ECN-based transport such as DCTCP may fall back on using packet loss signal, leading to high packet losses and long queueing delay.

\parab{Problems of delay-signal-only solutions.}
Delay-based transports use the delay signal~\cite{vegas, timely} that reflects the cumulative end-to-end network delay. Typically, they have a threshold to control the total amount of in-flight traffic. \textit{However, given different buffer depths in WAN and DCN, a dilemma arises when setting the delay threshold --- either inter-DC throughput or intra-DC latency is sacrificed.}

Cross-DC flows may face congestion either in WAN or DCN. Delay signal handles both indistinguishablly given its end-to-end nature. On the one hand, if we assume congestion occurs in WAN, the delay thresholds should be large enough (usually in proportion to the BDP) to fully utilize the WAN bandwidth. However, if the bottleneck resides in the DCN instead, the large thresholds (\eg, 10$\,$ms$\,\times\,$1$\,$Gbps = 1.25$\,$MB) can easily exceed the DC switch shallow buffers (\eg, 83$\,$KB per Gbps) and cause frequent packet losses.
On the other hand, if we assume congestion happens in DCN, the delay thresholds should be low enough (at least bounded by the DC switch buffer sizes) to avoid severe intra-DC packet losses. However, if the bottleneck resides in WAN instead, the low thresholds can greatly impair the bandwidth utilization.
In sum, the dilemma of setting delay thresholds arises.

\begin{figure}
    \centering
    \includegraphics[width=0.55\textwidth]{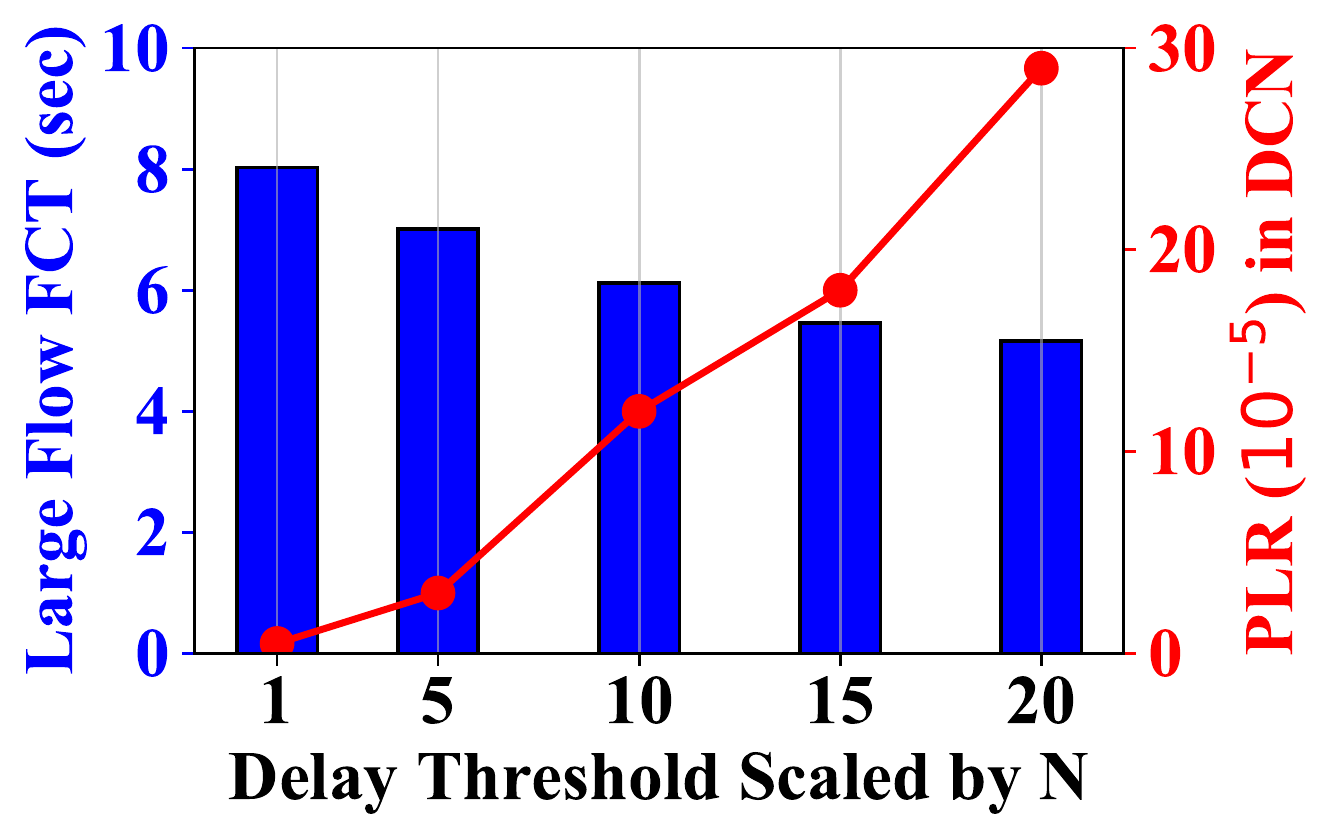}
    \caption{Dilemma in setting delay threshold. The left y-axis shows throughput by the flow completion time (FCT) of large flows. The right y-axis shows packet loss rate (PLR) inside DCN.}
    \label{fig:delay_dilemma}
\end{figure}

To demonstrate the problem, we run the same benchmark workloads used earlier in the section. We experiment Vegas with the default setting (\(\alpha=2, \beta=4\)) and scaled by $N$ settings (\(\alpha=2 \times N, \beta=4 \times N\)), where N is set to 1, 5, 10, 15, 20.
Results are shown in Figure~\ref{fig:delay_dilemma}. On the one hand, small delay thresholds degrade the inter-DC throughput, leading to high average FCT for large flows. On the other hand, large delay thresholds increase packet losses significantly in shallow-buffered DCN. Therefore, setting the delay thresholds are faced with a dilemma of either hurting inter-DC throughput or degrading intra-DC packet loss rate.

In addition, low delay thresholds impose harsh requirement over accurate delay measurement, for which extra device supports (\eg, NIC prompt ACK in~\cite{timely}) are needed.


\section{Design} \label{sec-gemini:design}

We introduce our design rationale in \S\ref{sec:design_rationale}, describe the detailed \sys congestion control algorithm in \S\ref{sec:detailed_design}, and provide guidelines for setting parameters in \S\ref{sec:parameters}.

\subsection{Design Rationale}\label{sec:design_rationale}
\parab{How to achieve persistent low latency in the heterogeneous network environment?}
Persistent low latency implies low end-to-end queueing delay and near zero packet loss. Obviously, ECN, as a per-hop signal, is not a good choice for bounding the end-to-end latency; not to mention, ECN has limited availability in WAN. If we use delay signal alone, small delay threshold is necessary for low loss given the DC switch shallow buffer. However, with a small amount of in-flight traffic, we may not be able to fill the network pipe of the WAN segment (demonstrated in \S\ref{sec:problem}).

Instead of using a single type of signal alone, we integrate ECN and delay signals to address this challenge. In particular, delay signal, given its end-to-end nature, is effectively used to bound the total in-flight traffic; and ECN signal, as a per-hop signal, is leveraged to control the per-hop queues. Aggressive ECN marking is performed at the DC switch to prevent shallow buffer overflow. Thus, the constraint of using small delay thresholds is removed, leaving more space to improve WAN utilization. In this way, the aforementioned dilemma of delay-based transport is naturally resolved.

\parab{How to maintain high throughput for inter-DC traffic in shallow-buffered DCN?}
A majority of transport (\eg, DCTCP) follow additive-increase multiplicative-decrease (AIMD) congestion control rule. The queue length they drain in each window reduction is proportionate to $BDP$ ($C \times RTT$)~\cite{buffer-sizing, dctcp, alizadeh2011analysis}. Essentially, the queue length drained each time should be smaller than the switch buffer size to avoid buffer empty and maintain full throughput. Thus, given large RTT range in cross-DC network, high buffers are required. In deep-buffered WAN, setting a moderately high delay threshold works well to balance throughput and latency. However, in shallow-buffered DCN, aggressive ECN marking is required for low queueing and low loss rate. With limited buffer space, sustaining high throughput gets extremely difficult (demonstrated in \S\ref{sec:problem}).

To address this buffer mismatch challenge, we modulate the aggressiveness of ECN-triggered window reduction by RTT. Maintaining high throughput, in effect, requires large RTT flows to drain queues as small as small RTT flows do during window reduction. Intuitively, we make larger RTT flows reduce rates more gently, thus resulting in smoother ``sawtooth'' window and queue length dynamics. In this way, bandwidth under-utilization can be effectively mitigated, while still using a small ECN marking threshold. The use of small ECN threshold leave enough headroom in the shallow buffer switches because it keeps the average buffer occupancy low, reducing the delay and packet drop.

Further, we adjust the window increase step in proportion to BDP. Conventional AIMD adopts fixed constant window increase step for all flows. This either hurts convergence speed of large-BDP inter-DC flows, or makes the system unstable for small-BDP intra-DC flows. When BDP is large, AIMD requires more RTTs to climb to the peak rate, leading to slower convergence. In contrast, when BDP is small, AIMD may frequently overshoot the bottleneck bandwidth, resulting in more frequent losses and thus less stable performance. Therefore, we adjust the window increase step in proportion to BDP for better robustness under heterogeneity.



\begin{figure}
    \centering
    \includegraphics[width=0.9\textwidth]{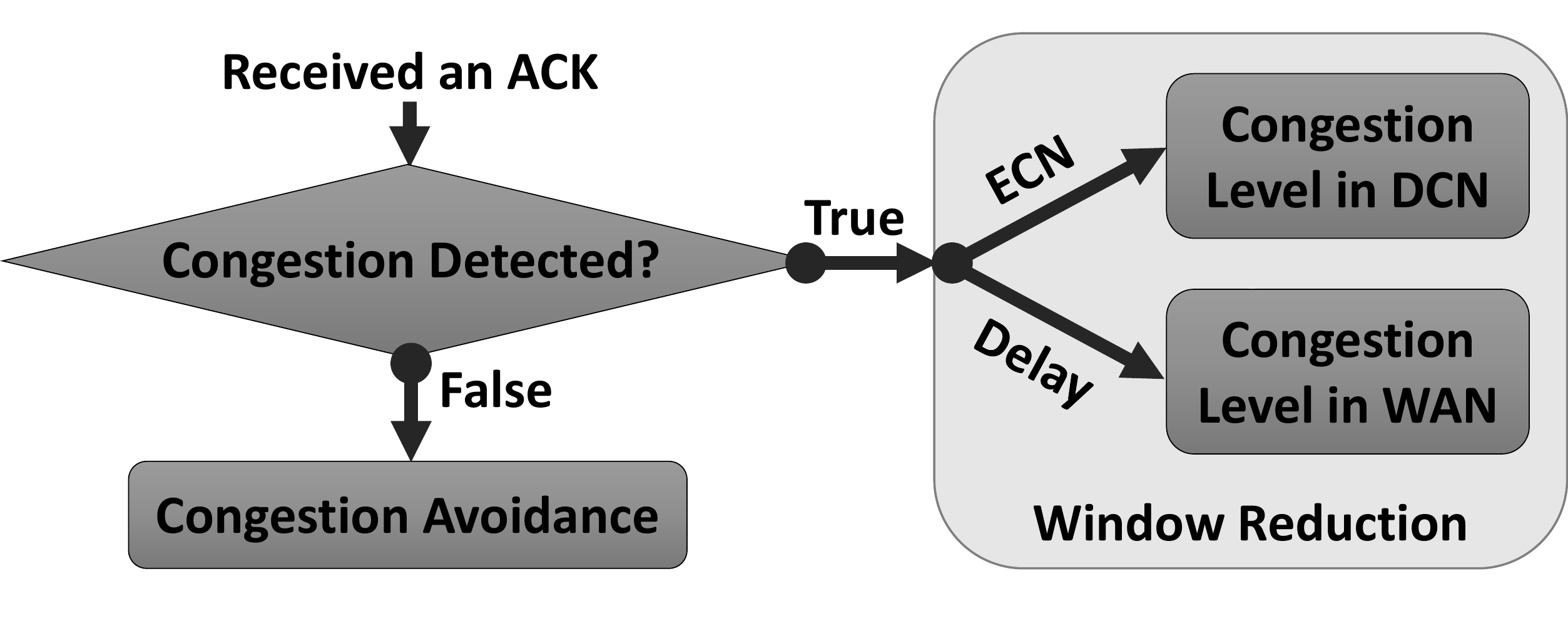}
    \caption{\sys Congestion Control Process.}
    \label{fig:design}
\end{figure}

\subsection{\sys Algorithm}\label{sec:detailed_design}

\sys is a window-based congestion control algorithm that uses additive-increase and multiplicative-decrease (AIMD). Following the design rationale above, \sys leverages both ECN and delay signals for congestion detection. It further adjusts the extent of window reduction as well as growth function based on RTTs of the flows to incorporate heterogeneity.
The \sys algorithm is summarized by flowchart in Figure~\ref{fig:design} and pseudocode in Algorithm~\ref{alg:cc}. Parameters and variables are summarized in Table~\ref{tab:parameter}.

\begin{figure}[htb]
\centering
\begin{minipage}{.8\linewidth}
\begin{algorithm}[H]
\caption[\sys Congestion Control Algorithm.]{\sys Congestion Control Algorithm.}
\label{alg:cc}

\SetKwInOut{Input}{Input}
\SetKwInOut{Output}{Output}

\SetInd{1.1em}{1.1em}

\nonl \tikzmk{A}
\Input{\hspace{0.1em} New Incoming ACK}
\Output{\hspace{0.1em} New Congestion Window Size}
\tikzmk{B} \boxit{mygrey}
\nonl \tikzmk{A} \nl
\tcc{Update transport states (\eg, $\alpha$)}
update\_transport\_state($\alpha$, rtt\_base, rtt\_min) \;
\tcc{When congested, set 1; else 0.}
congested\_dcn \hspace{0.36em} $\leftarrow$ \hspace{0.15em} ecn\_indicated\_congestion() \;
congested\_wan \hspace{0.1em} $\leftarrow$ \hspace{0.15em} rtt\_indicated\_congestion() \;
\tikzmk{B} \boxit{myblue}
\nonl \tikzmk{A} \nl
\eIf{congested\_dcn~$||$~congested\_wan}{
 	\If{time since last cwnd reduction $>$ 1$\,$RTT }{
 	  F \hspace{2.05em} $\leftarrow$ \hspace{0.15em} 4 $\times$ k / (c $\times$ rtt\_base + k) \;
       f\_dcn \hspace{0.3em} $\leftarrow$ \hspace{0.15em} $\alpha$ $\times$ F $\times$ congested\_dcn \;
       f\_wan \hspace{0.1em} $\leftarrow$ \hspace{0.15em} $\beta$ $\times$ congested\_wan \;
	  cwnd \hspace{0.45em} $\leftarrow$ \hspace{0.15em} cwnd $\times$ (1 - max(f\_dcn, f\_wan)) \;
} } {
h \hspace{1.8em} $\leftarrow$ \hspace{0.15em} H $\times$ c $\times$ rtt\_base \;
cwnd \hspace{0.1em} $\leftarrow$ \hspace{0.15em} cwnd + h / cwnd \;
}
\nonl \tikzmk{B} \boxend{mypink}
\end{algorithm}
\end{minipage}
\end{figure}

\parab{Integrating ECN and delay for congestion detection.} 
The congestion detection mechanism leverages both ECN and delay signals. Delay signal is used to bound the total in-flight traffic in the network pipe. ECN signal is used to control the per-hop queues inside DCN. By integrating ECN and delay signal, low latency can be achieved~\cite{zeng2017ear}. Specifically, DCN congestion is detected by ECN, so as to meet the stringent per-hop queueing control requirement imposed by shallow buffers. WAN congestion is detected by delay, because the end-to-end delay is dominated mostly in WAN than in DCN\footnote{The delay signal cannot exclude the DCN queueing delay. However, DCN queueing is often low due to DCN shallow buffer, much lower than that on WAN. Such a low DCN queueing delay has very limited impact with a relative large delay threshold as shown in Table~\ref{tab:throughput_rtt_noise}.}. 

DCN congestion is indicated by the ECN signal --- the ECN-Echo flag set in the ACKs received by the senders. The ECN signal is generated exactly the same as DCTCP. Data packets are marked with Congestion Experienced (CE) codepoint when instantaneous queueing exceeds marking threshold at the DC switches. Receivers then echo back the ECN marks to senders through ACKs with the ECN-Echo flags. Given shallow-buffered DCN, the ECN signal is leveraged with a small marking threshold for low packet losses.

WAN congestion is indicated by the delay signal --- ACKs returned after data sending with persistent larger delays: $\textit{RTT}_{min} > \textit{RTT}_{base} + T$, where $\textit{RTT}_{min}$ is the minimum RTT observed in previous RTT (window); $\textit{RTT}_{base}$, or simplified as $\textit{RTT}$, is the base RTT (minimum RTT observed during a long time); $T$ is the delay threshold. Inspired by~\cite{nichols2012controlling}, we use $\textit{RTT}_{min}$ instead of average or maximum RTTs, which can better detect persistent queueing and tolerate transient queueing possibly caused by bursty traffic. Given deep-buffered WAN, the delay signal is used with a moderately high threshold for high throughput and bounded end-to-end latency.


When either of the two signals indicate congestion, we react to the signal by reducing the congestion window correspondingly. When both ECN and delay signals indicate congestion, we react to the one of heavier congestion:
\begin{equation*}
CWND = CWND \times (1 - max(f\_dcn, f\_wan))
\end{equation*}
where $f\_dcn$ determines the extent of window reduction for congestion in DCN; and $f\_wan$ determines that of WAN. We show how to compute them later in the section.

\begin{table}[t]
  \centering
  \resizebox{0.8\linewidth}{!}{
  \begin{tabular}{|c|l|}
  \hline
    \textbf{Parameter} & \textbf{Description} \\
    \hline
    $K$ & ECN marking threshold\\
    \hline
    $T$ & Delay threshold\\
    \hline
    $\beta$ & Parameter for window reduction in WAN\\
    \hline
    $H$ & Parameter for congestion window increase\\
    \hline \hline
    \textbf{Variable} & \textbf{Description} \\
    \hline
    $CWND$ &  Congestion window\\
    \hline
    $f\_dcn$ & Extent of window reduction in DCN\\
    \hline
    $f\_wan$ & Extent of window reduction in WAN\\
    \hline
    $\textit{RTT}_{min}$ & Minimum RTT observed in previous RTT\\
     \hline
    $\textit{RTT}_{base}$ & Minimum RTT observed during a long time\\
    \hline
    $\textit{RTT}$ & Simplified notation of $\textit{RTT}_{base}$\\
    \hline
    $C$ & Bandwidth capacity (constant for given network)\\
    \hline
    $\alpha$ & Average fraction of ECN marked packets\\
    \hline
    $F$ & Scale factor for DCN congestion control\\
    \hline
    $h$ & Adaptive congestion window increase step\\
    \hline
  \end{tabular}
  }
  \caption{Parameters and Variables Used in \sys.}
  \label{tab:parameter}
\end{table}

\parab{Modulating the ECN-triggered window reduction aggressiveness by RTT.} 
The window reduction algorithm aims to maintain full bandwidth utilization while reducing the network queueing as much as possible. This essentially requires switch buffer never underflow at the bottleneck link. Given distinct buffer depths, \sys reduces congestion window differently for congestion in DCN and WAN.

In DCN, given shallow buffer, strictly low ECN threshold is used for low packet losses. We adopt the DCTCP algorithm, which works well under the low ECN threshold for the intra-DC flows. However, for large RTT inter-DC flows, the throughput drops greatly. This is because the buffer drained by a flow during window reduction increases with its RTT (\eg, the amplitude of queue size oscillations for DCTCP is $O(\sqrt{C \times \textit{RTT}})$~\cite{dctcp, alizadeh2011analysis}). Larger RTT flows drain queues more and easily empty the switch buffers, leading to low link utilization. Inspired by this, \sys extends DCTCP by modulating the window reduction aggressiveness based on RTT. This guides the design of $f\_dcn$ --- the extent of window reduction when congestion is detected in DCN. When ECN signal indicates congestion, we compute $f\_dcn$ as follows:
\begin{equation*}
f\_dcn = \alpha \times F
\end{equation*}
where $\alpha$ is the exponential weighted moving average (EWMA) fraction of ECN marked packets, $F$ is the factor that modulates the congestion reduction aggressiveness. We derive the scale factor $F=\frac{4K}{C \times \textit{RTT} + K}$ (see Theorem~\ref{theo:f}), where $C$ is the bandwidth capacity (a constant parameter for given network), $\textit{RTT}$ is the minimum RTT observed during a long time, $K$ is the ECN marking threshold. Thus, for intra-DC flows, following the guideline in DCTCP by setting $K=(C \times \textit{RTT})/7$, we have $F=\frac{1}{2}$, exactly matching the DCTCP algorithm. For inter-DC flows with larger RTTs, $F$ gets smaller, leading to smaller window reduction and smoother queue length oscillation.

In WAN, given much deeper buffer, high throughput can be more easily maintained than in DCN. In fact, window reduction based on a fixed constant, like standard TCPs~\cite{tcp1988, cubic} do, is enough for high throughput. There are potentially a wide range of threshold settings to effectively work with (see \S\ref{sec:parameters}). This guides the design of $f\_wan$ --- the extent of window reduction when congestion is detected in WAN. When RTT signal indicates congestion, we compute $f\_wan$ as follows:
\begin{equation*}
f\_wan = \beta
\end{equation*}
where $\beta$ is a window decrease parameter for WAN.

The window reduction is performed no more than once per RTT, which is the minimum time required to get feedback from the network under the new sending rate. Despite the congestion detection by ECN and delay, packet losses and timeouts may still occur. For that, we keep the same fast recovery and fast retransmission mechanism from TCP.


\parab{Window increase that adapts to RTT variation.} 
The congestion avoidance algorithm adapts to RTTs (or BDP when the bandwidth capacity is fixed) to help balance convergence speed and stability. For conventional AIMD, large BDP flows need more RTTs to climb to the peak rate, leading to slow convergence; while small BDP flows may frequently overshoot the bottleneck bandwidth, leading to unstable performance. Therefore, adjusting the window increasing step in proportion to BDP compensates the RTT variation, and makes the system more robust under diverse RTTs.
Further, it also mitigates RTT unfairness~\cite{lakshman1997tcp, brown2000resource}, which in turn helps to improve tail performance.
This leads to the adaptive congestion window increase factor $h$.
When there is no congestion indication, for each ACK,
\begin{equation*}
CWND = CWND + \frac{h}{CWND}
\end{equation*}
$h$ is a congestion avoidance factor in proportion to BDP:
$h = H \times C \times \textit{RTT}$, where $H$ is a constant parameter, $C$ is the bandwidth capacity, $\textit{RTT}$ is the minimum RTT observed during a long time. We prove that factor $h$ together with the scale factor $F$ guarantees bandwidth fair-sharing regardless of different RTTs in Appendix \S\ref{sec:rttfairness}.



\parab{Summary.}
\sys resolves the conflicting requirements imposed by network heterogeneity naturally by integrating ECN and delay signal, specifically, (1) in face of \textit{distinct buffer depths}, \sys handles congestion in WAN and DCN by delay and ECN signal respectively, simultaneously meeting the need of strictly low latency in DCN and high bandwidth utilization in WAN; (2) in face of \textit{mixed traffic} with large range of RTTs in shallow-buffered DCN, \sys maintains high throughput by modulating the window reduction aggressiveness based on RTTs. In particular, large RTT flows reduce the windows more gently, effectively avoiding buffer empty and bandwidth under-utilization. This is achieved by scale factor $F$, which guarantees full throughput under limited buffer space or small ECN threshold at steady state (see Theorem~\ref{theo:f} with detailed proof). Besides, Gemini adapts its window increase step in proportion to the RTT, achieving faster convergence speed and better fairness.
Further, window growth function is also adapted to RTTs. This leads to faster convergence when more bandwidths are available, especially for those large RTT flows.

\section{Theoretical Analysis} \label{sec-gemini:analysis}

\subsection{Derivation of the Scale Factor F} \label{sec:deductionf}
We analyze the steady state behavior and prove that \sys achieves full throughput with scale factor $F = \frac{4K}{C \times \textit{RTT} + K}$.

\begin{theorem}	\label{theo:f}
	Given a positive ECN marking threshold $K$, we can maintain 100\% throughput under DCN congestion if congestion window is reduced as follows,
	\begin{equation*}
	 CWND = CWND \times (1 - \alpha \times F)
	\end{equation*}
where $\alpha$ is the EWMA of ECN fraction and \(F \leq \frac{4K}{C \times \textit{RTT} + K}\).
\end{theorem}

\begin{figure}
\centering
\includegraphics[width=0.8\textwidth]{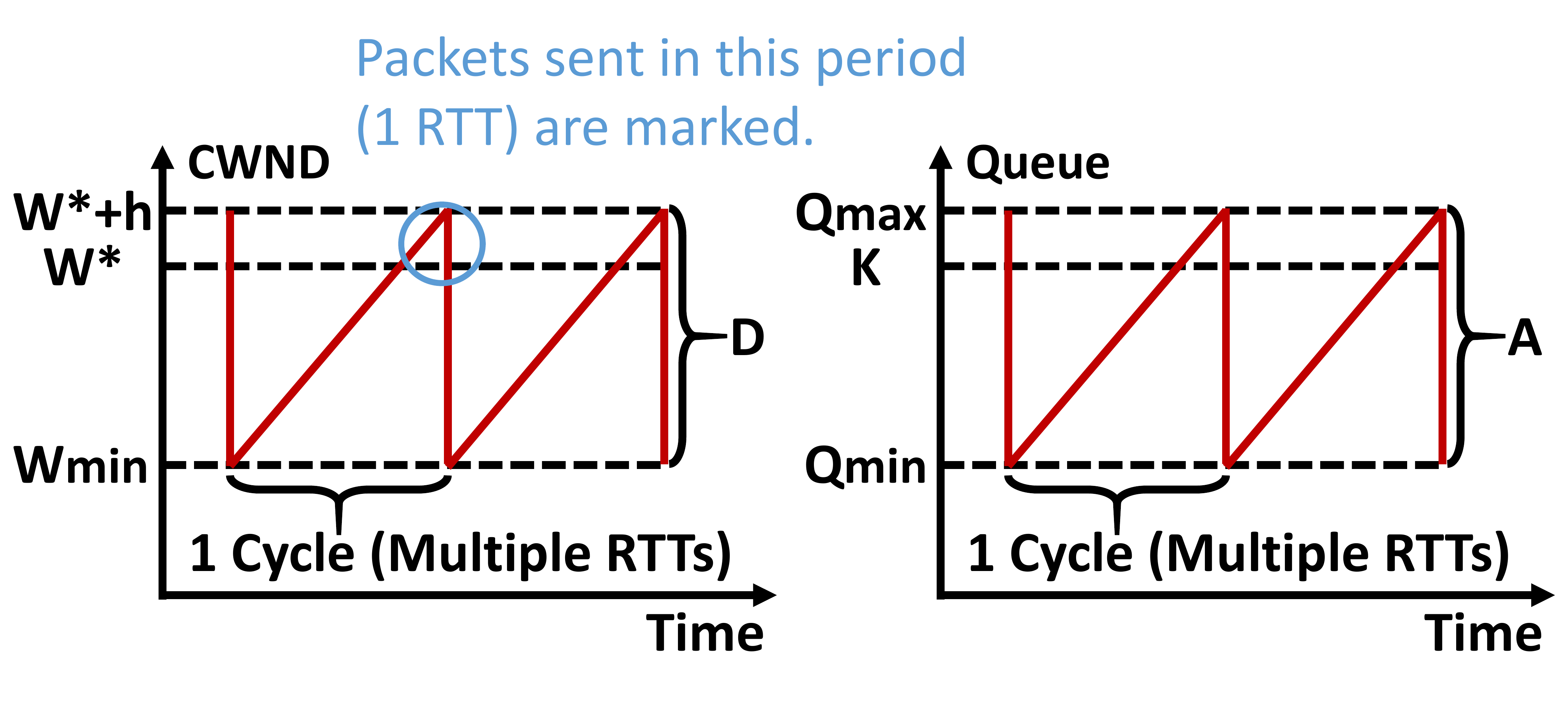}
\caption{AIMD Sawtooth Illustration.}
\label{fig:aimd}
\end{figure}

\emph{Proof:} Similar to prior work~\cite{dctcp, copa}, we assume all $N$ flows are long-lived, have identical round-trip times $RTT$, and share the same bottleneck link of capacity $C$. Assuming N window sizes are synchronized for the ease of analysis, the queue size is:
\begin{equation} \label{equ:q}
Q(t) = N \times W(t) - C \times \textit{RTT}
\end{equation}
where $W(t)$ is the dynamic window size. Therefore, the queueing dynamic also follows a similar sawtooth pattern as the window size. To achieve full link utilization, we need to guarantee: $Q_{min}\geq 0$ (see Figure~\ref{fig:aimd}).

The queueing will get higher than the marking threshold K and packets will get marked for exactly one RTT in each sawtooth cycle. Therefore, the fraction of marked packets, $\alpha$, can be calculated by dividing the packets sent in the last RTT over the packets sent in one sawtooth cycle.

For each sender, we use \(S(W_1, W_2)\) to represent the packets sent when the window changes from \(W_1\) to \(W_2 > W_1\). It takes \((W_2 - W_1) / h\) round trip times, during which the average window size is \((W_1 + W_2)/2\),
\begin{equation} \label{equ:s}
S(W_1, W_2) = (W_2^2 - W_1^2)/2h
\end{equation}

We use \(W^* = (C \times RTT + K)/N\) to represent the largest window that leads to queueing of K and kicks off ECN marking on the incoming packets. For the last RTT before the sender reacts, the window size peaks at \(W^* + h\). We have,
\begin{equation} \label{equ:alpha}
\alpha = S(W^*, W^*+h)/S((W^*+h)(1-\alpha F), W^*+h)
\end{equation}

Combining (\ref{equ:s}) and (\ref{equ:alpha}) and rearranging, we get:
\begin{equation} \label{equ:alphatow}
\alpha^2F(2-\alpha F) = (2W^*+h)h/(W^*+h)^2 \approx 2h/W^*
\end{equation}

We assume $\alpha F/2$ is small for approximation:
\begin{equation} \label{equ:alphatow2}
\alpha \approx \sqrt{h/FW^*}
\end{equation}

Therefore, the queueing amplitude $A$ in Figure~\ref{fig:aimd} can be obtained (N flows):
\begin{equation} \label{equ:d}
D = (W^*+h) - (W^*+h)(1-\alpha F) = (W^*+h)\alpha F
\end{equation}

\begin{equation} \label{equ:a}
\begin{split}
A = N \times D = N(W^*+h)\alpha F \approx N \sqrt{hFW^*} \\
= \sqrt{NhF(C \times \textit{RTT} + K)}
\end{split}
\end{equation}

With (\ref{equ:q}), we have:
\begin{equation} \label{equ:qmax}
Q_{max} = N \times (W^*+h) - C \times \textit{RTT} = K + Nh
\end{equation}

With (\ref{equ:a}) and (\ref{equ:qmax}), the minimum queue length is:
\begin{equation} \label{equ:qmin}
\begin{split}
Q_{min} = Q_{max} - A = K + Nh - \sqrt{NhF(C \times \textit{RTT} + K)}
\end{split}
\end{equation}

Finally, to find the relationship between the scale factor $F$ and the ECN marking threshold $K$, we minimize (\ref{equ:qmin}) over $N$ so that the value is no smaller than zero (\ie, no network under-utilization). We have:
\begin{equation} \label{equ:f}
F \leq \frac{4K}{C \times \textit{RTT} + K}
\end{equation}
As we can see, given a fixed ECN marking threshold $K$, the larger RTT a flow has, the smaller $F$ it gets. Therefore, the flows with larger RTTs adjust window more smoothly to achieve high throughput.

Note that the theoretical analysis here is a generalized form of that in the DCTCP paper~\cite{dctcp} and the result is consistent with it. Specifically, when following the DCTCP algorithm by setting a constant parameter $F=\frac{1}{2}$, we have $K \geq (C \times \textit{RTT})/7$, exactly matching the original DCTCP guideline.

\subsection{Proof of RTT-fairness} \label{sec:rttfairness}

We show \sys achieves fair-share of the bottleneck bandwidth in DCN where inter-DC and intra-DC flows coexist.
\begin{theorem} \label{theo:rtt}
\sys achieves ideal RTT-fairness with following AIMD rule:
\begin{icompact}
\item[] \textbf{Decrease:} When congestion indicated by ECN per RTT,
\begin{equation*}
CWND = CWND \times (1 - \alpha \times F)
\end{equation*}
where $\alpha$ is the ECN fraction and \(F = \frac{4K}{C \times \textit{RTT} + K}\).
\item[] \textbf{Increase:} When there is no congestion indication per ACK,
\begin{equation*}
CWND = CWND + \frac{h}{CWND}
\end{equation*}
where $h$ is an adaptive congestion avoidance function in proportion to BDP: \(h \propto \textit{RTT}\).
\end{icompact}
\end{theorem}

\parab{Proof:} From previous subsection, we know that the average window size is:
\begin{equation} \label{equ:averagew}
\overline{W} = \frac{W^*+h + (W^*+h) \times (1-\alpha F)}{2}
\end{equation}

Therefore, when two flows competing for one bottleneck link reach the steady state:
\begin{equation} \label{equ:fracw}
\frac{\overline{W_1}}{\overline{W_2}} = \frac{(W_1^*+h_1) \times (1-\frac{\alpha_1 F_1}{2})}{(W_2^*+h_2) \times (1-\frac{\alpha_2 F_2}{2})} \approx \frac{W_1^*}{W_2^*}
\end{equation}
when assuming that \(1>>\frac{\alpha F}{2}\) and \(W^* >> h\).

When two flows F1 and F2 with different RTTs (assuming \(\textit{RTT}_1 < \textit{RTT}_2\)) are competing on one bottleneck link, Equation \ref{equ:alphatow2} is still valid for the small RTT flow F1, in other form:
\begin{equation} \label{equ:wtoalpha}
W_1^* = 2h_1/(F_1 \alpha_1^2)
\end{equation}

However, Equation \ref{equ:wtoalpha} does not hold for large RTT flow F2. When small RTT flow F1 reduces its CWND as soon as it gets the ECN feedback after $RTT_1$, the bottleneck queue length drops immediately and packets of large RTT flow F2 will stop being marked with ECN. So flow F2 will get only around \(S(W_2^*, W_2^*+h_2) \frac{\textit{RTT}_1}{\textit{RTT}_2}\) packets marked with ECN. Following same approach from Equation \ref{equ:alpha} to \ref{equ:wtoalpha}, for F2,
\begin{equation} \label{equ:wtoalpha2}
W_2^* = 2h_2/(F_2 \alpha_2^2) \times \frac{\textit{RTT}_1}{\textit{RTT}_2 }
\end{equation}

Packets traversing the same link have the same probability to be ECN marked.
Thus, we get:
\begin{equation} \label{equ:alphaequal}
\alpha_1 = \alpha_2
\end{equation}

Plugging Equation \ref{equ:f}, \ref{equ:wtoalpha}, \ref{equ:wtoalpha2}, \ref{equ:alphaequal} into Equation \ref{equ:fracw}, we have:
\begin{equation} \label{equ:fracw2}
\frac{\overline{W_1}}{\overline{W_2}} = \frac{F_2}{F_1} = \frac{C \times \textit{RTT}_1 + K}{C \times \textit{RTT}_2 + K}
\end{equation}

When assuming the average queue length is around K. We have the average RTT:
\begin{equation} \label{equ:avertt}
\overline{\textit{RTT}} \approx \textit{RTT}+\frac{K}{C}
\end{equation}

Therefore, we have the bandwidth sharing ratio:
\begin{equation} \label{equ:frackr}
\frac{\overline{R_1}}{\overline{R_2}} = \frac{\overline{W_1}}{\overline{\textit{RTT}_1}} / \frac{\overline{W_2}}{\overline{\textit{RTT}_2}} \approx 1
\end{equation}
where $R_i$ denotes the sending rate of flow $i$. 

\subsection{Guidelines for Setting Parameters} \label{sec:parameters}
Default \sys parameter settings are shown in Table~\ref{tab:parametersetting}. We adopt the default parameter settings throughout all our experiments unless otherwise specified. We provide the following rules of thumbs for setting the parameters, but leave finding the optimal threshold settings to the future work.

\begin{table}[H]
\centering
\begin{tabularx}{0.5\textwidth}{|Y|Y|}
\hline
Parameter   & Default Value              \\ \hline
$K$         & $50~pkts\,/\,Gbps$      \\ \hline
$T$         & $5~ms$                     \\ \hline
$\beta$     & $0.2$                    \\ \hline
$H$ & $1.2 \times 10^{-7}$ \\ \hline
\end{tabularx}
\caption{Default \sys Parameter Settings.}
\label{tab:parametersetting}
\end{table}

\parab{ECN Marking Threshold ($K$).}
The scaling factor $F$ ensures full link utilization given an ECN threshold ($K$). As a lower $K$ indicates a smaller queue, setting $K$ as low as possible may seem desirable. However, there is actually a trade-off here. When $K$ is small, the scaling factor $F$ is also small, making the flows reduce their congestion window slowly, leading to slower convergence.
Therefore, we recommend a moderately small threshold of 50 packets per Gbps. In addition, to mitigate the effect of packet bursts (especially for large BDP inter-DC traffic), we use a per-flow rate limiter at the sender to evenly pace out each packet.

\parab{Queueing Delay Threshold ($T$).}
 $T$ should be sufficiently large to achieve high throughput in the cross-DC pipe. It should also leave enough room to filter out the interference from the DCN queueing delay. In practice (\S\ref{sec:background}), RTTs (include queueing) in production DCNs are at most 1ms. We recommend to set \(T = 5\,ms\) that is higher enough to remove the potential DCN queueing interference (see Table~\ref{tab:throughput_rtt_noise}).

\parab{Window Decrease Parameter ($\beta$).}
\sys reduces the window size by $\beta$ multiplicatively when WAN congestion is detected. To avoid bandwidth under-utilization, we need to have queueing headroom \(T > \frac{\beta}{1-\beta}\textit{RTT}\), or \(\beta < \frac{T}{T+\textit{RTT}}\) based on the buffer sizing theory~\cite{buffer-sizing}. Thus, we have $\beta < 0.33$, assuming $\textit{RTT}=10ms$ and $T=5ms$. We recommend to set \(\beta = 0.2\) (the same reduction factor as Cubic and Vegas) for smoother 'sawtooth'. In practice, this is stricter than necessary as competing flows are often desynchronized~\cite{buffer-sizing}. We show that the recommended $T$ and $\beta$ settings can well serve the cross-DC networks in a wide range of RTTs in \S\ref{sec:static_exp}.

\parab{Window Increase Parameter ($H$).}
In congestion avoidance phase, \sys grows its congestion window size by $h$ MSS every RTT. In our implementation, we actually scale $h$ with BDP (\(C \times \textit{RTT}\)) instead of RTT only, that is, \(h = H \times C \times \textit{RTT}\). This is reasonable as large BDP means potentially large window size. Scaling $h$ with BDP achieves better balance between convergence speed and stability.
We recommend to set \(H = 1.2 \times 10^{-7}\) with bounded minimum/maximum increase speed of 0.1$\,$/$\,$5 respectively as a protection. This leads to $h=1$ when $C=1Gbps$ and $\textit{RTT}=8ms$, a middle ground between large BDP inter-DC traffic and low BDP intra-DC traffic.

\section{Evaluation} \label{sec-gemini:evaluation}

In this section, we present the detailed \sys Linux kernel implementation and evaluation setup in \S\ref{sec:exp_setup}, and conduct extensive experiments to answer the following questions:

\begin{icompact}
	\item[]
\hspace{-1em}\textbf{\S\ref{sec:static_exp} Does \sys achieve high throughput and low latency?} We show that \sys achieves higher throughput (1$-$1.5$\times$) and equally low delay compared to DCTCP under DCN congestion; lower delay ($>\,$7$\times$) and equally high throughput compared to Cubic under WAN congestion.
	\item[]
\hspace{-1em}\textbf{\S\ref{sec:convergence_exp} Does \sys converge quickly, fairly and stably?} In static traffic experiments, we show that \sys converges to the bandwidth fair-sharing point quickly and stably under both DCN congestion and WAN congestion, regardless of distinct RTTs differed by up to 64 times.
	\item[]
\hspace{-1em}\textbf{\S\ref{sec:realistic_exp} How does \sys perform under realistic workload?} In realistic traffic experiments, we show that under both cases (intra-DC heavy or inter-DC heavy traffic pattern), \sys persistently achieves the one of the best flow completion times for both short and large flows.
\end{icompact}


\subsection{Implementation and Experiment Setup} \label{sec:exp_setup}
\parab{\sys Implementation:}
\sys is developed based on Linux kernel 4.9.25. Linux TCP stack has a universal congestion control interface defined in struct $tcp\_congestion\_ops$, which supports various pluggable congestion control modules. The congestion window reduction algorithm is implemented in $in\_ack\_event()$ and $ssthresh()$. The congestion avoidance algorithm is implemented in $cong\_avoid()$.

\parab{Testbed:}
Experiments are conducted in 2 testbeds with 1Gbps and 10Gbps capacity respectively. The 1Gbps testbed has a larger scale than the 10Gbps one. Both testbeds share the same topology as shown in Figure~\ref{fig:testbed}. There are 2 data centers connected by an inter-DC WAN link. Each data center has one border router, two DC switches and multiple servers. Border routers are emulated by servers with multiple NICs, so that we can use \textsc{netem}~\cite{netem} to emulate WAN propagation delay. Intra-DC (under single ToR) and inter-DC base RTTs are $\sim\,$200$\,\mu$s and $\sim\,$10$\,$ms, respectively. Dynamic buffer allocation~\cite{dt} at the DC switches is enabled like most operators do in real deployments to absorb bursts.
\begin{icompact}
\item\textit{Large-scale 1Gbps Testbed:}  There are 50 Dell PowerEdge R320 servers and 4 Pica8 P-3297 switches. Pica8 P-3297 switches have 4MB buffer shared by 48 ports. The WAN buffer is set to 10,000 1.5$\,$KB-MTU-sized packets per port. All network interfaces are set to 1Gbps full duplex mode.
\item\textit{Small-scale 10Gbps Testbed:} There are 10 HUAWEI RH1288 V2 servers and 1 Mellanox SN2100 switch (divided into multiple VLANs). Mellanox SN2100 switches have 16MB buffer shared by 16 ports. The WAN buffer is set to 80,000 1.5$\,$KB-MTU-sized packets per port. All network interfaces are set to 10Gbps  full duplex mode.
\end{icompact}
Remark: We show results of the large-scale testbed by default.

\parab{Benchmark Workloads:}
We generate realistic workloads based on traffic patterns that have been observed in a data center supporting web search~\cite{dctcp}. Flows arrive by the Poisson process and the source and destination is chosen randomly from a configured IP pool. The workload is heavy-tailed with about 50\% small flows (size $<\,$100$\,$KB) while 95\%  of all bytes belong to the the larger 30\% of the flows of size greater than 1$\,$MB. We run the workload with a publicly available traffic generator that has been used by other work~\cite{clicknp, mqecn}.

\parab{Performance Metrics:}
We measure flow throughput and packet round-trip time in \S\ref{sec:static_exp} to quantify the throughput and latency. We measure throughput in \S\ref{sec:convergence_exp} to demonstrate the convergence, stability and fairness. For realistic experiments in \S\ref{sec:realistic_exp}, similar to prior work~\cite{dctcp,pfabric}, we use flow completion time (FCT) as the main performance metric.

\parab{Schemes Compared:}
We experiment Cubic~\cite{cubic}, Vegas~\cite{vegas}, BBR~\cite{bbr}, DCTCP~\cite{dctcp} and \sys. All these protocols have implementations in Linux kernel TCP and are readily deployable in practice. \textit{Cubic} is the default loss-based congestion control algorithm used in Linux system. It is experimented with and without ECN. \textit{DCTCP} is an ECN-based congestion control algorithm designed to achieve high throughput, low latency and high burst tolerance in DCN. The ECN marking threshold is set to 300 packets to guarantee high throughput for inter-DC traffic. \textit{Vegas} uses two parameters $\alpha$ and $\beta$ to control the lower and upper bound of excessive packets in flight. We experiment the default setting (\(\alpha=2, \beta=4\)) and scaled by 10 settings (\(\alpha=20, \beta=40\)) to show the throughput and latency trade-off. \textit{BBR} is designed primarily for the enterprise WAN. It tries to drive the congestion control to the theoretical optimal point~\cite{kleinrock1979power} with maximized throughput and minimized latency, based on accurate bandwidth and RTT estimation. \textit{\sys} is the transport design proposed in this work. Default \sys parameter settings are shown in Table \ref{tab:parametersetting}. We adopt the default parameter settings throughout all experiments in this work if not specified. The ECN marking is configured only at the DC switches.

\begin{figure}[!t]
\centering
\includegraphics[width=0.6\textwidth]{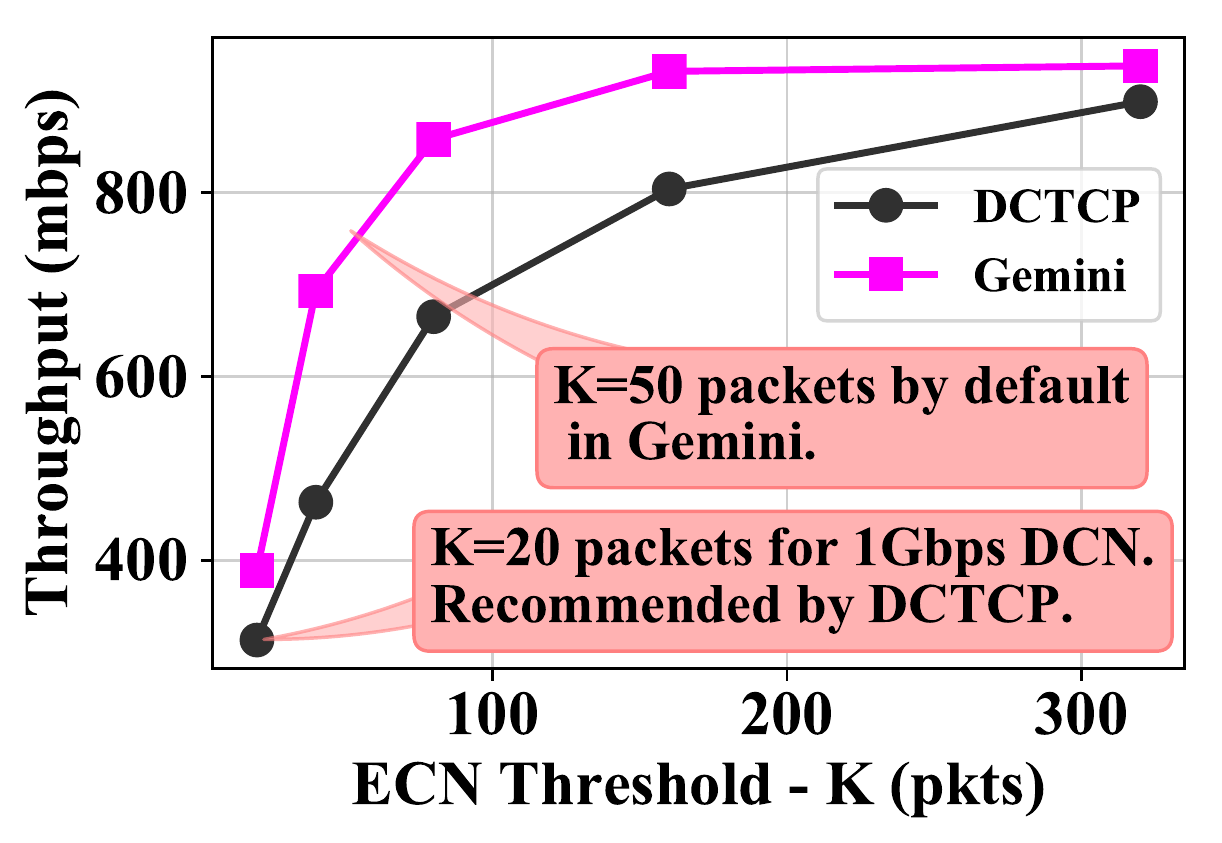}
\caption{Aggregate throughput of inter-DC flows that bottlenecked at a DCN link. \sys is less buffer-hungry (requires 0-76\% smaller $K$) than DCTCP when achieving similar throughput.}
\label{fig:throughput_10ms2}
\end{figure}

\subsection{Throughput and Latency} \label{sec:static_exp}
We show that \sys achieves high throughput and low latency under both DCN congestion and WAN congestion.

\parab{Handling Congestion in DCN.} 
ECN-based DCN congestion control module needs to cope with the mismatch between DC switch shallow buffer and high-BDP inter-DC traffic so as to strike a good balance between latency and throughput. We show that, by adding BDP-aware scale factor $F$, the mismatch issue can be mitigated to a great extent. 

To demonstrate that, we generate many-to-one long flows sharing one DC switch bottleneck link. We perform two experiments, with all intra-DC flows in the first one and all inter-DC flows in the second. The RTT-based WAN congestion control module is disabled here for \sys (the module will not work even if we enable it, because the RTT threshold itself will filter our the DCN congestion). We set the ECN marking threshold $K$ to 20, 40, 80, 160, 320 packets.

Results show that there is little gap between \sys and DCTCP for the intra-DC flows. The average RTTs of inter-DC flows are also similar (so the results are neglected here). The throughputs of inter-DC flows are shown in Figure~\ref{fig:throughput_10ms2}. \sys maintains slightly higher throughput (938$\,$mbps) than DCTCP (899$\,$mbps) when setting $K$ as high as 320 packets. Setting a higher threshold is prohibitive given limited buffer left to avoid packet losses under bursty traffic.

\parab{Handling Congestion in WAN.} 
\sys leverages delay signal for WAN congestion control. To quantify the signal error, we measure RTTs in our testbed under a quiescent scenario. The standard deviation of the measured intra-rack and inter-DC RTT are 17$\,\mu$s and 58$\,\mu$s, respectively. To show how noisy RTT can impact \sys, we add RTT estimation errors deliberately in our \sys kernel module and run many-to-one static flows sharing one bottleneck link in WAN. The random noise is added to each RTT sample with uniform distribution in the range of [$\,$0,$\,$x$\,$]$\,$ms, where x is set to 0, 0.2, 1, 2. The results are listed in Table~\ref{tab:throughput_rtt_noise}. The noise from the variable kernel processing time ($<\,$0.1$\,$ms) has no impact to the \sys throughput. When considering the potential DCN queueing delay interference ($<\,$1$\,$ms), the aggregate throughput of \sys only drops slightly. This verifies that setting $T = 5\,ms$ can well filter out the interference from DCN queueing delay. We also attribute the robustness partially to the design that uses $\textit{RTT}_{min}$ in each window time to detect persistent congestion.

\begin{table}[h]
\centering
\begin{tabularx}{0.6\textwidth}{|Y|c|c|c|c|}
\hline
Avg. RTT Noise (ms) & 0   & 0.1 & 0.5 & 1   \\ \hline
Throughput (Mbps)   & 944 & 944 & 936 & 927 \\ \hline
\end{tabularx}%
\caption{Impact of RTT Noise on Throughput.}
\label{tab:throughput_rtt_noise}
\end{table}

\begin{figure}[!t]
\centering
\includegraphics[width=0.8\textwidth]{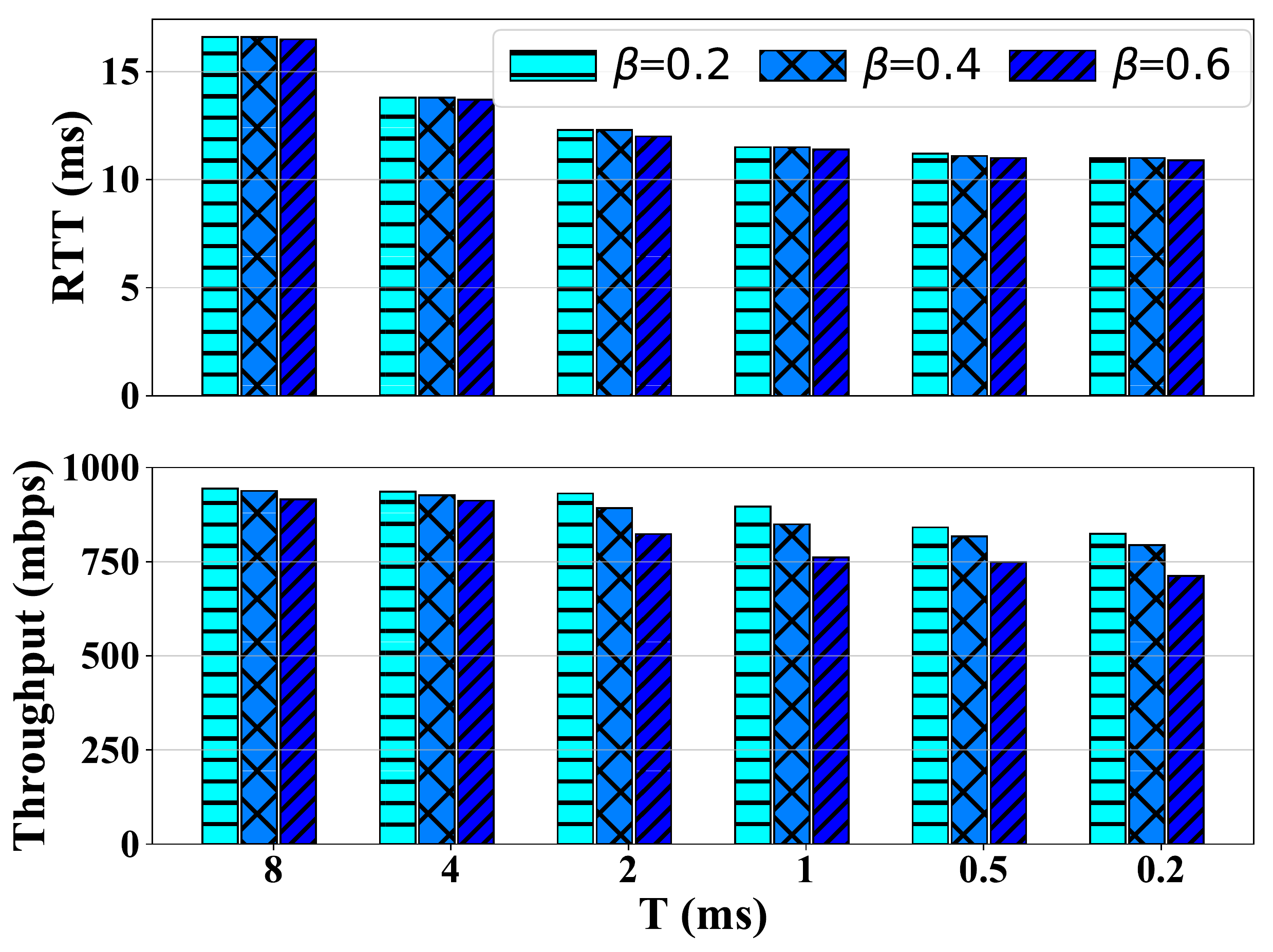}
\caption{RTT and throughput of inter-DC flows bottlenecked at a WAN link.}
\label{fig:t_beta}
\end{figure}

To further demonstrate the effectiveness of the congestion control in WAN, we run many-to-one static flows sharing one bottleneck link in WAN, with varying $T$ and $\beta$ settings. The results are shown in Figure~\ref{fig:t_beta}. In general, \sys maintains near full throughput with average queueing-delayed RTTs of 16$\,$ms (highest when $T=8ms$ and $\beta=0.2$). Compared to the transport protocols that leverage loss signals in WAN, \sys achieves similar high throughput at the cost of much lower latency. For example, in another experiment with same setup, Cubic suffers from 7$\times$ higher average RTTs ($\sim\,$100$\,$ms).

\textbf{Parameter Sensitivity.} 
The \sys performance mainly relies on two parameters, \ie, ECN marking threshold $K$ and queueing delay threshold $T$ for congestion control in DCN and WAN, respectively. We now analyze the performance of \sys under various parameter settings.

Figure~\ref{fig:throughput_10ms2} shows the \sys performance with varying $K$ under DCN congestion. We can see that \sys performs better than DCTCP in a large range of parameter settings. In fact, the \sys throughput is not degraded until the threshold $K$ is set to as low as 100 packets. This means that \sys is less buffer-hungry (requires 0-76\% smaller $K$) than DCTCP when achieving similar throughput, leaving enough space to improve burst tolerance and latency of intra-DC traffic.

Figure~\ref{fig:t_beta} shows the \sys performance with varying $T$ and $\beta$ under WAN congestion. As expected, a lower $T$ leads to lower RTT latency at the cost of slightly decreased throughput. Reducing $\beta$ improves throughput but may hurt convergence speed at the same time. We recommend to set \(T = 5\,ms\) and \(\beta = 0.2\). $T$ that is a  bit higher than needed in this case, leaving more room for higher RTT networks. In practice, this is also necessary to filter out the interference from DCN queueing delay (usually within 1$\,$ms). We repeat the previous experiments under higher RTTs with unchanged parameter settings. Results show that \sys can still achieve 857$\,$mbps throughput (within 10\% of the highest throughput) under 100$\,$ms base RTT. This verifies the default settings can work well under a wide RTT range.

\begin{figure}[!t]
\centering
\hspace{1em}
\begin{minipage}[t]{0.45\textwidth}
\includegraphics[width=\textwidth]{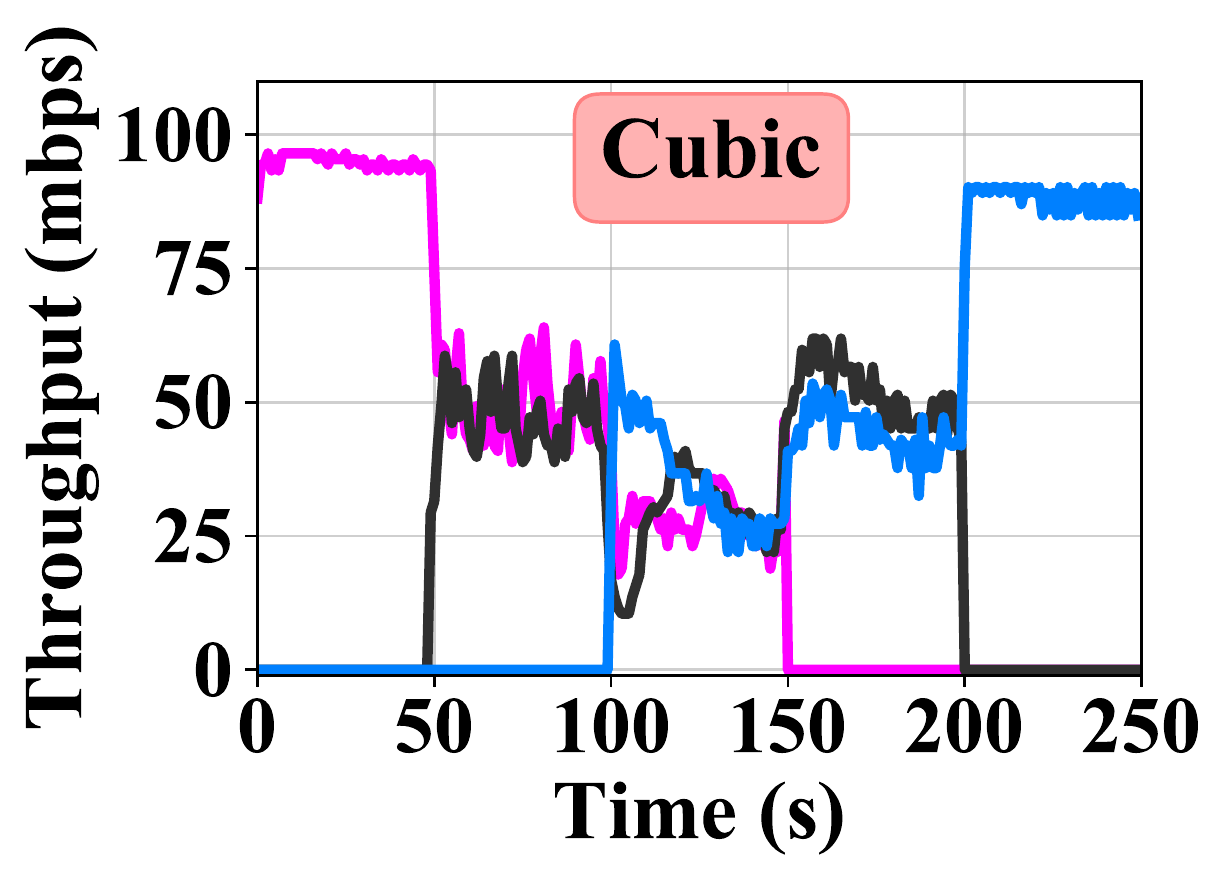}
\end{minipage}
\begin{minipage}[t]{0.45\textwidth}
\includegraphics[width=\textwidth]{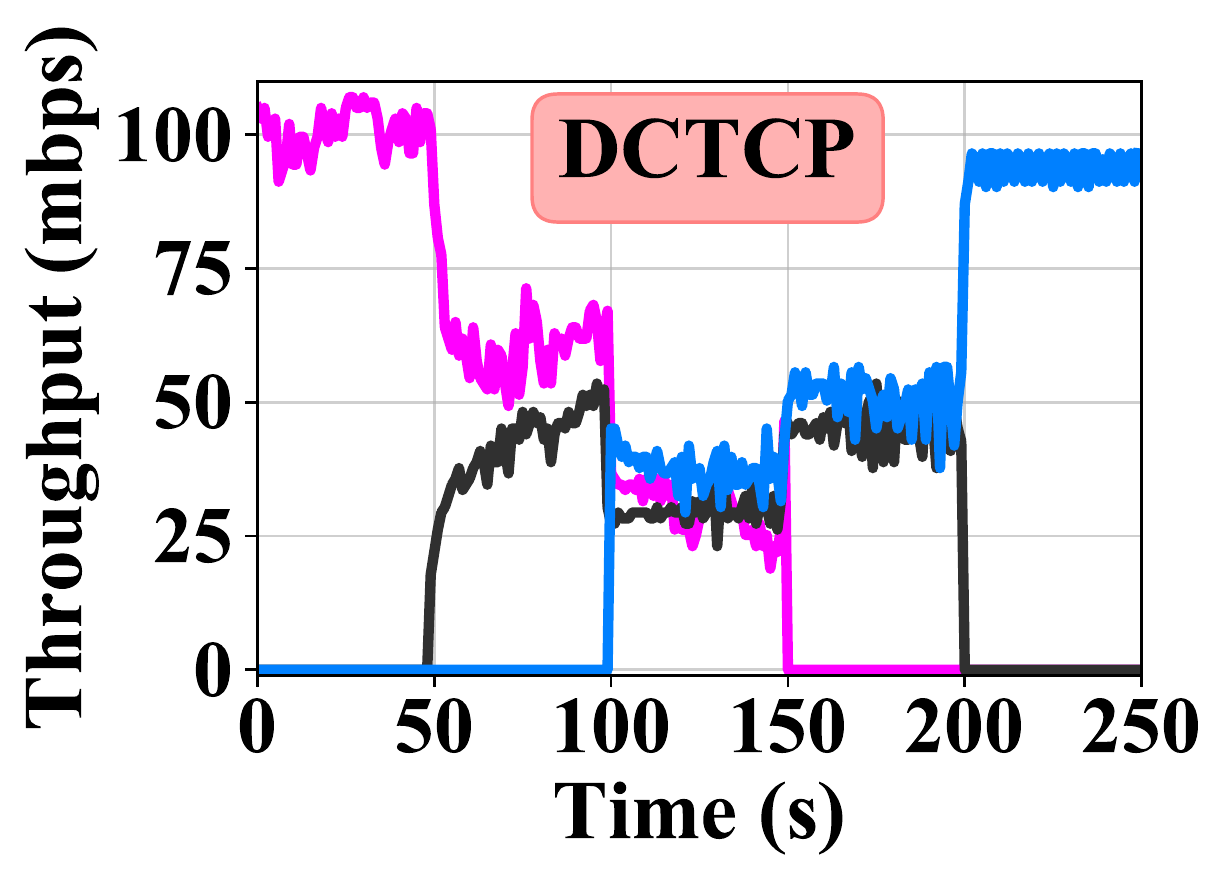}
\end{minipage}
\newline
\begin{minipage}[t]{0.45\textwidth}
\includegraphics[width=\textwidth]{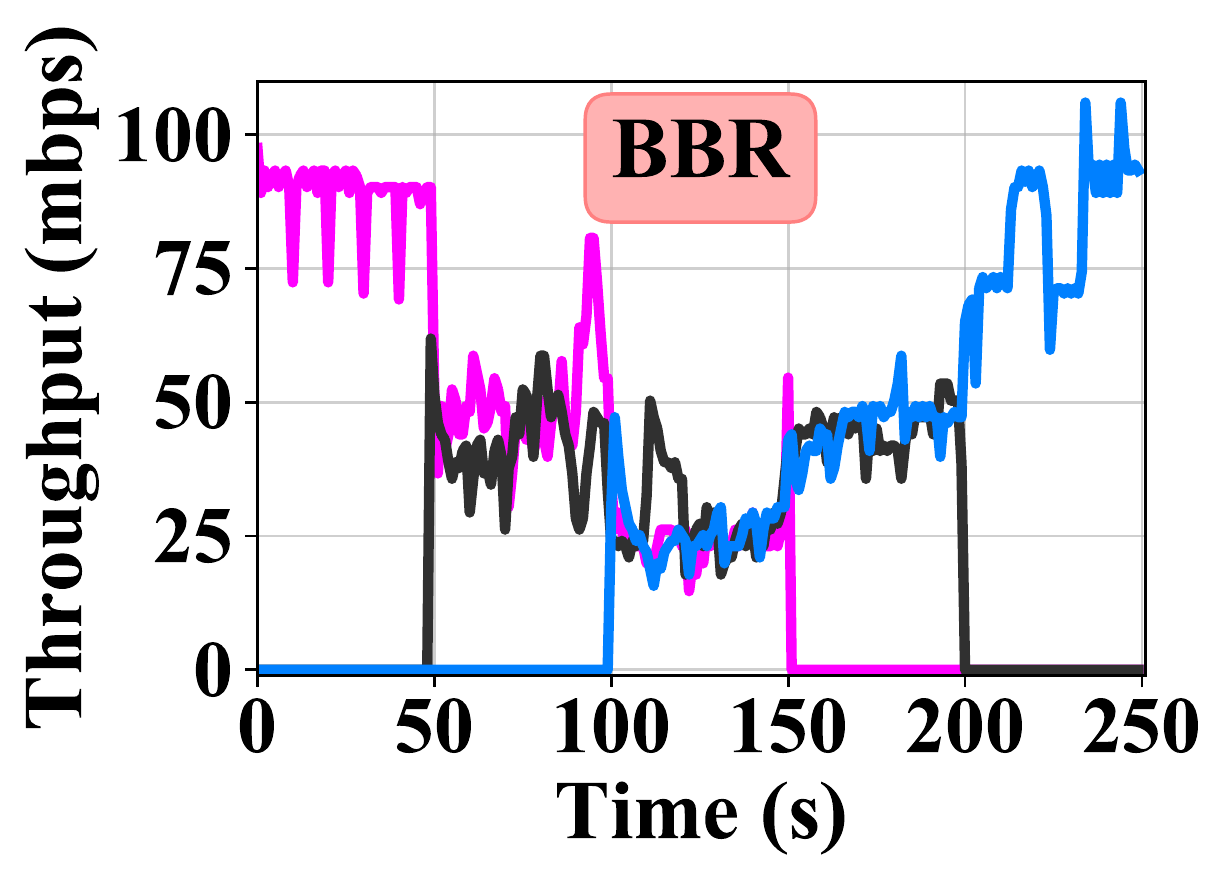}
\end{minipage}
\begin{minipage}[t]{0.45\textwidth}
\includegraphics[width=\textwidth]{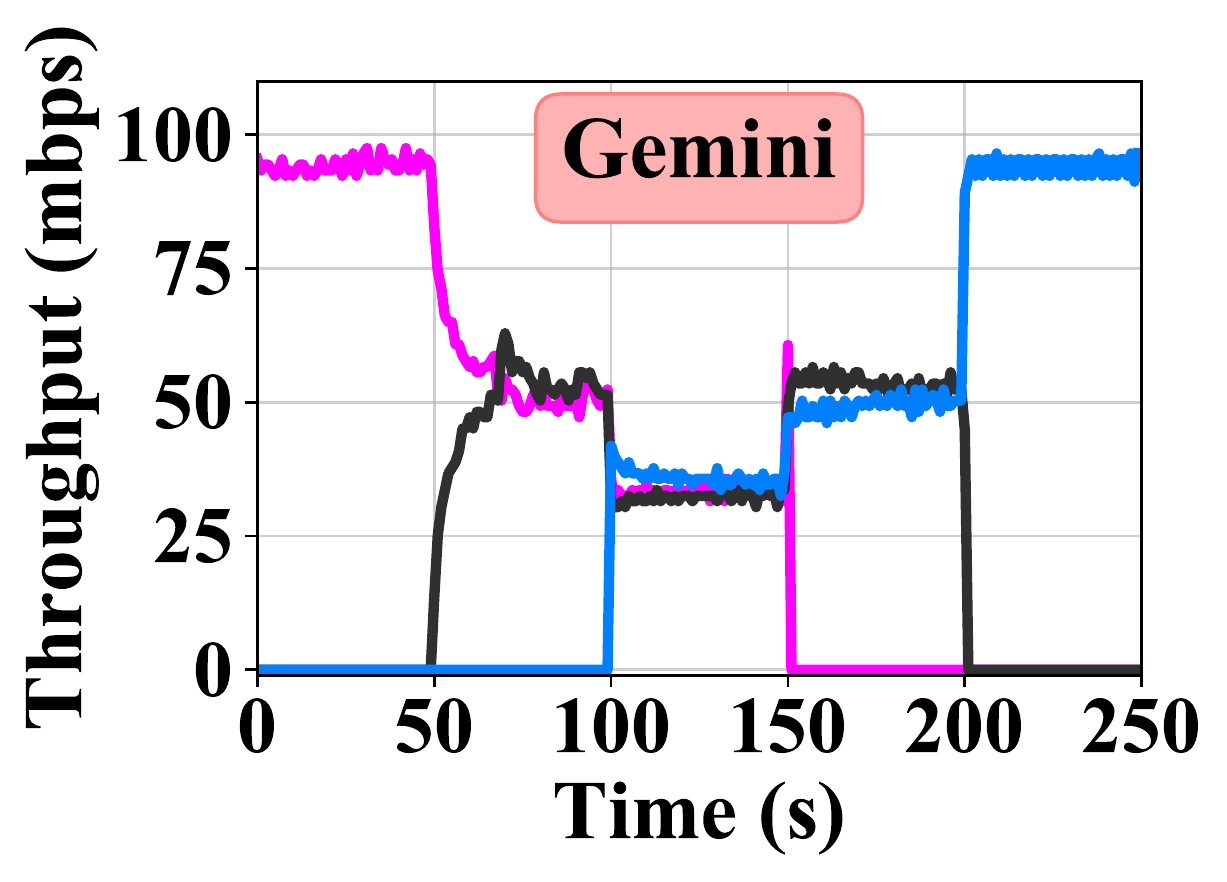}
\end{minipage}
\caption{\sys converges quickly, fairly and stably.}
\label{fig:convergence}
\end{figure}

\subsection{Convergence, Stability and Fairness} \label{sec:convergence_exp}
To evaluate the convergence and stability of \sys, we first start a group of 10 flows from one server. At 50 seconds, we start a second group of flows from another server in the same rack. At 100 seconds, we start a third group of flows from another rack in the same DC. All flows run for 150 seconds and share the same destination server in a remote DC.

Figure~\ref{fig:convergence} shows the throughput dynamics (one flow is shown for each flow group). \sys guarantees fair convergence given its AIMD nature. In fact, \sys converges quickly and stably under both DCN congestion (50--100 secs) and WAN congestion (100--200 secs). For example, during 100--150 secs, the average Jain's fairness index~\cite{jain1984quantitative} of \sys is 0.996, much better than the other protocols (0.926, 0.975, 0.948 for Cubic, DCTCP and BBR, respectively).

\begin{figure}[!t]
\centering
\includegraphics[width=0.6\textwidth]{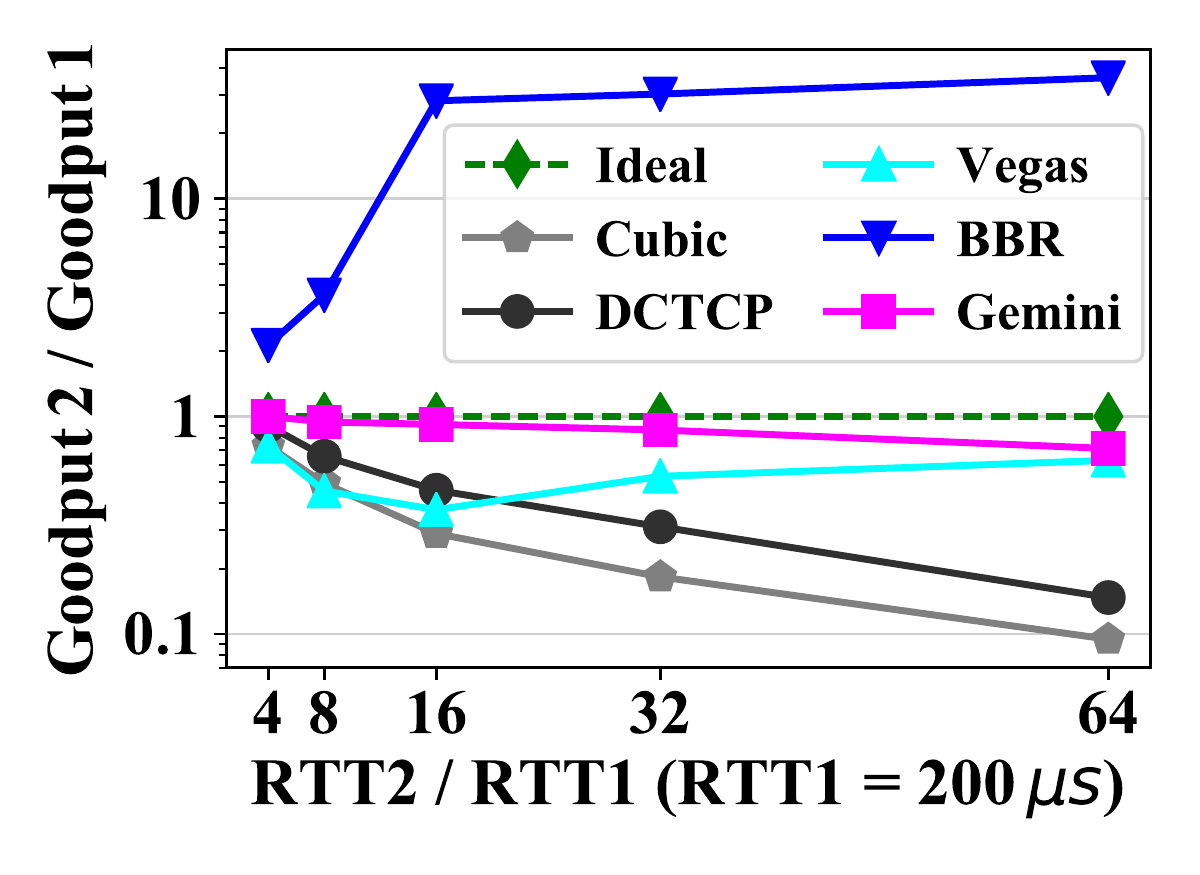}
\caption{RTT-fairness. RTT1 and RTT2 are the intra-DC and inter-DC RTTs.\sys achieves bandwidth fair-sharing regardless of different RTTs.}
\label{fig:rttfairness2}
\end{figure}

Fairness is important for good tail performance. RTT unfairness~\cite{lakshman1997tcp, brown2000resource} is the major challenge in achieving per-flow bandwidth fair-sharing in cross-DC networks, where intra-DC and inter-DC traffic with different RTTs coexists. We show that, good RTT-fairness can be achieved by \sys with the factor $h$ and the scale factor $F$.
To demonstrate that, we generate 4 inter-DC flows and 4 intra-DC flows sharing the same bottleneck link inside DC. The intra-DC RTT is $\sim\,$200$\,\mu$s. With tc \textsc{netem}~\cite{netem}, the inter-DC RTT is set to 4$\times$, 8$\times$, 16$\times$, 32$\times$, 64$\times$ the intra-DC RTT. All ECN-enabled protocols adopt the same ECN threshold of 300 packets for fair comparison. 
The experiment result is shown in Figure~\ref{fig:rttfairness2}. While Cubic and DCTCP achieve proportional RTT-fairness and BBR skews towards large RTT flows, \sys maintains equal bandwidth fair-sharing regardless of the varying RTTs.

\subsection{Realistic Workloads} \label{sec:realistic_exp}
We evaluate \sys under realistic workloads. The workloads are generated based on traffic patterns that have been observed in a data center supporting web search~\cite{dctcp}. Flows arrive by the Poisson process. The source and destination is chosen uniformly random from a configured IP pool. The workload is heavy-tailed with about 50\% small flows (size $<\,$100$\,$KB) while 80\%  of all bytes belong to the the larger 10\% of the flows of size greater than 10$\,$MB. We run the workload with a publicly available traffic generator that has been used by other work~\cite{clicknp, mqecn}. Similar to prior work~\cite{dctcp, pfabric, cloudburst-arxiv, cloudburst}, we use flow completion time (FCT) as the main performance metric.

\begin{figure*}[!t]
\centering
\subfigure[Small Flow - Average]
{
\centering
\includegraphics[width=0.45\textwidth]{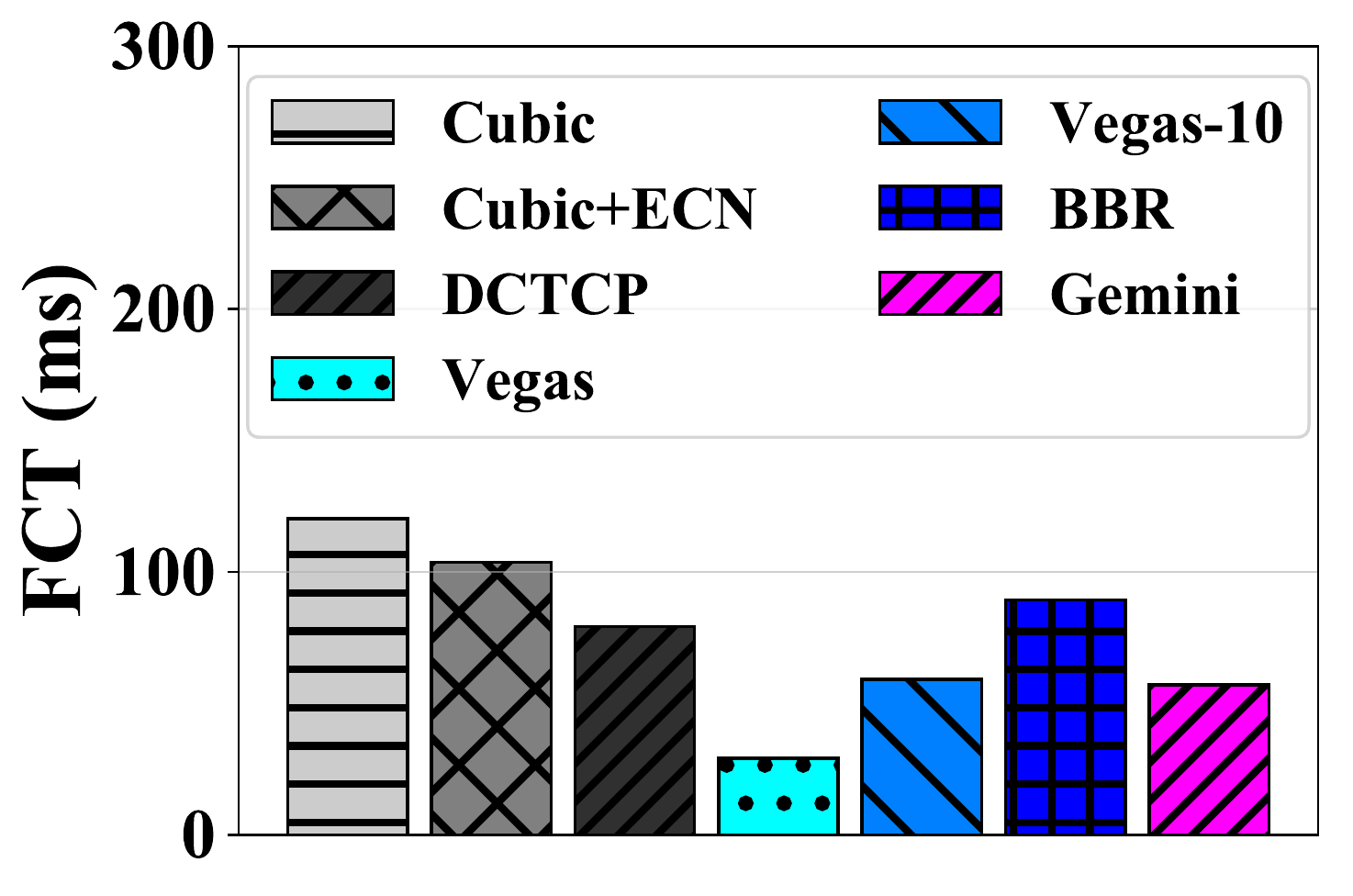}
\label{fig:interdcsmallave2}
}
\subfigure[Small Flow - 99th Tail]
{
\centering
\includegraphics[width=0.45\textwidth]{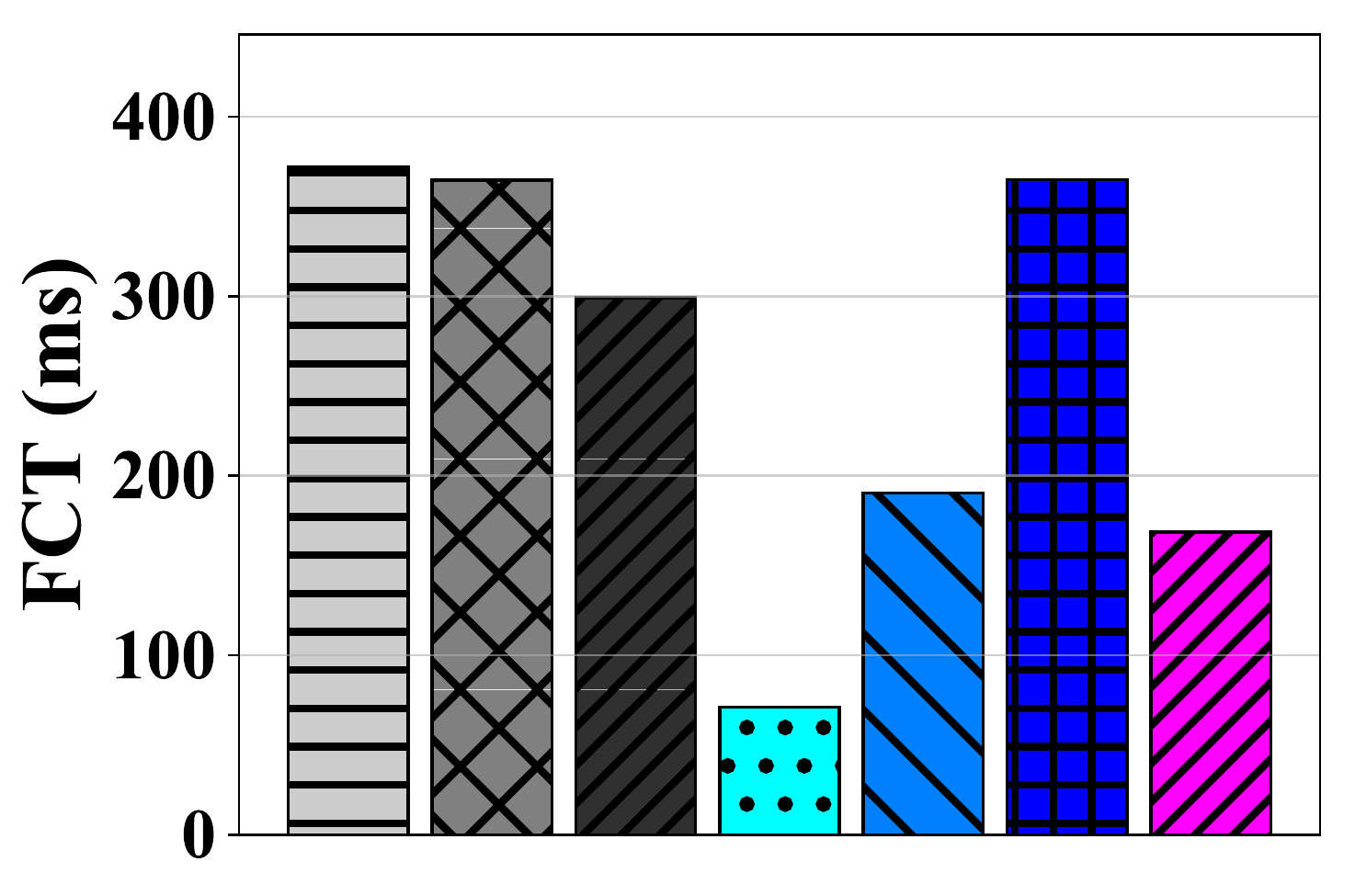}
\label{fig:interdcsmalltail2}
}
\subfigure[Large Flow - Average]
{
\centering
\includegraphics[width=0.45\textwidth]{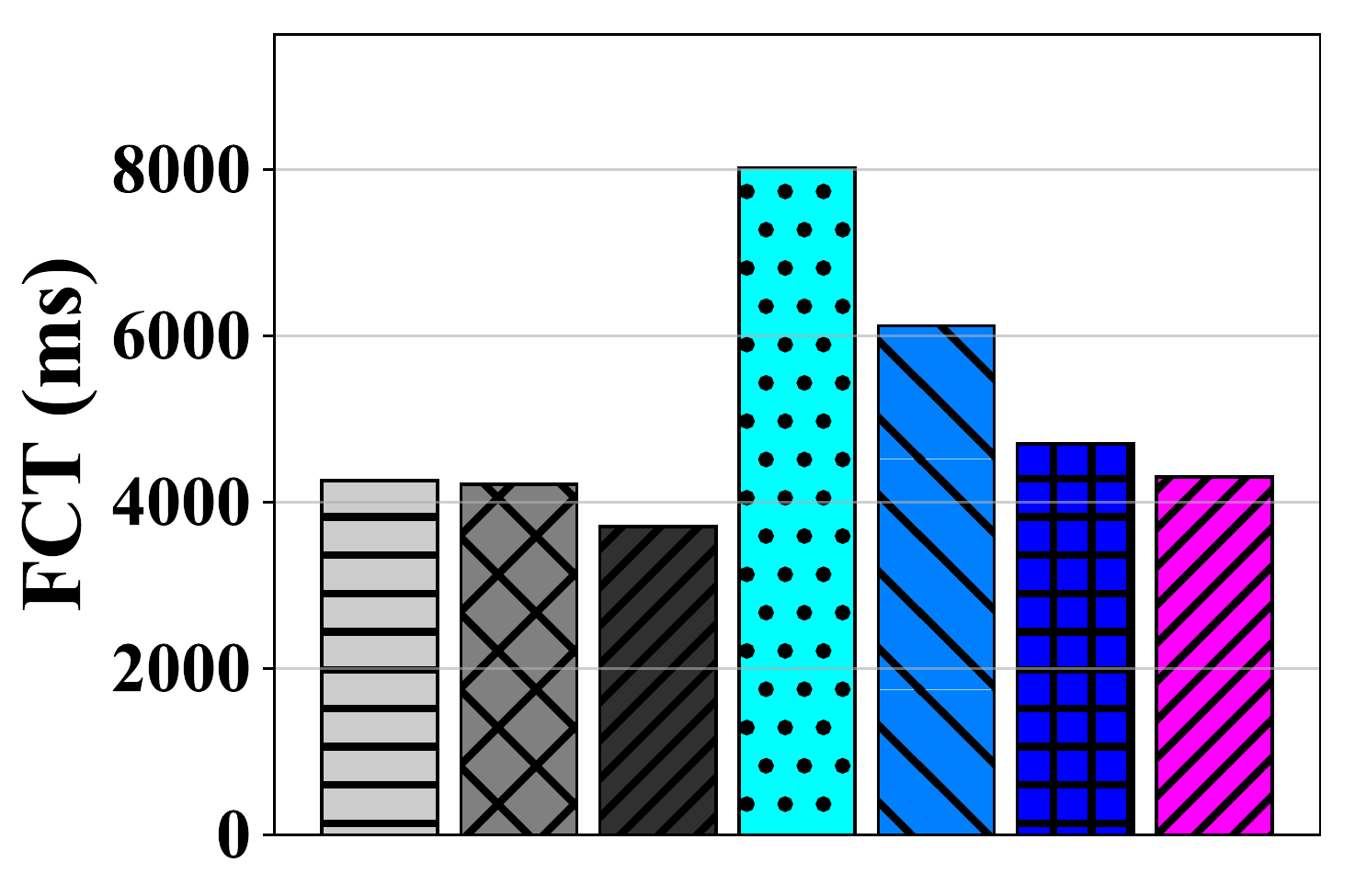}
\label{fig:interdclarge2}
}
\subfigure[All Flow - Average]
{
\centering
\includegraphics[width=0.45\textwidth]{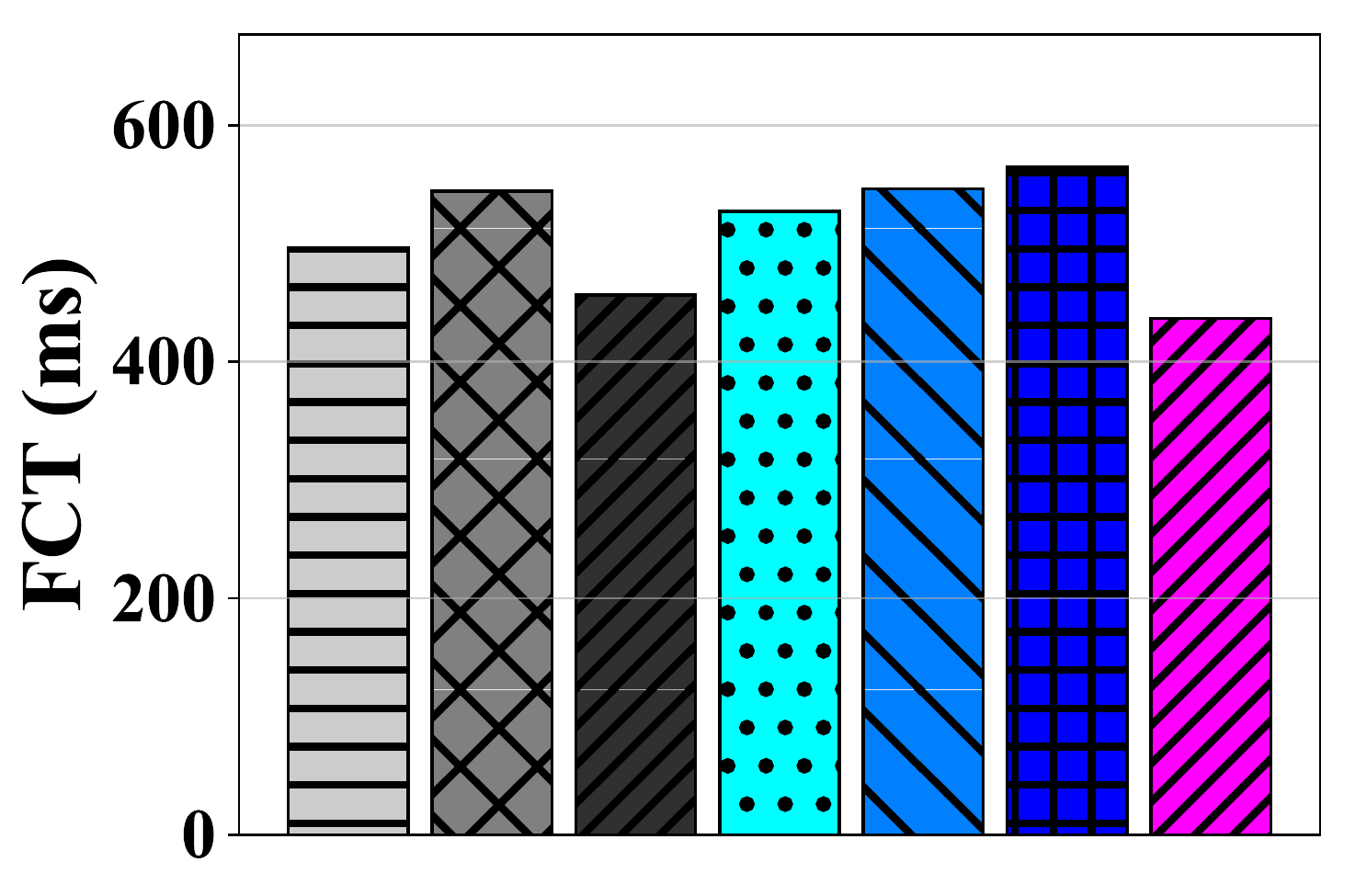}
\label{fig:interdcaverage2}
}
\caption{[Large-scale 1Gbps Testbed] FCT results under traffic pattern 1: Inter-DC traffic, highly congested in WAN. Small flow: Size $<\,$100$\,$KB. Large flow: Size $>\,$10$\,$MB.
\sys achieves the best or second best results in most cases of Figure~\ref{fig:interdc2}-~\ref{fig:mixedtraffics2_10Gbps} (within 11\% of the second best scheme in the worst case).}
\label{fig:interdc2}
\end{figure*}

\begin{figure*}[!t]
\centering
\subfigure[Small Flow - Average]
{
\centering
\includegraphics[width=0.45\textwidth]{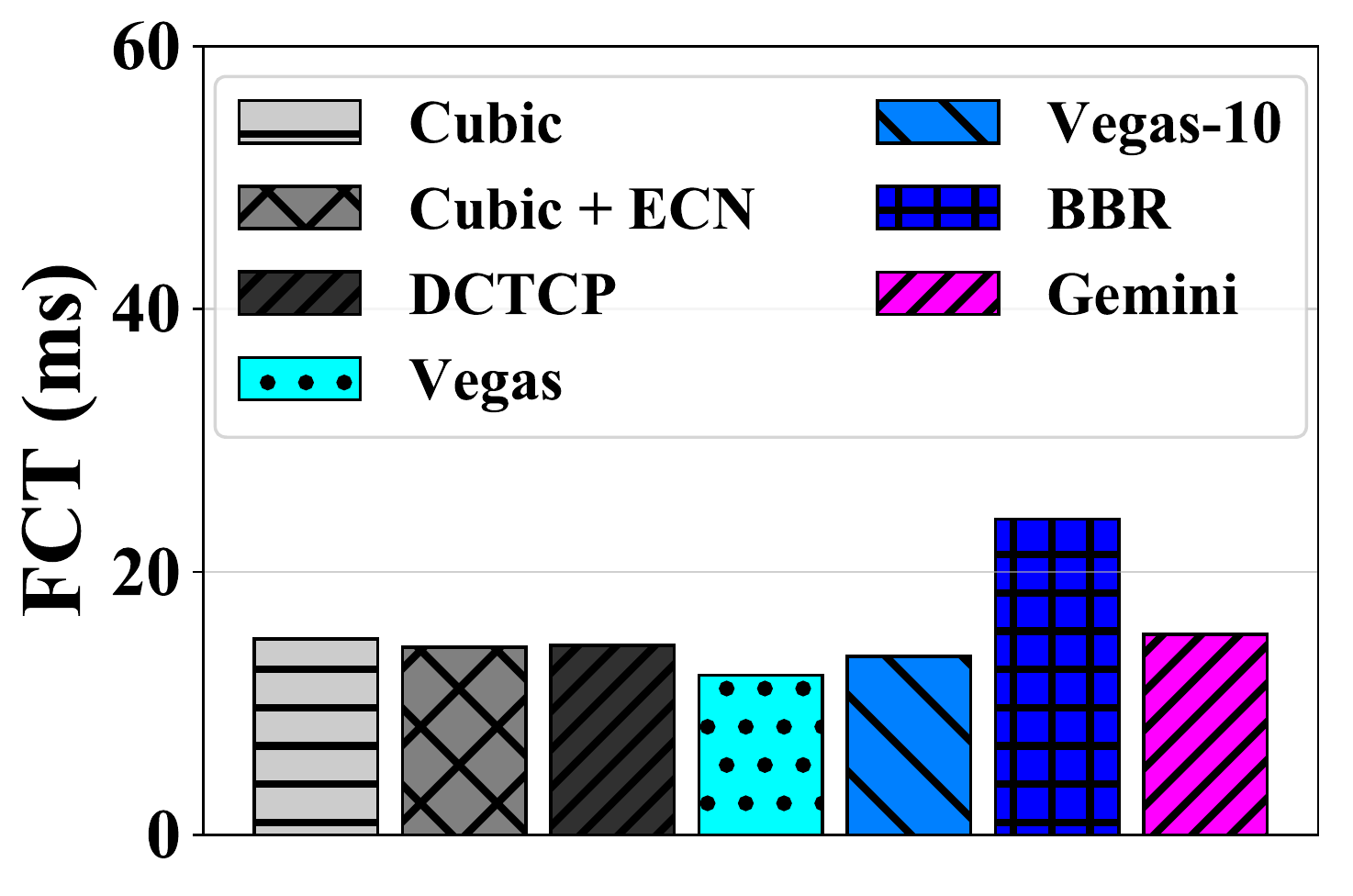}
\label{fig:interdcsmallave2_10Gbps}
}
\subfigure[Small Flow - 99th Tail]
{
\centering
\includegraphics[width=0.45\textwidth]{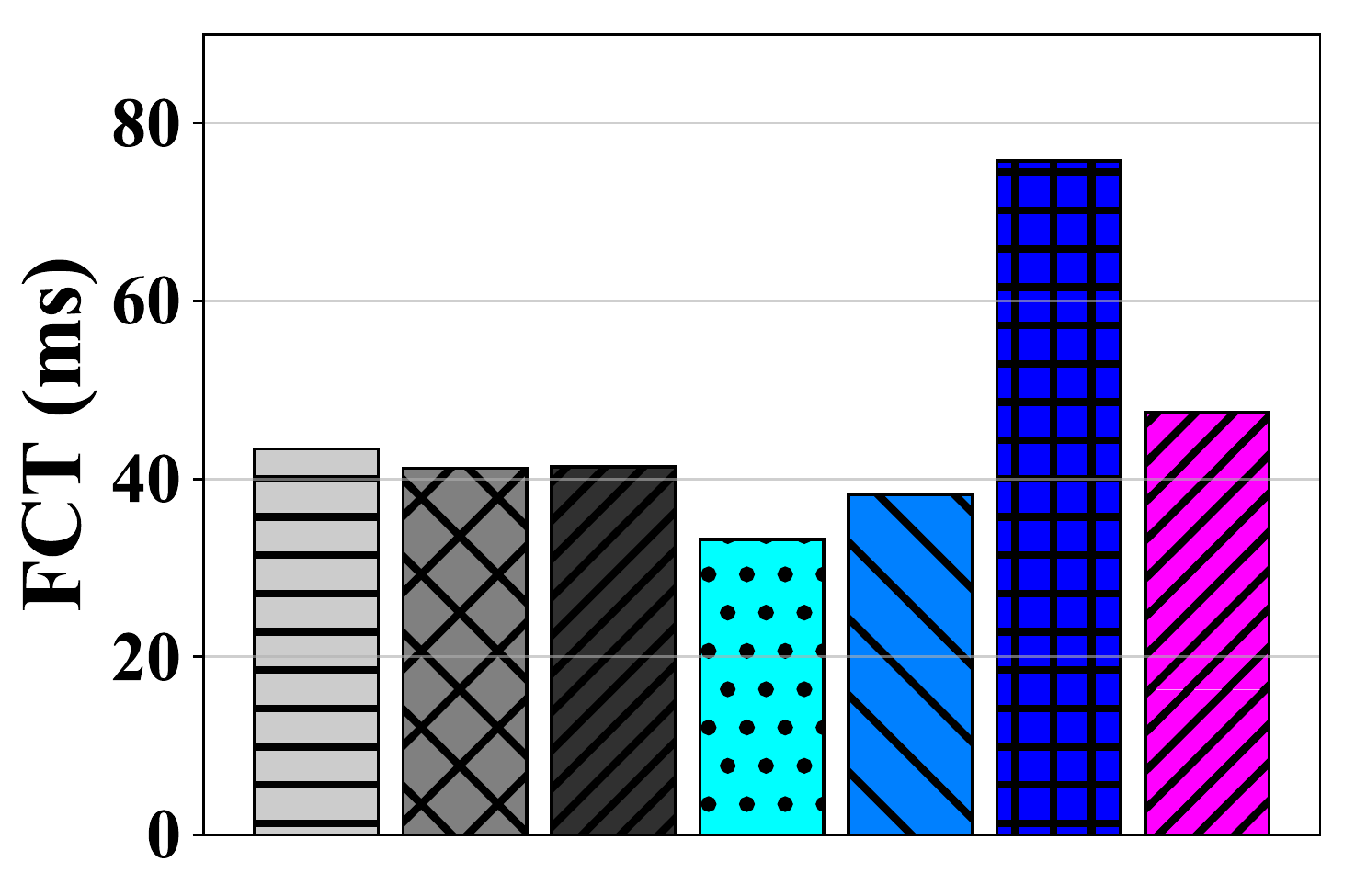}
\label{fig:interdcsmalltail2_10Gbps}
}
\subfigure[Large Flow - Average]
{
\centering
\includegraphics[width=0.45\textwidth]{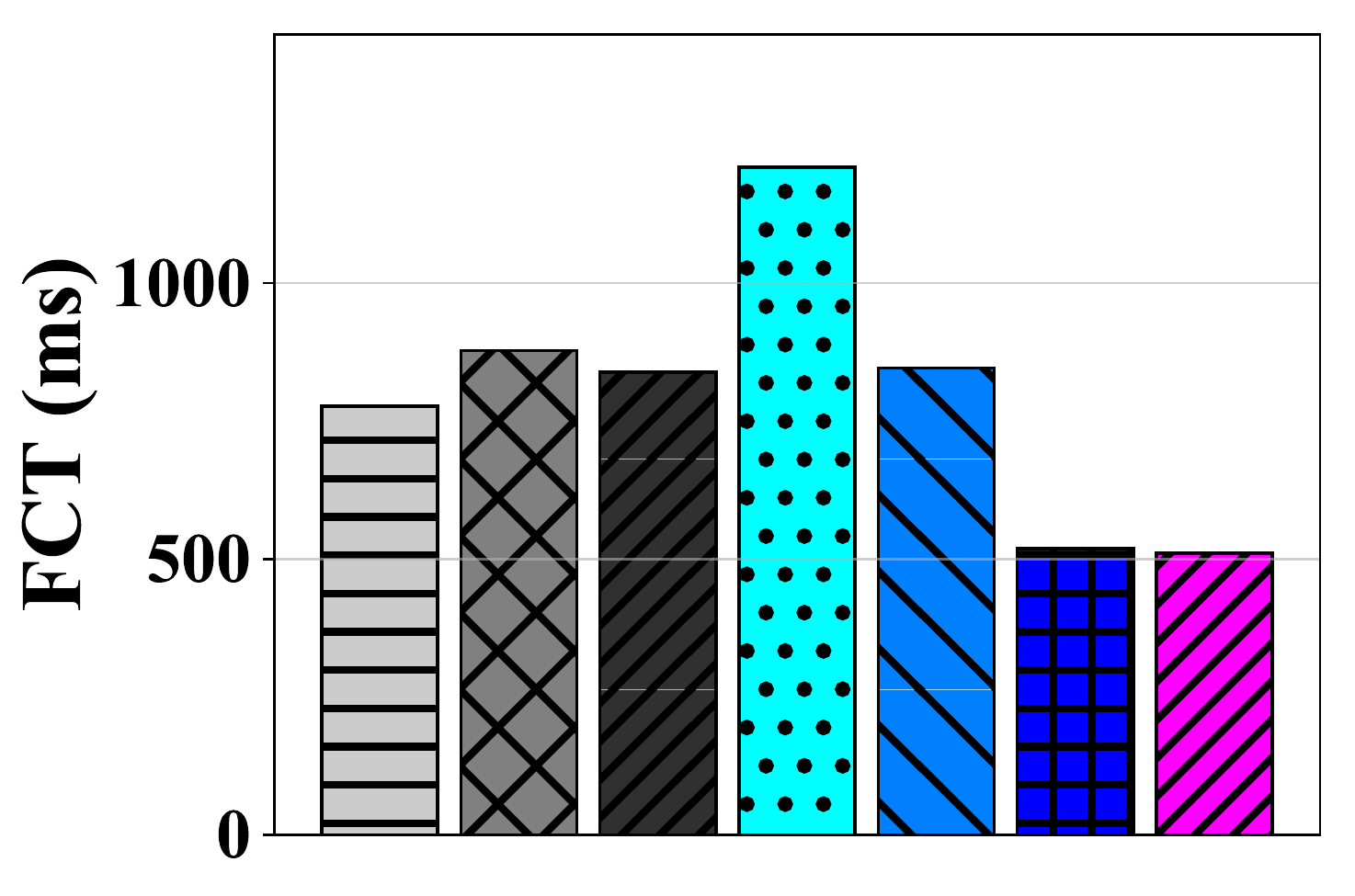}
\label{fig:interdclarge2_10Gbps}
}
\subfigure[All Flow - Average]
{
\centering
\includegraphics[width=0.45\textwidth]{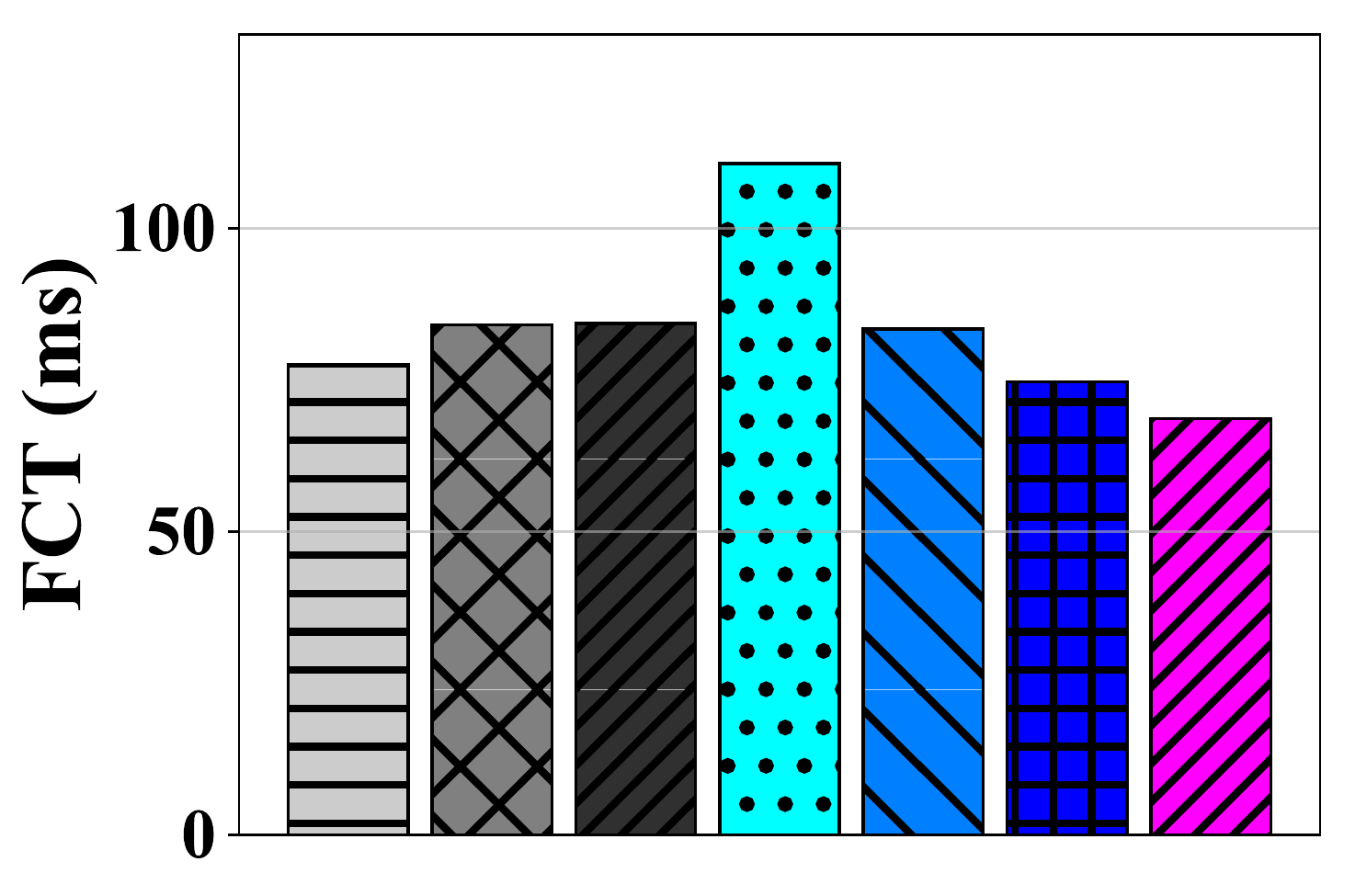}
\label{fig:interdcaverage2_10Gbps}
}
\caption{[Small-scale 10Gbps Testbed] FCT results under traffic pattern 1: Inter-DC traffic, highly congested in WAN. Small flow: Size $<\,$100$\,$KB. Large flow: Size $>\,$10$\,$MB.}
\label{fig:interdc2_10Gbps}
\end{figure*}

\parab{Traffic Pattern 1: Inter-DC traffic, highly congested in WAN.}
In this experiment, all flows cross the WAN segment. The average utilization of the inter-DC WAN link is $\sim$90\%. The DC border routers are highly congested, while intra-DC links have much lower utilization ($\sim$11.25--45\%).

The experiment results are shown in Figure~\ref{fig:interdc2} and~\ref{fig:interdc2_10Gbps}: (1) For small flow FCT, \sys performs better than Cubic, DCTCP and BBR on both average and 99th tail. This is because Cubic and DCTCP suffer from the large queueing delay in WAN segment while \sys well handles that with RTT signal. BBR suffers a lot from loss as the misestimates of bandwidth and RTT are magnified by high congestion. BBR does not react to loss events explicitly until loss rate $> 20\%$ (as a protection). This design choice benefits the long-term throughput while hurts short-term latency. (2) For large flow FCT, \sys performs much better than Vegas. The default parameter setting for Vegas is very conservative (\(\alpha=2, \beta=4\)), leading to poor throughput of large flows. Setting larger thresholds in Vegas-10 (\(\alpha=20, \beta=40\)) improves throughput but hurts latency of small flows. (3) For overall FCT, \sys performs the best among all experimented transport protocols.

\parab{Traffic Pattern 2: Mixed traffic, highly congested both in WAN and DCN.}
In this experiment, the source and the destination of each flow is chosen uniformly random among all servers. Intra-DC and inter-DC traffic coexists in the network. The average utilization of the inter-DC WAN link is $\sim$90\%. The average utilization of the link from the DC switch to the border router is $\sim$67.5\%. Therefore, both WAN and DCN are highly congested.

The experiment results are shown in Figure~\ref{fig:mixedtraffics2} and~\ref{fig:mixedtraffics2_10Gbps}: (1) For small flow FCT, \sys performs one of the best among experimented transport protocols. In fact, \sys has consistently low packet loss rates ($<\,$10$\times10^{-5}$) under both traffic patterns. This is because Gemini handles DCN congestion adaptively with different RTTs, thus allowing a low ECN threshold with larger buffer headroom to absorb burst. Besides, it also enforces pacing to reduce burst losses. (2) For large flow FCT, \sys performs better than Cubic and Vegas. Vegas does not perform well because it cannot control congestion in WAN and DCN simultaneously. \sys can identify and react to congestion in DCN and WAN differently using ECN and RTT signals respectively. (3) For overall FCT, \sys performs one of the best among all experimented transport protocols.

{
\begin{figure*}[!t]
\centering
\subfigure[Small Flow - Average]
{
\centering
\includegraphics[width=0.45\textwidth]{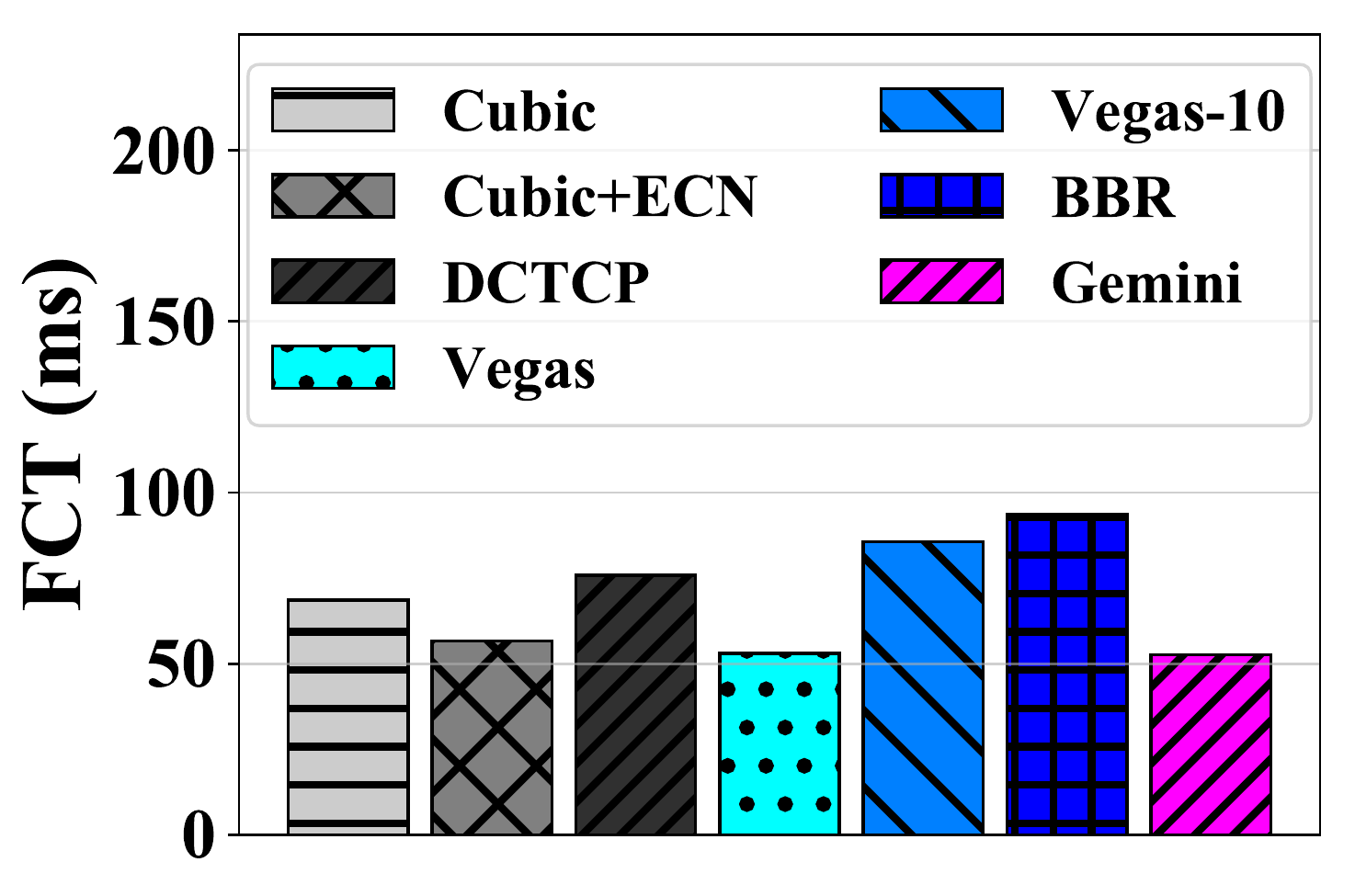}
\label{fig:mixedtrafficssmallave2}
}
\subfigure[Small Flow - 99th Tail]
{
\centering
\includegraphics[width=0.45\textwidth]{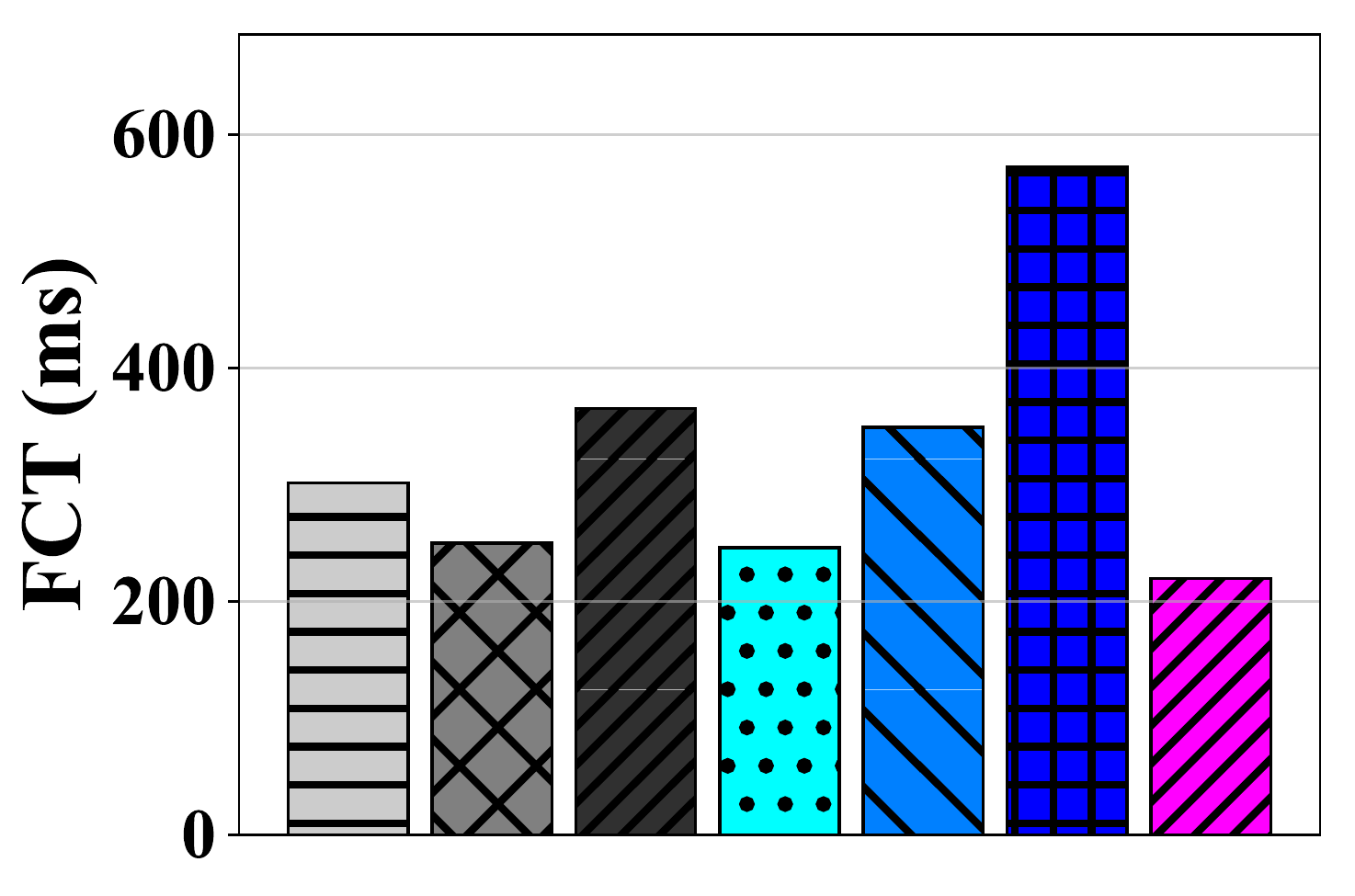}
\label{fig:mixedtrafficssmalltail2}
}
\subfigure[Large Flow - Average]
{
\centering
\includegraphics[width=0.45\textwidth]{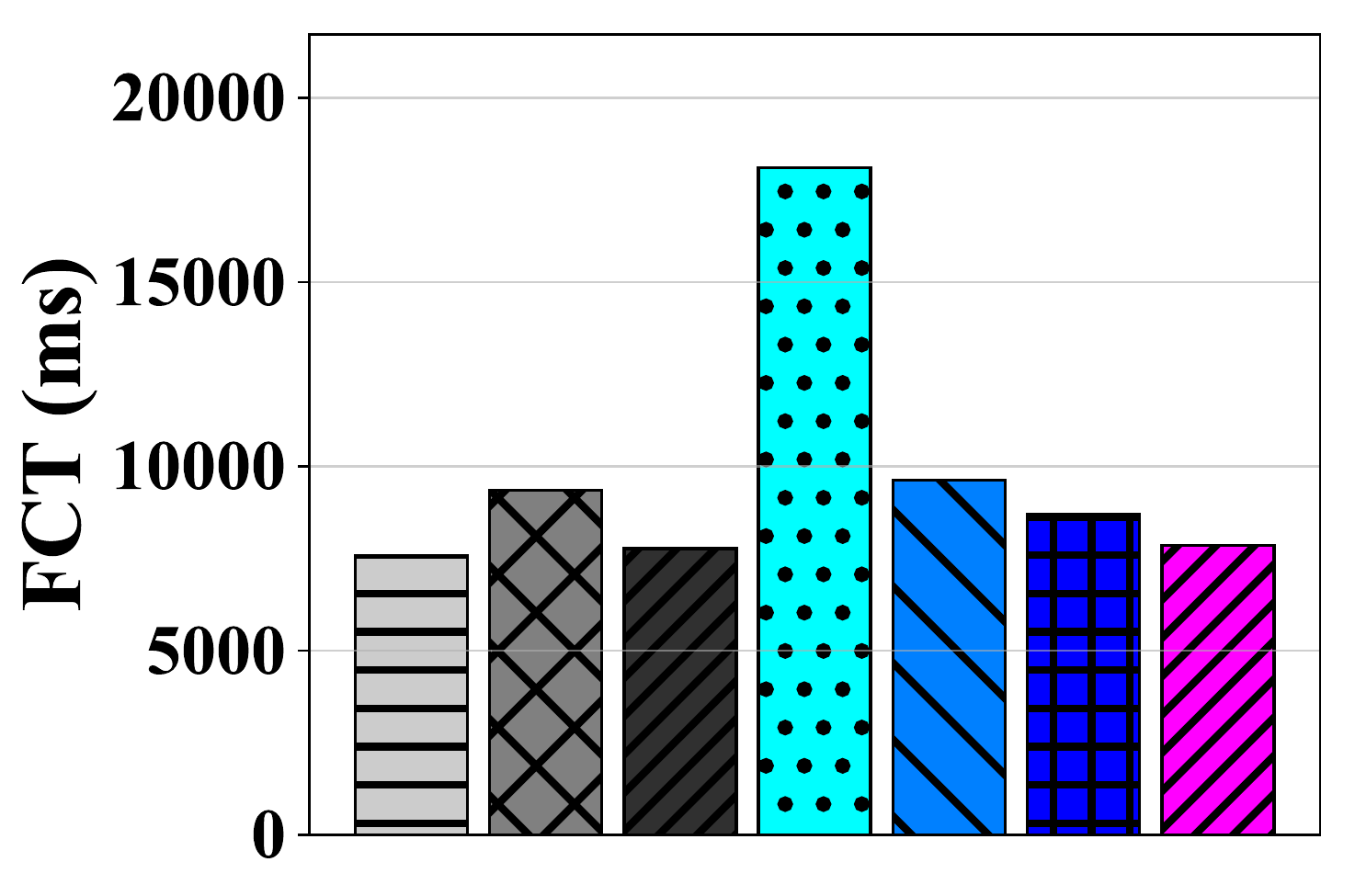}
\label{fig:mixedtrafficslarge2}
}
\subfigure[All Flow - Average]
{
\centering
\includegraphics[width=0.45\textwidth]{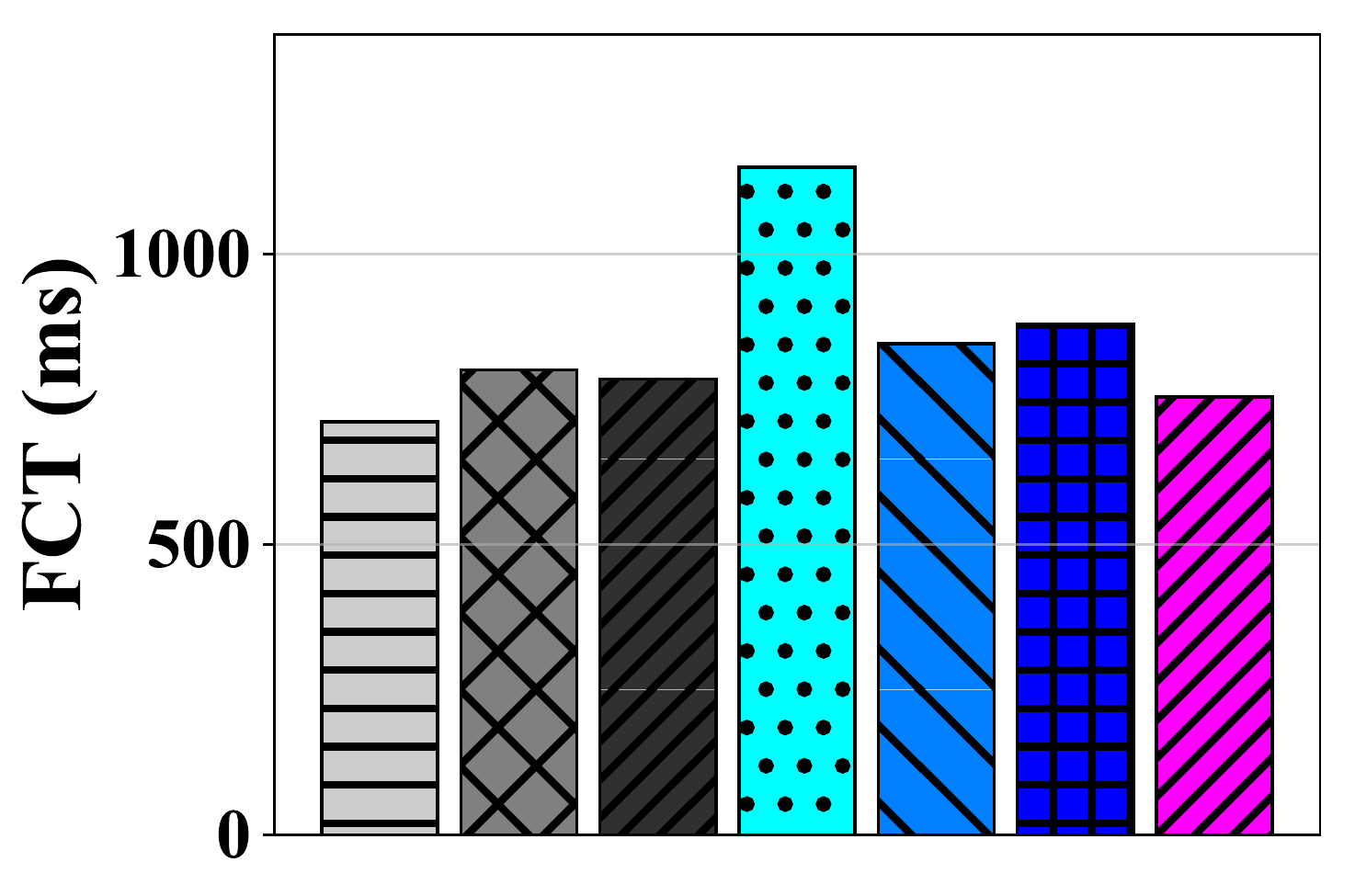}
\label{fig:mixedtrafficsaverage2}
}
\caption{[Large-scale 1Gbps Testbed] FCT results under traffic pattern 2: mixed inter-DC and intra-DC traffic, highly congested both in WAN and DCN. Small flow: Size $<\,$100$\,$KB. Large flow: Size $>\,$10$\,$MB.}
\label{fig:mixedtraffics2}
\end{figure*}

\begin{figure*}[!t]
\centering
\subfigure[Small Flow - Average]
{
\centering
\includegraphics[width=0.45\textwidth]{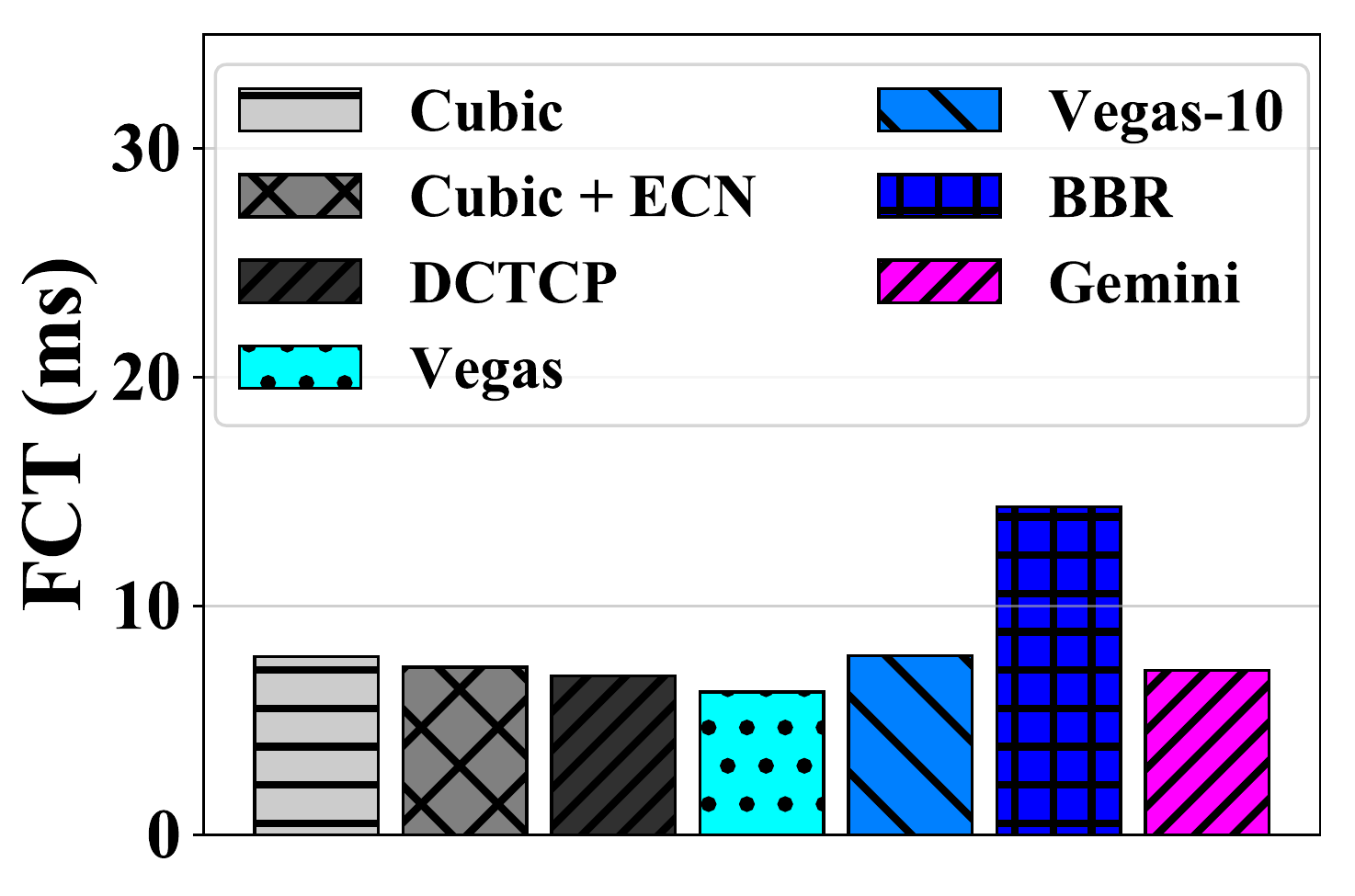}
\label{fig:mixedtrafficssmallave2_10Gbps}
}
\subfigure[Small Flow - 99th Tail]
{
\centering
\includegraphics[width=0.45\textwidth]{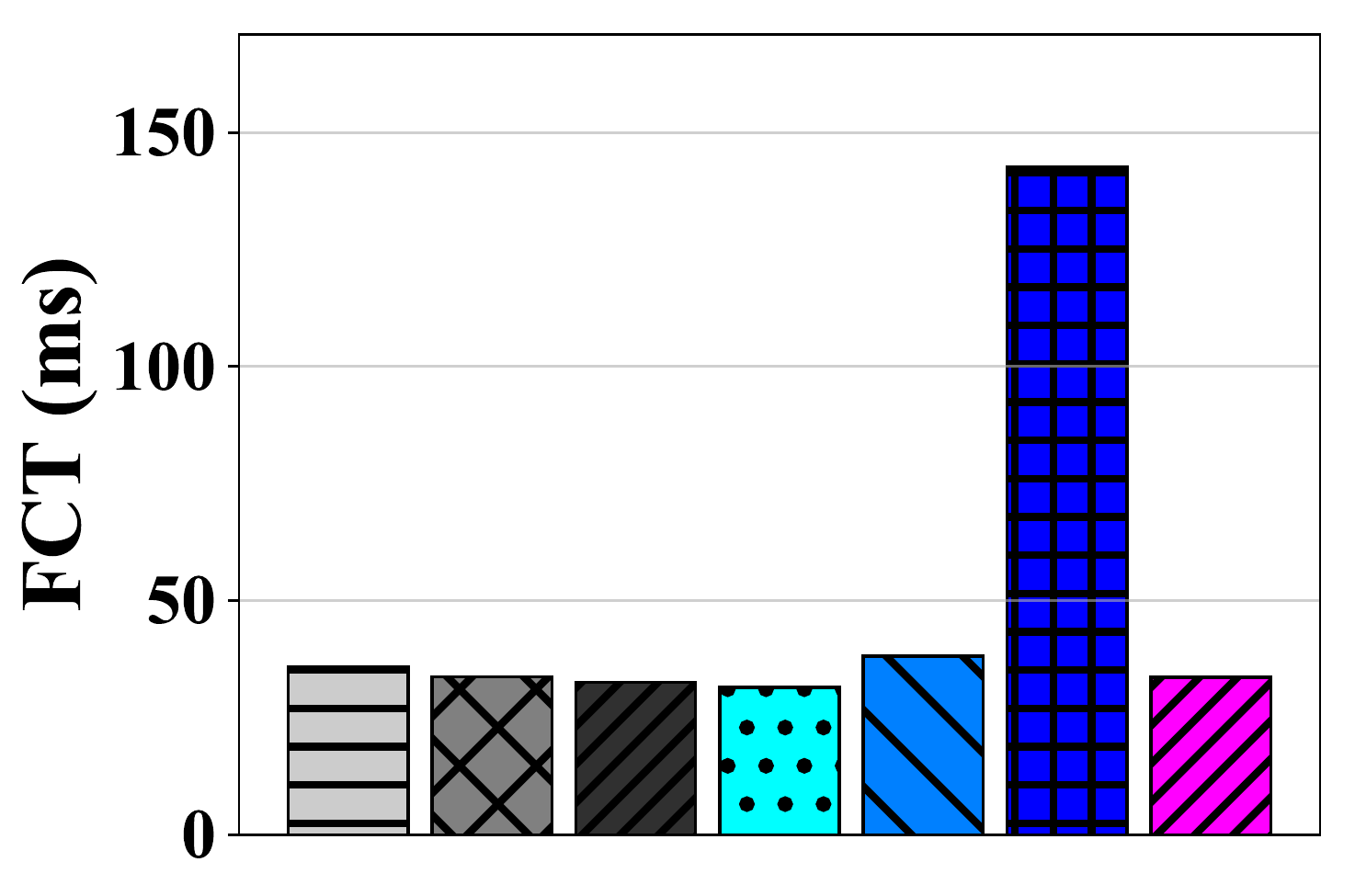}
\label{fig:mixedtrafficssmalltail2_10Gbps}
}
\subfigure[Large Flow - Average]
{
\centering
\includegraphics[width=0.45\textwidth]{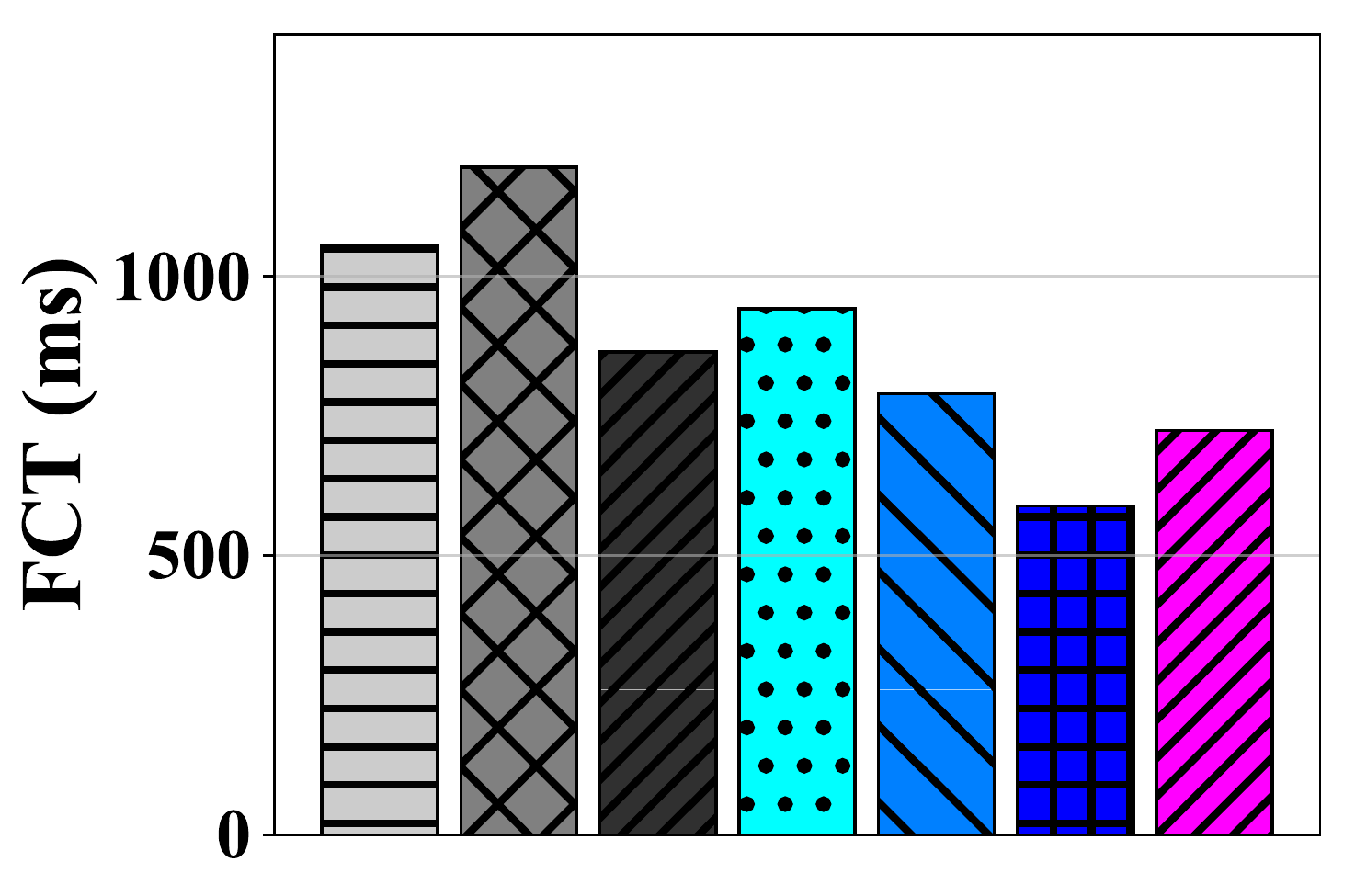}
\label{fig:mixedtrafficslarge2_10Gbps}
}
\subfigure[All Flow - Average]
{
\centering
\includegraphics[width=0.45\textwidth]{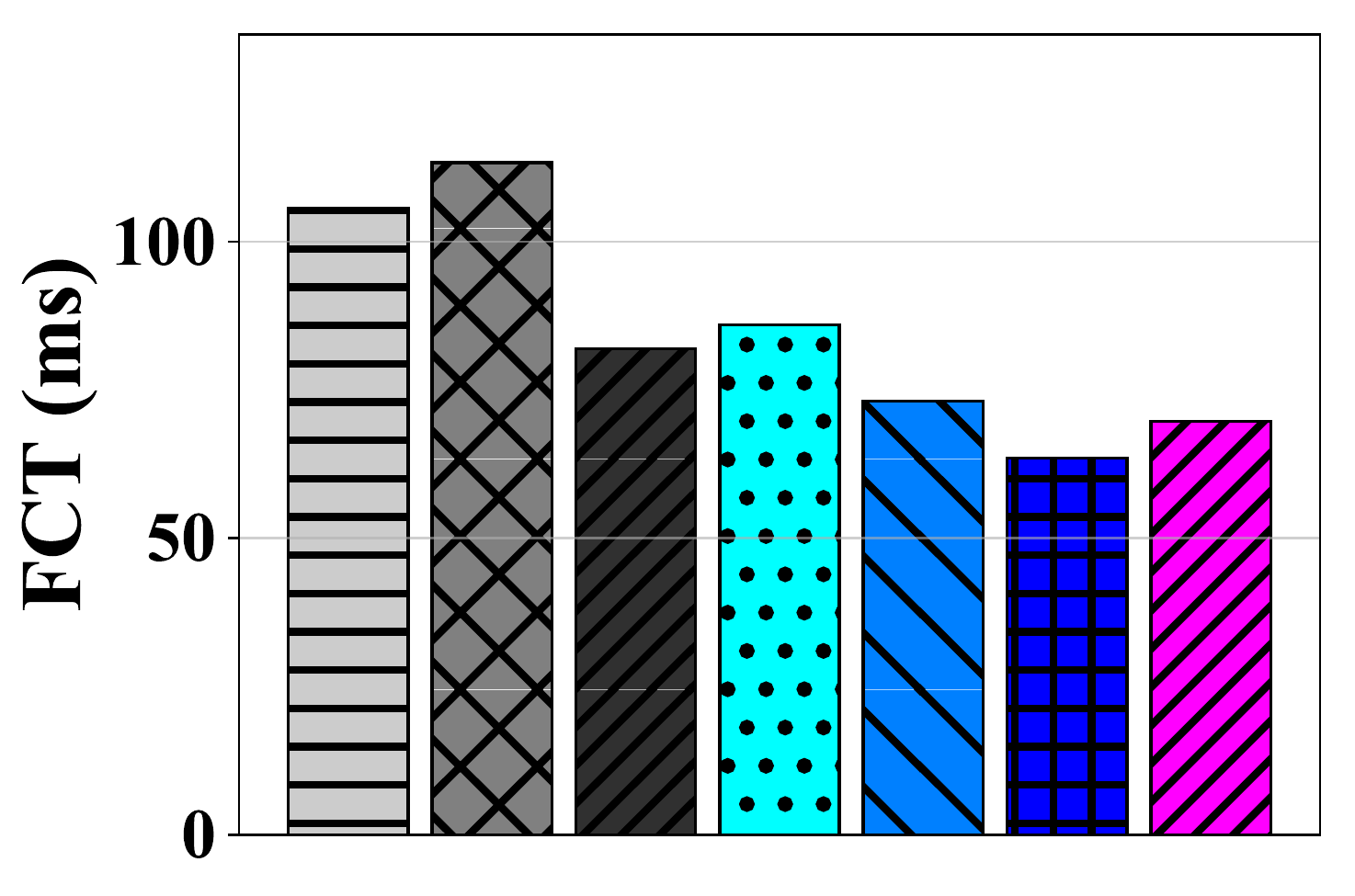}
\label{fig:mixedtrafficsaverage2_10Gbps}
}
\caption{[Small-scale 10Gbps Testbed] FCT results under traffic pattern 2: mixed inter-DC and intra-DC traffic, highly congested both in WAN and DCN. Small flow: Size $<\,$100$\,$KB. Large flow: Size $>\,$10$\,$MB.}
\label{fig:mixedtraffics2_10Gbps}
\end{figure*}
}

\section{Discussion} \label{sec-gemini:discussion}

\subsection{Practical Considerations} \label{sec:considerations}
\parab{TCP Friendliness.}
\sys is not TCP-friendly. For example, like all ECN-based protocols, \sys has fairness issues if it coexists with non-ECN protocols.
In fact, recent work \cite{copa, goyal2018elasticity} shows that it is fundamentally difficult to achieve high performance while attaining perfect friendliness to buffer-filling protocols. Thus, they sacrifice performance to guarantee friendliness when detecting the buffer-fillers. \sys can adopt similar approaches (\eg, switching to TCP-competitive mode when buffer-fillers are detected).
We believe advance switch support like ideal fair queueing \cite{sharma2018approximating} would be a better solution.
We do not focus on the problem in this work.

\parab{Real-world Deployment.}
For private clouds, deploying \sys requires the cloud owners to add the new congestion control kernel module at the end-hosts and configure ECN at the DC switches.
In terms of partial deployment, \sys can work with common TCP remote ends since it relies
on exactly the same ACK mechanism as TCP Cubic.
For public clouds, cloud users and cloud operators are of different entities. On one hand, cloud users control the VMs and thus can deploy \sys with minimal support (ECN only) from cloud operators. On the other hand, cloud operators control the hypervisors at the end-hosts and the underlying network devices. \sys can be enforced similarly like \cite{vcc, acdc}.

\subsection{Alternative Solutions} \label{sec:alternative}
\vspace{-0.2em}
\parab{Multiple Queues + Different Protocols.} \label{sec:multiple_q} 
A seemingly straightforward solution under cross-DC network is to use different transport protocols for intra-DC and inter-DC traffic, like what major infrastructure owners, \eg, Google~\cite{jupiter, bbr} and Microsoft~\cite{dctcp, swan} do for their first-party applications. However, a few issues make this approach less attractive: (1) classifying inter-DC and intra-DC flows based on IPs is nontrivial. In cloud network, virtual subnets extend across DCs for the ease of management, decoupling the IPs from the DC locations. Maintaining an always-up-to-date IP-to-DC mapping globally is daunting and gives away the management benefit of running virtual subnets; (2) different transport protocols are unlikely to fair-share the network bandwidth, thus requiring the switches to allocate different queues for them. General cloud users usually do not have this luxury; (3) even if we pay the cost to adopt such a solution, some desired performance goals are still missed: (a) The inter-DC traffic may encounter congestion on either WAN or DCN. The existing protocols do not address both at the same time. The inter-DC traffic that tends to use wide-area transport will suffer from low throughput and high loss rate when experiencing congestion at the shallow-buffered DC switch; (b) Flow-level bandwidth fair-sharing cannot be guaranteed by coarse-grained traffic isolation. Static bandwidth allocation may even lead to bandwidth under-utilization.

\parab{TCP Proxy.} \label{sec:tcp_proxy} 
Another possible solution is to terminate and relay the TCP flows with proxies at the border of each network, or the so-called Split TCP~\cite{kopparty2002splittcp, flach2013reducing, le2015experiences}. In this way, traffic can be transported using the best-suited protocol for each network. The TCP proxy way has flaws in practice:
\begin{icompact}
\item Proxies in the middle add extra latencies. The latency overhead can greatly impair network performance, especially for short flows that can finish in one RTT.
\item Relaying every inter-DC flow is impractical. As a rule of thumb, Google~\cite{jupiter} allocates 10\% of aggregate intra-DC bandwidth for external connectivity, requiring prohibitive amount of relay bandwidth for peak WAN usage.
\item Configuring and managing proxy chains is complex and error-prone. Single fault from one of the intermediate relays may tear down the whole communication.
\end{icompact}

\section{Related Work} \label{sec-gemini:related}

To facilitate cross-DC network communication, there are three lines of work in general, each operating on a different granularity. First, WAN traffic engineering~\cite{b4, swan, kumar2018semi} works on the datacenter level. It distributes network traffic to multiple site-to-site paths (usually hundreds of updates per day). Second, bandwidth allocation~\cite{bwe} applies to the tenant or flow group level. It re-allocates the site-to-site bandwidths and split them among all competing flow groups. Third, transport protocol regulates the per-flow sending rate in realtime. We focus on the transport design in this work.

To the best of our knowledge, transport design under heterogeneous cross-DC network is unexplored in literature. However, there are vast transport protocols under wide area network and datacenter network that are highly related:

\parab{Wide Area Network Transport.} Cubic~\cite{cubic} is the default TCP congestion control in the Linux system. It achieves high scalability and proportional RTT-fairness by growing window with a cubic function of time. Vegas~\cite{vegas} is the seminal transport protocol that uses delay signal to avoid intrinsic high loss and queueing delay of loss-based transport. After that, many WAN protocols~\cite{compound, bbr, copa} are proposed to use delay signal. For example, BBR~\cite{bbr} is proposed primarily for the enterprise WAN. The core idea is to work at the theoretically optimal point~\cite{kleinrock1979power} with the aid of sophisticated network sensing (\eg, precise bandwidth and RTT estimation). Copa~\cite{copa} adjusts sending rate towards $1/(\delta d_q)$, where $d_q$ is the queueing delay, by additive-increase additive-decrease (AIAD). It detects buffer-fillers by observing the delay evolution and switches between delay-sensitive and TCP-competitive mode.
These transport protocols consider WAN only and usually suffer a lot from the intra-DC congestion in cross-DC network.

\parab{Datacenter Network Transport.} DCTCP~\cite{dctcp} detects the network congestion with ECN and react in proportion to the measured extent of congestion. Following that, many ECN-based protocols~\cite{hull, d2tcp, l2dct, dcqcn} are proposed for DCN congestion control. The other line of work leverages delay signal with microsecond-level accuracy for congestion feedback, which is enabled by recent advances~\cite{softnic, timely, dx} in NIC technology. For example, TIMELY~\cite{timely} and RoGUE~\cite{rogue} use RTT signal for congestion control in RDMA communication.
DX~\cite{dx} detects congestion and reduces window size based on the average queuing delay.
We show that ECN or delay signal alone is insufficient for cross-DC congestion control.

vCC~\cite{vcc} and AC/DC~\cite{acdc} attempt to optimize the transport protocols in the cloud networks, from the cloud provider's perspective with no interference to the user's VMs. They enforce the advanced congestion control mechanisms (\eg, DCTCP) in the hypervisors or the virtual switches. On public cloud environments, \sys can be enforced by cloud operators in a similar way to serve the users for the cross-DC network communication.

Explicit rate control~\cite{xcp, rcp, d3, pdq, fcp} allocates network resources with in-network assistance. Centralized rate control~\cite{fastpass, flowtune} regulates traffic rate with a centralized controller. Multi-path transport~\cite{mptcp, mptcp-dcn, mp-rdma} creates subflows along multiple paths to achieve high aggregate throughput and traffic load balance. Proactive congestion control~\cite{expresspass, ndp, homa, aeolus} leverages receiver-side credits to trigger new traffic sending. Learning-based congestion control~\cite{tcpexmachina, pcc, pcc-vivace} learns the congestion control strategy by machine either online or offline. These novel transport protocols often have less comprehensive and unpredictable performance, and may require advanced network support (\eg, cutting payload~\cite{cp}) that are unavailable or bad supported in cross-DC network facilities.

\section{Final Remarks} \label{sec-gemini:conclusion}

As geo-distributed applications become prevalent, cross-DC communication gets increasingly important. We investigate existing transport and find that they leverage either ECN or delay signal alone, which cannot accommodate the heterogeneity of cross-DC networks.
Motivated by this, we design \sys, a solution for cross-DC congestion control that integrates both ECN and delay signal. \sys uses the delay signal to bound the total in-flight traffic end-to-end, while ECN is used to control the per-hop queues inside a DCN. It further modulates ECN-triggered window reduction aggressiveness with RTT to achieve high throughput under limited buffer.
We implement \sys with Linux kernel and commodity switches.
Experiments show that \sys achieves low latency, high throughput, fair and stable convergence, and delivers lower FCTs compared to various transport protocols (\eg, Cubic, Vegas, DCTCP and BBR) in cross-DC networks.


\chapter{\sysB: Proactive Congestion Control for Inter-Datacenter Networks} 
\label{sec:flashpass}


With the prevalence of the geo-distributed web applications and services, wide-area network (WAN) is becoming an increasingly important cloud infrastructure~\cite{b4, swan, bwe, b4after}. For example, Google~\cite{b4after} reveals that its inter-datacenter wide-area traffic has been growing exponentially with a doubling of every 9 months in recent 5 years. This pushes the WAN facility to evolve much faster than the rest of its infrastructure components. 

To scale the wide-area network cost-effectively and flexibly, large enterprises such as Google and Alibaba have been building and deploying their customized wide-area routers based on shallow-buffered commodity switching chips~\cite{dc-silicon, b4, jupiter, clos-router, buffer} (\S\ref{motivation-wan}). 
However, conventional wisdom~\cite{buffer-sizing} dictates that buffer size of one bandwidth-delay product (BDP) is required to achieve full link utilization in the worst (synchronized) case. Thus, the cheap shallow-buffered WAN gear imposes stringent requirements on the underlying transport protocols. 

We revisit the buffer sizing problem with the newly evolved TCP-style reactive congestion control (RCC)~\cite{reno, vegas, cubic, copa, bbr} algorithms, and find that shallow buffer can easily get overwhelmed by the wide-area traffic (\S\ref{motivation-reactive}). Specifically, running legacy reactive transport protocols over shallow-buffered WAN may lead to either high packet losses or degraded throughput or both. 
To mitigate these problems, current practice seeks help from global traffic engineering~\cite{b4, swan, b4after}, endhost rate limiting~\cite{bwe}, or traffic scheduling with differentiated services.

\begin{figure}[!t]
\centering
\includegraphics[width=0.95\textwidth]{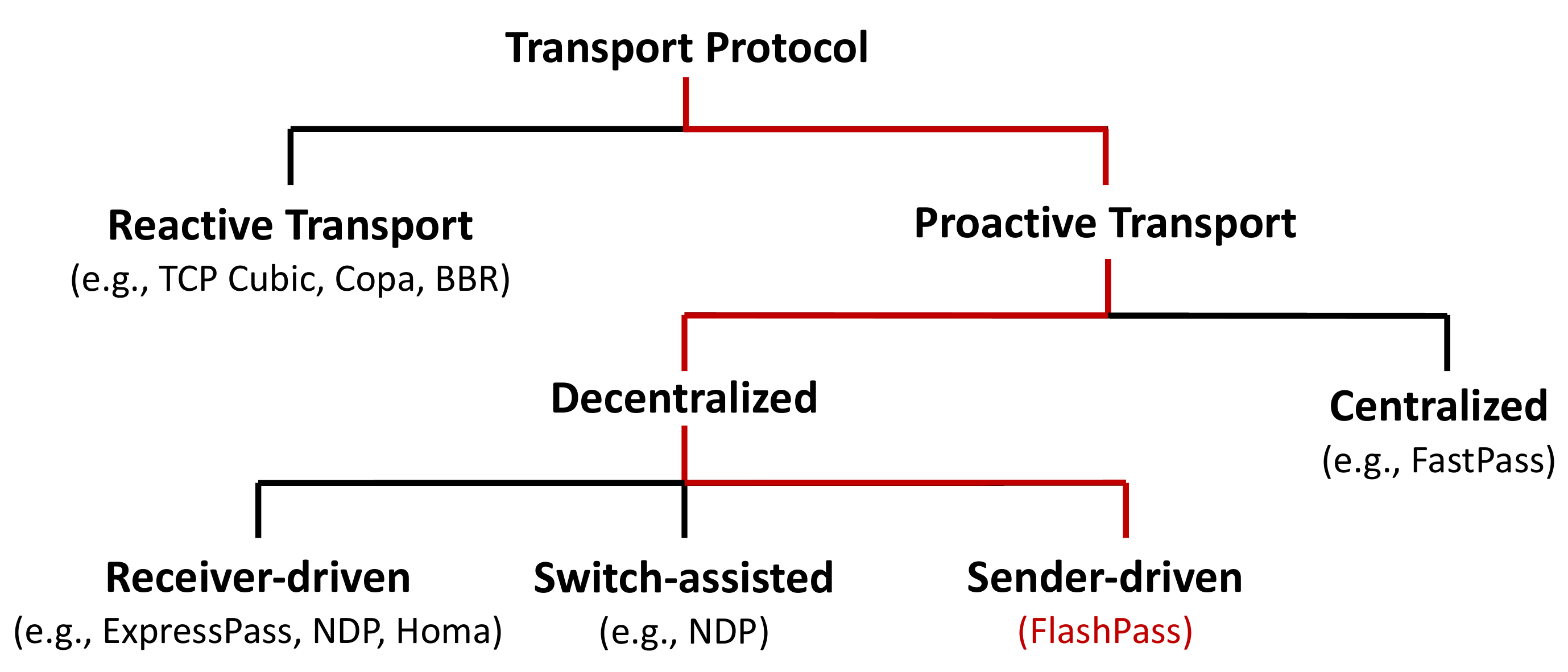}
\caption{Design space of transport protocols.} 
\label{fig:design-space}
\end{figure}

Instead, we ask the question: can we design a transport to simultaneously achieve low loss rate and high throughput under shallow-buffered WAN? Inspired by the emerging proactive congestion control (PCC)~\cite{fastpass, expresspass, ndp, homa, aeolus} design (Figure~\ref{fig:design-space}), we answer this question affirmatively by taking the initiative to extend the PCC idea for shallow-buffered WAN.

However, while PCC has been proven to work well in datacenter network (DCN), we find several practicality issues to make it work on WAN (\S\ref{motivation-proactive}). First of all, centralized PCC~\cite{fastpass} and switch-assisted PCC~\cite{ndp} protocols impose high requirements on network facilities (e.g., cutting payload~\cite{cp}) and hence are either unscalable or impractical. Besides, some receiver-driven protocols like Homa~\cite{homa} are based on the assumption of single bottleneck between the top-of-rack (ToR) switch and receiver, which does not hold on WAN. 

While credit-emulated PCC like ExpressPass~\cite{expresspass} seems to work, we find it may suffer from efficiency issues on WAN. Firstly, unlike homogeneous DCN, wide-area traffics have much diverse RTTs. The well scheduled credits on the reverse path may still trigger data packet crush on the data forward path due to different turn-around times.
Secondly, there is one RTT delay for credits to trigger data sending for both starting phase (to start data sending quicker) and finishing phase (to stop credit generation in time) in PCC. Such an overhead is prohibitive given much higher RTT on WAN.

Therefore, we propose a novel PCC solution - \sysB to address these challenges (\S\ref{sec-flashpass:design}). Firstly, to address the imperfect credit scheduling issue, \sysB leverages a sender-driven emulation process together with send time calibration. Unlike receiver-driven protocols, sender-driven \sysB can exactly rehearse the future data sending in the same direction on the emulation network. With the addition of the timestamp information on the emulation and credit packets, \sysB can strictly schedule data sending in the time space.
Secondly, to mitigate the impact of one RTT delay for credits to trigger data sending, \sysB leverages Aeolus~\cite{aeolus} to enable early data transmission in the starting phase, and further incorporates a credit over-provisioning mechanism together with a selective dropping discipline for efficient credit or bandwidth allocation in the flow finishing phase.

We measured realistic workload on a production wide-area network, and experimented \sysB with NS2 simulation to demonstrate its superior performance (\S\ref{sec-flashpass:evaluation}). 
In static workload experiments, \sysB achieves near full throughput (9.14Gbps) with zero packet loss persistently across various settings under the shallow-buffered WAN. The throughput is up to 55.9\% higher than that of TCP Cubic. While comparing to ExpressPass, \sysB achieves the similar throughput with zero packet loss rate (up to 0.12\% losses for ExpressPass).
In realistic dynamic workload experiments, \sysB (with Aeolus enhancement) reduces the overall flow completion times of TCP Cubic and ExpressPass (also with Aeolus enhancement) by up to 32\% and 11.4\%; and the reduction of small flow 99-th tail completion times can get up to 49.5\% and 38\%, respectively.
We also presented a practical deployment analysis (\S\ref{sec-flashpass:discussion}) for implementing \sysB, however, building a fully functional prototype is beyond the scope of this work. 

\section{Background and Motivation} \label{sec-flashpass:motivation}

\subsection{Shallow-buffered Network} \label{motivation-wan}

The last decade has witnessed an exponential growth of web applications and services (e.g., web search, cloud computing, social networking, etc.). This drives the large Internet companies (e.g., Google~\cite{google-dc}, Microsoft~\cite{microsoft-dc}, Facebook~\cite{facebook-dc}, and Alibaba~\cite{alibaba-dc}, etc.) to build the modern data centers (DCs) at an unforeseen speed and scale across the globe. 
With the ever-increasing communication demand, traditional network infrastructure built with commercial switches~\cite{dc-computer} fall short to meet the scale, management, and cost requirements. 

To address it, inspired by the then-emerging merchant switching silicon industry~\cite{dc-silicon}, large enterprises start to build and deploy their own customized networking hardware both on WAN~\cite{b4, clos-router} and DCN~\cite{dc-silicon, jupiter}.
However, while the cutting-edge merchant silicon provides the highest bandwidth density in a cost effective way, the shallow buffer (Table~\ref{tab:buffer}~\cite{buffer}) shipped with it can degrade the network performance to a great extent. For example, Google has reported its experience of high packet losses on the shallow-buffered WAN~\cite{b4} and DCN~\cite{jupiter}. 

\begin{table*}[!t]
\centering
\resizebox{\textwidth}{!}{%
\begin{tabular}{|l|l|l|l|l|}
\hline
\textbf{Switching Chips} 	& \textbf{BCM Trident+} & \textbf{BCM Trident2} & \textbf{BCM Trident3} & \textbf{BCM Trident4} \\ \hline
\textbf{Capacity (ports$\times$BW)}    & 48$\times$10Gbps           & 32$\times$40Gbps           & 32$\times$100Gbps          & 32$\times$400Gbps          \\ \hline
\textbf{Total Buffer}             & 9MB                   & 12MB                  & 32MB                  & 132MB                 \\ \hline
\textbf{Buffer per port}          & 192KB                 & 384KB                 & 1MB                   & 4.125MB               \\ \hline
\textbf{Buffer / Capacity} & 19.2KB                & 9.6KB                 & 10.2KB                & 10.56KB               \\ \hline
\end{tabular}
}

\vspace{1em}

\resizebox{0.7\textwidth}{!}{%
\begin{tabular}{|l|l|l|}
\hline
\textbf{Switching Chips} 	& \textbf{BF Tofino} & \textbf{BF Tofino2} \\ \hline
\textbf{Capacity (ports$\times$BW)}    & 64$\times$100Gbps       & 32$\times$400Gbps        \\ \hline
\textbf{Total Buffer}             & 22MB               & 64MB                \\ \hline
\textbf{Buffer per port}         & 344KB              & 2MB                 \\ \hline
\textbf{Buffer / Capacity} & 3.44KB             & 5.12KB              \\ \hline
\multicolumn{3}{l}{$^{\mathrm{a}}$BCM is short for Broadcom. BF is short for (Intel) Barefoot.}
\end{tabular}
}
\caption{Buffer size for commodity switching chips.}
\label{tab:buffer}
\end{table*}

The buffer pressure is especially high for WAN communication. Conventional wisdom on buffer sizing problem~\cite{buffer-sizing} dictates that one bandwidth-delay product (BDP) buffering is required to achieve full link utilization in the worst case (i.e., with synchronized flows). However, the commodity switching chips provide shallow buffer of less than 20KB per port per Gbps according to Table~\ref{tab:buffer}. That is even lower than 0.1\% of WAN BDP (25MB per Gbps assuming 200ms RTT). Thus, it is extremely challenging to deliver low loss rate and high throughput simultaneously on shallow-buffered WAN.


\subsection{Reactive Congestion Control (RCC) is Insufficient} \label{motivation-reactive}
We revisit the buffer sizing~\cite{buffer-sizing}  problem, and see if the state-of-art reactive congestion control (RCC) protocols~\cite{reno, vegas, cubic, copa, bbr} perform well for shallow-buffered WAN. The theoretical analysis is summarized in Table~\ref{tab:rcc}:
\begin{itemize}
\item \textbf{TCP NewReno}\cite{reno} is the seminal loss-based congestion control algorithm. It follows the additive-increase and multiplicative-decrease (AIMD) control rule. The conventional buffer sizing theory~\cite{buffer-sizing} indicates one BDP buffer is required for full link utilization.
\item \textbf{TCP Cubic}\cite{cubic} is loss-based congestion control and enabled by default in Linux sysBtem. It increases window size based on a cubic function of time and decreases multiplicatively by a fraction of $\beta$=0.2 by default on loss. The resulting buffer requirement is BDP/4. 
\item \textbf{TCP Vegas}\cite{vegas} reacts to both delay and loss signal. It applies additive-increase and additive-decrease (AIAD) control rule based on delay signal to control the lower and upper bound of excessive packets in flight, and also performs multiplicative decrease on loss signal. The buffer requirement is 5$\times$flow\# packets.
\item \textbf{Copa}\cite{copa} is based on delay signal. It adjusts sending rate towards $1/(\delta d_q)$ by AIAD control rule, where $d_q$ is the measured queueing delay. With default $\delta=0.5$, the buffer requirement is 5$\times$flow\# packets.
\item \textbf{BBR}\cite{bbr} is model-based congestion control. It tries to drive the transport to the theoretical optimal point~\cite{kleinrock1979power} based on accurate bandwidth and RTT estimation. BBR bounds the inflight packets to cwnd\_gain$\times$BDP by default. 
However, accurate bandwidth estimation is difficult to achieve, often leading to high buffering in practice.
\end{itemize}

\begin{table*}[!t]
\centering
\resizebox{0.8\textwidth}{!}{
\begin{tabular}{|l|l|l|}
\hline
\textbf{Protocol}   & \textbf{Signal}  & \textbf{Algorithm} \\ \hline
\textbf{TCP NewReno}~\cite{reno} & Loss             & AIMD                \\ \hline
\textbf{TCP Cubic}~\cite{cubic}   & Loss             & AIMD                                     \\ \hline
\textbf{TCP Vegas}~\cite{vegas}   & Delay \& loss    & AIAD (MD on loss)                                        \\ \hline
\textbf{Copa}~\cite{copa}        & Delay only       & AIAD                \\ \hline
\textbf{BBR}~\cite{bbr}         & No direct signal & Not incremental                  \\ \hline
\end{tabular}
}

\vspace{1em}

\resizebox{\textwidth}{!}{
\begin{tabular}{|l|l|l|l|}
\hline
\textbf{Protocol}   &  \textbf{Buffer requirement for high throughput and low loss} \\ \hline
\textbf{TCP NewReno}~\cite{reno}  &  $\beta$ / (1 - $\beta$) $\times$ BDP = BDP   ~~($\beta$ = 0.5)                        \\ \hline
\textbf{TCP Cubic}~\cite{cubic}   &  $\beta$ / (1 - $\beta$) $\times$ BDP = BDP/4   ~~($\beta$ = 0.2)                      \\ \hline
\textbf{TCP Vegas}~\cite{vegas}   & ($\beta$ + 1) $\times$ n = 5n pkts ~~(n = flow\#, $\alpha$ = 2 pkts, $\beta$ = 4 pkts)                                      \\ \hline
\textbf{Copa}~\cite{copa}         & 2.5n / $\delta$ = 5n pkts ~~(n = flow\#, $\delta$ = 0.5 / pkt)              \\ \hline
\textbf{BBR}~\cite{bbr}          & (cwnd\_gain-1)$\times$BDP in Probe\_BW phase (cwnd\_gain=2)              \\ \hline
\multicolumn{2}{l}{$^{\mathrm{a}}$Copa and BBR are both insensitive to loss signal, i.e., no fast loss recovery.}
\end{tabular}
}
\caption{Reactive congestion control (RCC) protocols for WAN.}
\label{tab:rcc}
\end{table*}

Based on the analysis, we find that the reactive congestion control protocols all require non-negligible buffering for high throughput and low loss rate. Even worse, the buffer requirement is unscalable (in proportion to either network capacity or flow number) as network evolves and traffic demand grows. Note that this problem is fundamental to RCC protocols, because they can detect congestion only after the formation of queues and require one RTT delay before taking reaction\footnote{We do not consider RCC protocols using advanced switch features such as in-band network telemetry (INT) that are often unavailable on WAN. We leave the discussion of these protocols to \S\ref{sec-flashpass:related}.}.

We run NS3 simulation to illustrate the performance degradation of reactive protocols under shallow buffer. We generate 200 parallel flows from different senders to single receiver sharing the same 10Gbps bottleneck link with 40ms RTT (thus BDP=50MB) for 5 seconds. We experiment buffer size of 0.2, 0.5, 1, 2, 5, 10, 20, 50, 100, and 200~MB. Commodity switching chips should have shallow buffer of no more than 0.2MB per port in 10Gbps network (according to Table~\ref{tab:buffer}). We enable selective acknowledgement (SACK) for efficient loss detection and retransmission.

\begin{figure}[!t]
\centering
\includegraphics[width=0.9\textwidth]{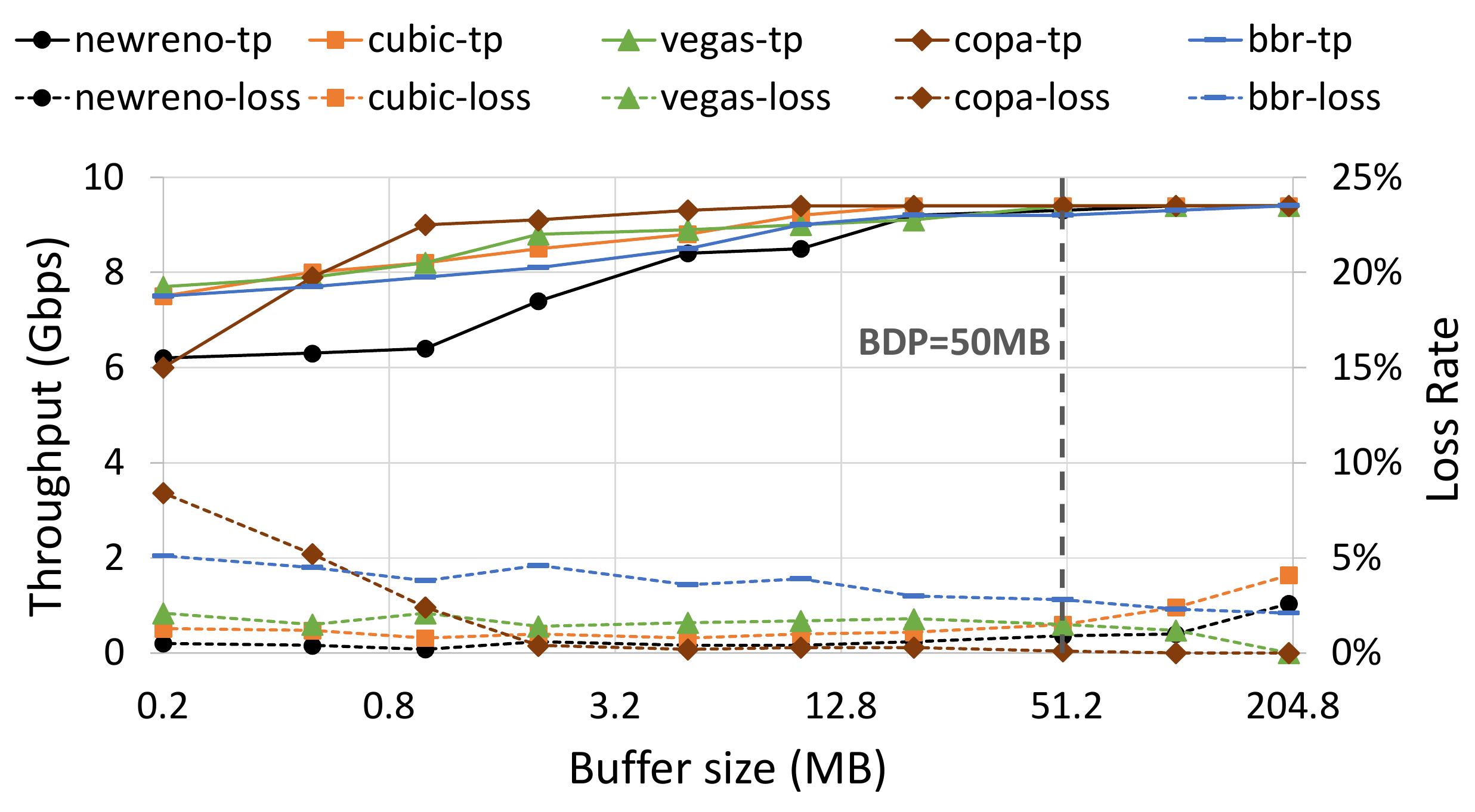}
\caption{Performance of reactive transports under shallow buffer. The solid lines indicate throughput (or tp) on the left y-axis, and the dash lines indicate loss rate on the right y-axis.} 
\label{fig:reactive-cc}
\end{figure}

Experimental results are shown in Figure~\ref{fig:reactive-cc}. We find that the total throughput drops by 18\%-37\% as buffer decreases to the shallow size (i.e., 200KB under 10Gbps). The loss rate also increases dramatically for some protocols (e.g., 8.4\% for Copa and 5.1\% for BBR). We observe significant losses at the end of slow start (time at $\sim$0.5 second). This is because in the slow start phase, window sizes are doubling every RTT, resulting in packet losses of roughly half a window size at the end. 
Therefore, we conclude that with RCC protocols, shallow buffer can easily get overwhelmed by large BDP wide-area traffic, leading to high packet losses and degraded throughput.

\subsection{Proactive Congestion Control (PCC) as a Solution} \label{motivation-proactive}

Inspired by the emerging proactive congestion control  (PCC)~\cite{fastpass, expresspass, ndp, homa, aeolus} design (Figure~\ref{fig:design-space}), we now explore the possibility of employing PCC protocols for shallow-buffered WAN. 
Unlike RCC that uses a ``try and backoff" approach, PCC operates in a ``request and allocation" style. The key conceptual idea is to explicitly allocate the bandwidth of bottleneck link(s) among active fows and proactively prevent congestion. As a result, ultra-low buffer occupancy and (near) zero packet loss can be achieved. Furthermore, PCC reaches peak rate in roughly one RTT, avoiding the long convergence time and high losses of RCC slow start.   

However, as most of existing PCC protocols are designed for DCN, many of them cannot work practically on WAN. Centralized PCC~\cite{fastpass} and switch-assisted PCC~\cite{ndp} protocols impose high requirements on network facilities (e.g., cutting payload~\cite{cp}) and hence are either unscalable or impractical. Some receiver-driven protocols like Homa~\cite{homa} adopt simple credit scheduling on the receiver side, assuming that there is single bottleneck link between the top-of-rack (ToR) switch and receiver. However, this assumption does not hold on WAN. 

\begin{figure}[!t]
\centering
\includegraphics[width=0.9\textwidth]{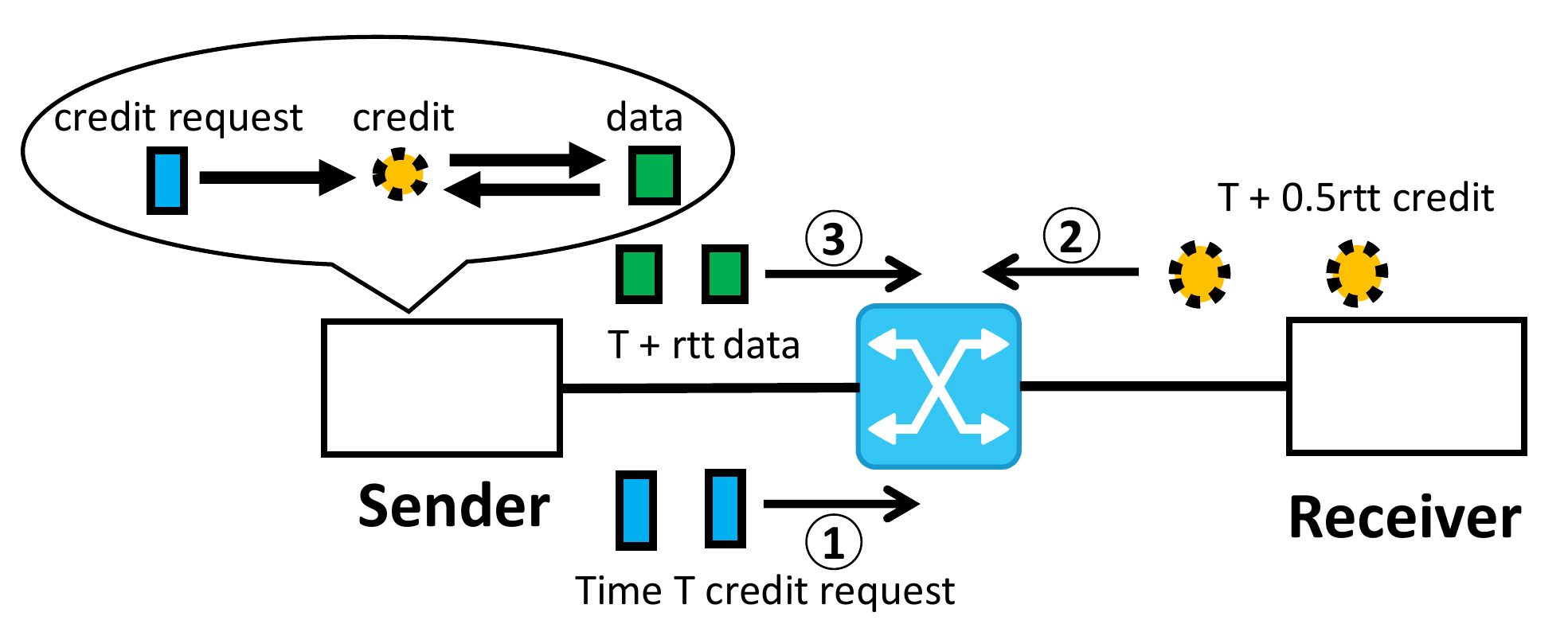}
\caption{Receiver-driven proactive congestion control (PCC).} 
\label{fig:receiver-driven}
\end{figure}

There are other receiver-driven protocols like ExpressPass~\cite{expresspass} that leverage sophisticated credit emulation process on a separate queue for credit allocation. Figure~\ref{fig:receiver-driven} shows an overview of receiver-driven credit-emulated PCC protocols. Specifically, they generate credit packets on the reverse rate-limited path to emulate data sending. Each minimum-sized credit packet (e.g., 84B) passing through the network triggers the sender to transmit a MTU-sized data packet (e.g., 1538B). 
Thus, the credit queue is rate-limited to 84/(84+1538) $\approx$ 5\% of the link capacity, and the remaining 95\% is used for transmitting data packets.  
This makes them possible to work on WAN without assuming non-congested network core.

Therefore, we select ExpressPass~\cite{expresspass} as our baseline PCC design for WAN. We validate its superior performance over RCC protocols following the same static workload experiment in \S\ref{motivation-reactive} with the open-sourced NS2 simulation code~\cite{expresspass-code}. The results show that, ExpressPass achieves 9.5Gbps throughput and zero packet loss when running static workload with same RTTs on WAN  (see more results in Figure~\ref{fig:static_exp}).

However, we find that existing PCC protocols suffer from several efficiency issues that are particularly serious on WAN with different RTTs, especially under dynamic workload. 

\textbf{Problem 1:} Unlike homogeneous DCN, wide-area traffics have much diverse RTTs (not as simple as cases in \S\ref{motivation-reactive}). The well scheduled credit on the reverse path may still trigger data packet crush or bandwidth under-utilization on the data forward path due to different turn-around times of flows. 

\begin{figure}[!t]
\centering
\includegraphics[width=0.8\textwidth]{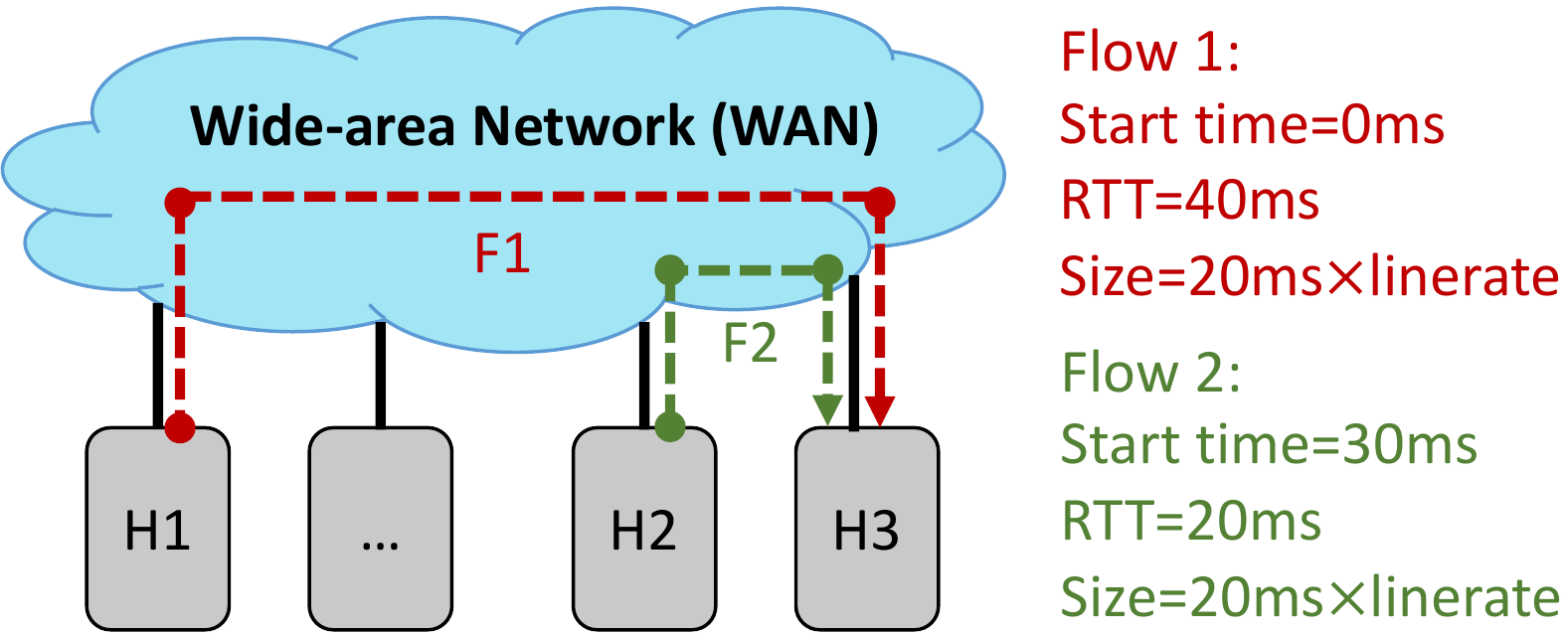}
\caption{Competing flows have different RTTs on WAN.} 
\label{fig:case1}
\end{figure}

We illustrate the problem with a case shown in Figure~\ref{fig:case1}. There are two flows from host H1 and host H2 to host H3, competing on the same bottleneck link that connects down to H3. The RTTs are 40ms and 20ms, respectively. Both have a traffic demand size of 20ms$\times$linerate. Table~\ref{tab:case1-design} shows the running process of receiver-driven PCC protocols. Flow~1 from H1 starts at 0ms, i.e., the credit request send time. After half a RTT, i.e., at time 20ms, the credit request reaches the receiver side H3 and triggers the credit generation. The credit generation lasts for 20ms and stops at 40ms, which is based on the traffic demand size. Flow~2 from H2 starts at 30ms and triggers credit generation from H3 during time 40-60ms. Thus, the credit generation of Flow~1 and Flow~2 are well interleaved. All credits then pass through the network successfully. They trigger out the real data sending, and the data arrival time are approximately 1 RTT after the credit send time. Ideally (implicitly assuming the same RTT for different flows in DCN), we expect the data packets to get through the network without crush with each other. However, in this case, giving different RTTs for Flow~1 and Flow~2 on WAN, we find the data arrival times of two flows are the same, i.e., 60-80ms. This leads to severe congestion, causing to large queueing and packet losses in the shallow-buffered WAN. 

\begin{table}[!t]
\centering
\begin{tabular}{|c|c|c|c|c|}
\hline
       & Start time & Credit send & Data send & Data arrive   \\ \hline
Flow 1 & 0ms        & 20-40ms     & 40-60ms   & {\color{magenta} $\sim$60-80ms}\\ \hline
Flow 2 & 30ms       & 40-60ms     & 50-70ms   & {\color{magenta} $\sim$60-80ms} \\ \hline
\multicolumn{5}{l}{$^{\mathrm{a}}$Simultaneous data arrival leads to high buffering or packet loss.}
\end{tabular}
\caption{Data crush of receiver-driven PCC (Figure~\ref{fig:case1}).}
\label{tab:case1-receiver-driven}
\end{table}

In theory, maximum queueing $Q_{max}$ can be calculated as:
\begin{equation}
\label{equ:queue}
Q_{max} = \Delta RTT \times linerate
\end{equation}
where $\Delta RTT$ is the maximum RTT difference among flows.

We run simulation under the Figure~\ref{fig:case1} scenario with 10Gbps links. We set a very large buffer size in the bottleneck link down to the host H3. We set initial credit rate to maximum and terminate credit generation as soon as the flow demand is reached. Results show that the maximum queueing reaches 8532 1.5KB-MTU-sized packets or 12.5MB, which exactly matches our theoretical analysis (i.e., Equation~\ref{equ:queue}). This will lead to severe packet losses on the shalow-buffered WAN. While in another simulation, we change the RTT of Flow 2 to be 40ms (the same RTT as Flow 1) and vary the start time to be 0ms, 10ms, 20ms or 30ms; and repeat the same experiment. We find that the maximum queueing length drops dramatically to no more than 6 packets. This explains why ExpressPass does not work well on WAN with large RTT difference, even if it may work for homogeneous network such as DCN.

\textbf{Problem 2:} There is one RTT delay for credits to trigger data sending. This impacts both starting phase (to start data sending quicker) and finishing phase (to stop credit generation in time) of a flow in PCC. Such an overhead is prohibitive given large RTT on WAN. While we find that recent work (e.g., Aeolus~\cite{aeolus,aeolus-ton}) has addressed the problem of starting phase and can be extended on WAN similarly, the problem of finishing phase remains unsolved. 

For the flow finishing phase, the delayed data sending may either lead to low network utilization with default aggressive credit generation, or possibly increase the flow completion time by early termination of credit generation.
As mentioned earlier, to work for the core-congested WAN, existing PCC protocols should leverage receiver-driven credit emulation. Then, receiver cannot determine the exact amount of credit successfully granted to each flow. This is problematic as it means that the receiver also cannot determine the exact time to stop credit generation. 
The receiver has two choices. 
On one hand, if it keeps sending credits until getting the last data packets, then there will be roughly one RTT wastage of credits, leading to network under-utilization.  
On the other hand, if it stops credit generation immediately when in-flight credits are enough to cover the flow traffic demand, the flow may need to request for more credits when some credits are dropped in the emulation process, leading to higher network latency.

\begin{figure}[!t]
\centering
\includegraphics[width=0.9\textwidth]{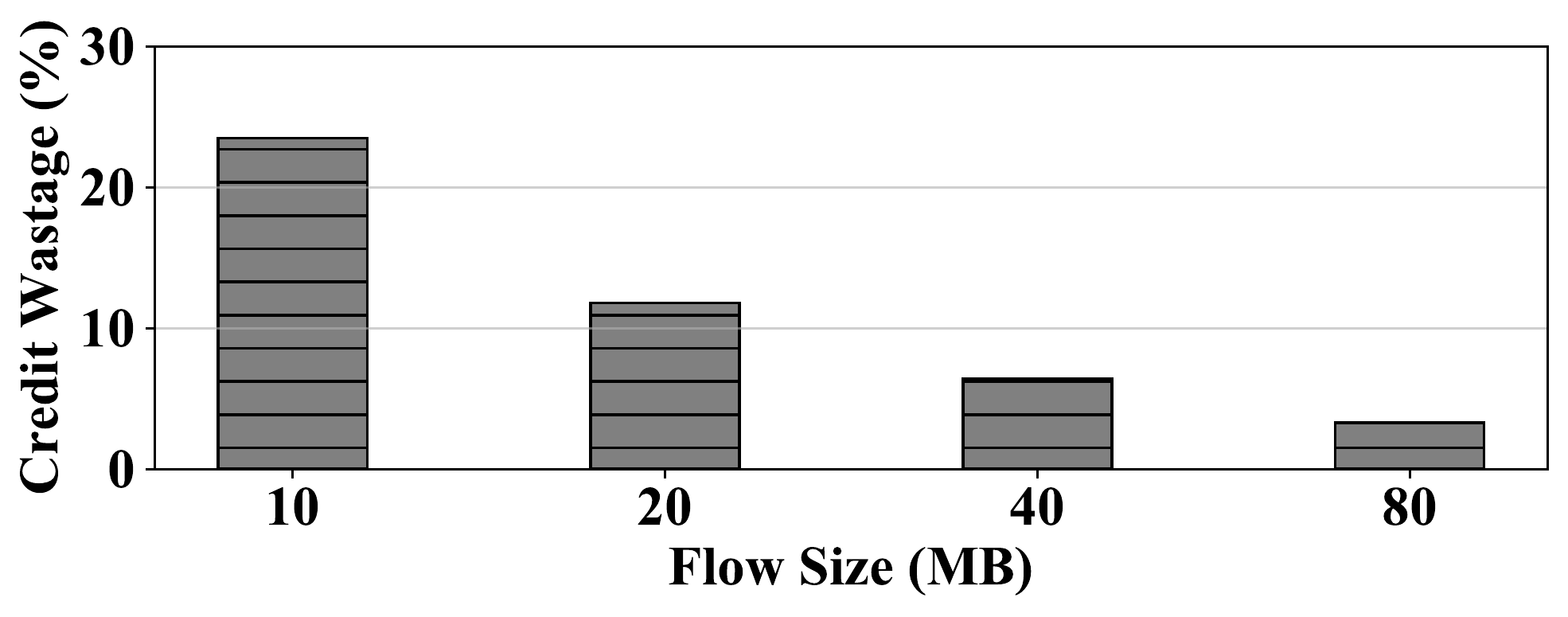}
\caption{Credit wastage of ExpressPass.} 
\label{fig:credit-wastage}
\end{figure}

Figure~\ref{fig:credit-wastage} shows our experimental result of running realistic workload on a network with an average flow RTT of 60ms and bottleneck link of 1Gbps (same setup as \S\ref{dive-overprovisioning}). We generate synthetic workload~\cite{annulus}. Flow sizes are varied from 10MB to 80MB (based on our WAN measurement in ~\S\ref{evaluation-setup}). The average network load is set to 0.8 of the full network capacity. We find that the credit wastage\footnote{The credit wastage is measured by the received but not used credits.} of ExpressPass can get up to 23.5\% of the total successfully received credits. Notice that the ExpressPass paper~\cite{expresspass} shows even higher credit wastage up to 60\%  in a workload with many small flows.
\section{Design} \label{sec-flashpass:design}

\subsection{Design Overview} \label{design-overview}
In this work, we aim to design a practical and efficient transport protocol to simultaneously achieve high throughput and low loss rate for shallow-buffered WAN (\S\ref{motivation-wan}). Our investigation over reactive transport shows its inherent insufficiency, requiring non-negligible buffering (\S\ref{motivation-reactive}). 
We then turn to the emerging proactive transport. While revealing the promising potential of the receiver-driven PCC protocols, we also find two key technical challenges (\S\ref{motivation-proactive}):
\begin{enumerate}
\item How to schedule credits effectively without triggering data crush even if network traffic has diverse RTTs?
\item How to generate credits sufficiently while not wasting credits even if the granted amount is unpredictable?
\end{enumerate}

The first challenge is fundamental to the receiver-driven protocols. This is because receiver-driven protocols uses the credit sending on the reverse path to emulate forward path data sending. The network delay between receiver and bottleneck link is different from that between sender and bottleneck. Thus, interleaved credits passing through the reverse path cannot guarantee well interleaved incoming data at the bottleneck. To address it, we propose to adopt a sender-driven emulation mechanism. This ensures that the emulation follows the same direction as the real data sending. Besides, strict timing should be enforced to the data sending. This ensures that the delay for emulation/credit packet to trigger data packet out keeps constant, instead of dependent on the different RTTs.

To address the second challenge, we should keep generating credits even if the expected incoming credits are enough to cover the flow traffic demand (we call it \textit{over-provisioning}). This is to ensure that flows can finish quickly with sufficient credits, even if the amount of credits successfully passing through the emulation network and granted to the sender is unknown. However, we should also make sure these over-provisioned credits waste no bandwidth, i.e., they should only occupy the leftover bandwidth by the ordinary traffic. To this end, we incorporate a selective dropping mechanism to grant the over-provisioned credits in a ``best-effort'' manner.

Therefore, we propose a novel PCC solution - \sysB to simultaneously deliver high throughput and low loss rate for shallow-buffered WAN. \sysB is the first sender-driven PCC protocol. It leverages a sender-driven emulation process together with send time calibration to effectively allocate credits for bandwidth and eliminate the data packet crush in time (\S\ref{design1}). It enables early data transmission in the flow starting phase, and incorporates credit over-provisioning together with a selective dropping mechanism to efficiently utilize network bandwidth in the flow finishing phase (\S\ref{design2}). 

\begin{figure}[!t]
\centering
\includegraphics[width=0.9\textwidth]{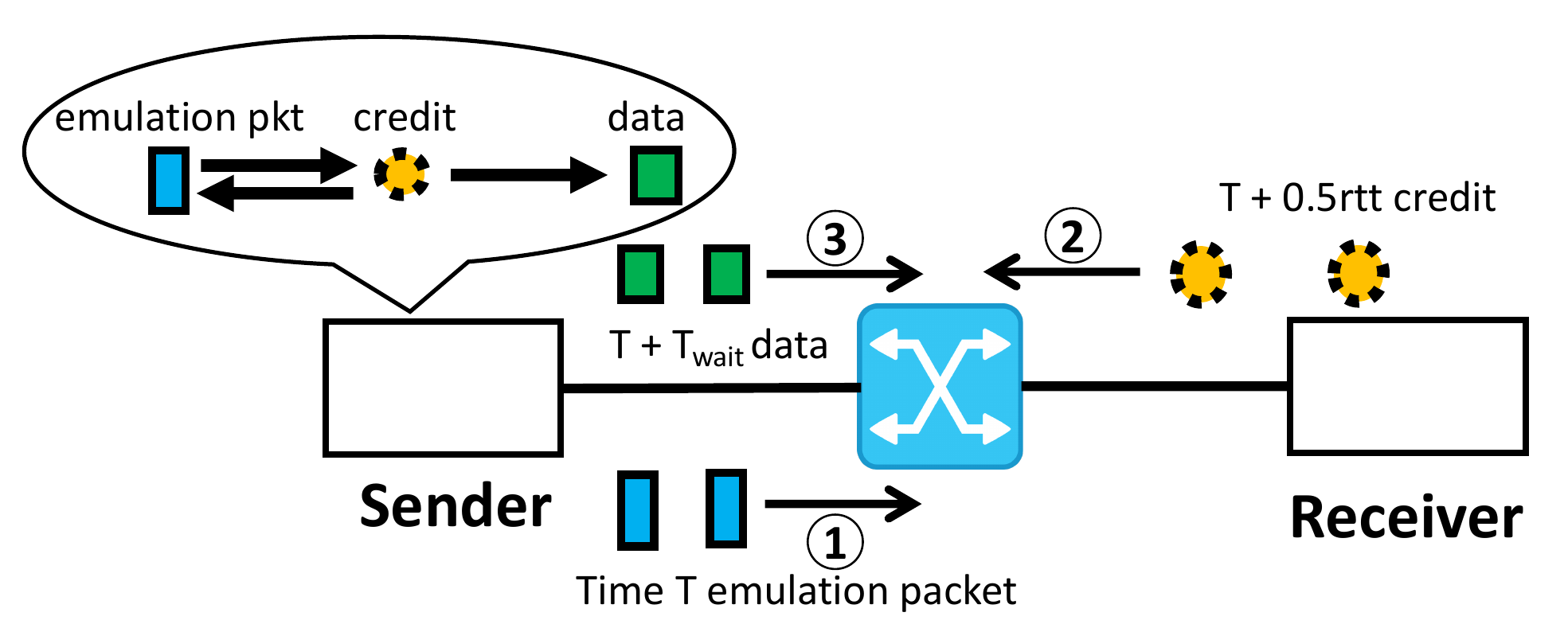}
\caption{Sender-driven emulation mechanism.} 
\label{fig:design}
\end{figure}

\subsection{Sender-driven Emulation Mechanism} \label{design1}
\sysB leverages a sender-driven emulation mechanism (Figure~\ref{fig:design}) to allocate network bandwidth without data packet crush.
The emulation process requires to separate the network into two parts: an emulation network running on the emulation queue, and a real data communication network on the data communication queue. Emulation packet is set to the minimum size, i.e., 84B Ethernet frame. The emulation packet passing through the network triggers a credit return from receiver and then a sender data transmission up to a maximum size Ethernet frame (e.g., 1538B). Thus, in Ethernet, the emulation network is rate-limited to 84/(84+1538) $\approx$ 5\% of the link capacity, and the remaining 95\% is used for the data communication. Unlike existing PCCs that send data packets as soon as credits arrive, \sysB encodes timestamps ($T$) in its emulation packets and credit packets, and grants data sending at exact time of $T+T_{wait}$ to eliminate data crush at the bottleneck link. This send time calibration mechanism effectively helps in avoiding data packet crush in time as shown below.

Figure~\ref{fig:design} shows the running process of the sender-driven emulation mechanism. 
\circled{1} At time $T$, a new flow arrives and immediately generates emulation packet at full rate of the emulation network. The emulation packet will compete on the emulation network and get dropped if queues form. The emulation network is configured with a low buffer of several packets.
\circled{2} At around time $T+0.5rtt$, emulation packet passes through the network and triggers the credit feedback from the receiver side. And around half a RTT later, the sender receives and records the credit packet.
\circled{3} At time $T+T_{wait}$,  the sender injects the data packet correspondingly into the data communication network. Note that $T_{wait}$ should be larger than the maximum RTT of the network. 
In this way, emulation network exactly mimics the bandwidth competition of the future data communication network. If network is over-utilized, only emulation packets get dropped on the emulation network and data communication network remains zero buffering.

To further illustrate the effectiveness of sender-driven emulation mechanism, we compare the result of \sysB (Table~\ref{tab:case1-design}) with that of the receiver-driven PCC (Table~\ref{tab:case1-receiver-driven}) - the running process both in the case of Figure~\ref{fig:case1}. In \sysB, the $T_{wait}$ is set to the maximum RTT of the network, i.e., 40ms. We find that the Flow~2 adjusts its data send time to 70-90ms, and thus lead to $\sim$80-100ms data arrival time at the receiver side. This effectively avoids the data crush with Flow~1 during time 60-80ms as with the receiver-driven solution. We should also notice that receiver-driven solution cannot resolve the data crush problem in general by simply adding a constant waiting time $T_{wait}$, even if it seems to work in this case (bottleneck at the ToR-to-receiver link). This is because in practice, bottleneck link may reside in any hop of the network, and the time or distance to reach the bottleneck is different and unknown for both senders and receivers. Thus, it is fundamentally difficult to avoid data crush without faithfully mimic the data sending from a sender-driven approach.

\begin{table}[!t]
\centering
\begin{tabular}{|c|c|c|c|c|}
\hline
       & Start time & Credit send & Data send & Data arrive   \\ \hline
Flow 1 & 0ms        & 20-40ms     & 40-60ms   & {\color{magenta} $\sim$60-80ms} \\ \hline
Flow 2 & 30ms       & 40-60ms     & 70-90ms   & {\color{magenta} $\sim$80-100ms} \\ \hline
\multicolumn{5}{l}{$^{\mathrm{a}}$Interleaved data arrival avoids high buffering and packet loss.}
\end{tabular}
\caption{Interleaved data arrival of \sysB (Figure~\ref{fig:case1}).}
\label{tab:case1-design}
\end{table}

\textit{Emulation feedback control:} \sysB uses an emulation feedback control to regulate the sending rate of emulation packets. This is to ensure high bandwidth utilization and fairness, which is similar to ExpressPass\footnote{More detailed illustration can be found in Figure~4 of ExpressPass~\cite{expresspass}}. For example, in the parking lot scenario, naively sending out emulation packets at linerate will lead to link under-utilization. To address it, a feedback control is required for the emulation process. 

Unlike data forwarding, the emulation feedback control has low cost in its emulation packet losses and thus can be aggressive in probing for more bandwidth. Observing this characteristic, \sysB adopts a simple yet effective loss-based feedback control as shown in Algorithm~\ref{alg:design}. The emulation packet loss rate is controlled between $min\_target\_loss$ and $max\_target\_loss$ (by default 1\% and 10\%, respectively). The rate adjustment follows a multiplicative-increase and multiplicative-decrease (MIMD) control rule. With the record of current loss rate and the target loss range, the increase and decrease adjustment can be calculated precisely. In this way, \sysB emulation process can achieve fast convergence to full link utilization. Notice that \sysB emulation process allows fast start for new flows while still guaranteeing low credit waste with the selective dropping mechanism.

\begin{figure}[htb]
\centering
\begin{minipage}{.85\linewidth}
\begin{algorithm}[H]
\caption{Emulation Feedback Control at \sysB Sender.} \label{alg:design}
\SetKwInOut{Input}{Input}
\SetKwInOut{Output}{Output}
\SetKwInOut{Initialize}{Initialize}
\SetInd{1.1em}{1.1em}

\nonl \tikzmk{A}
\Input{\hspace{0.1em} New Incoming Credit Packet}
\Output{\hspace{0.1em} Emulation Packet Sending Rate $cur\_rate$}
\Initialize{\hspace{0.1em} $cur\_rate \leftarrow max\_emulation\_rate$} 

\tikzmk{B} \boxit{mygrey}
\nonl \tikzmk{A} \nl
\tcc{Update the emulation loss rate}
$loss\_rate = 1 - \#\_credit\_in / \#\_emulation\_out$ \;

\tikzmk{B} \boxit{myblue}
\nonl \tikzmk{A} \nl
\tcc{Update sending rate every RTT}
\If{loss\_rate $>$ max\_target\_loss}{
  $cur\_rate = cur\_rate \times
  (1-loss\_rate) \times (1+min\_target\_loss)$ \;
} 
\ElseIf{loss\_rate $<$ min\_target\_loss}{
  $cur\_rate = cur\_rate \times (1+max\_target\_loss)$ \;
}
\nonl \tikzmk{B} \boxend{mypink}
\end{algorithm}
\end{minipage}
\end{figure}

\textit{Sender-driven vs receiver-driven?} There are both pros and cons for sender-driven approach. On one hand, sender-driven approach can emulate the data forwarding with precisely interleaved arrival time at the bottleneck link, and thus can practically achieve near zero buffering. This is the most desirable feature in our targeting shallow-buffered WAN scenario, and explains why we choose the sender-driven approach in our design. Besides, it does not need to ensure path symmetry like the receiver-driven protocols, where data packet must exactly follow the reverse path of the credit emulation. 

On the other hand, sender-driven approach introduces half a RTT longer delay between emulation and data sending. Such a longer delay can have negative impact without careful consideration. In \sysB, we have well handle this problem for both the starting phase and finishing phase. Besides, it also adds one-way more control packets. To reduce this overhead, \sysB adopts delayed credit sending. Specifically, it waits for more credits of the same flow and feeds them back in single packet if timestamp info indicates a long waiting time for data sending. The delayed feedback credit packets should include timestamps of all corresponding emulation packets.

\subsection{Over-provisioning with Selective Dropping Mechanism} \label{design2}
To handle the credit delay problem in the flow starting phase, \sysB adopts similar idea as in Aeolus~\cite{aeolus}. Specifically, new flows start at line rate in the first RTT before receiving any credit (i.e., \textit{pre-credit phase}). The unscheduled data packets in the pre-credit phase are selectively dropped if meeting the (non-first-RTT) scheduled packets in the network. In case of first-RTT packet losses, a tail loss probing is employed together with credit-scheduled retransmission. 

To handle the credit wastage problem in the flow finishing phase (unsolved by Aeolus), \sysB incorporates a credit over-provisioning mechanism with selective dropping discipline (a major contribution in this work). Specifically, \sysB keeps generating emulation packets and triggering out credit packets even if the expected incoming credits are enough to cover the flow traffic demand (we call it \textit{over-provisioning}). While during the packet emulation process, the over-provisioned emulation packets are selectively dropped if conflicting with the ordinary emulation packets in the network. This ensures that the over-provisioned credits only occupy the leftover bandwidth not used by the ordinary traffic demand.

However, it is also challenging to determine the right time to switch emulation packets from ordinary to over-provisioned. To this end, we need to maintain an estimation of the incoming credits. If the estimate cannot cover the remaining traffic demand, emulation packets are set to be ordinary. Otherwise, emulation packets are over-provisioned.  

\begin{figure}[!t]
\centering
\includegraphics[width=0.9\textwidth]{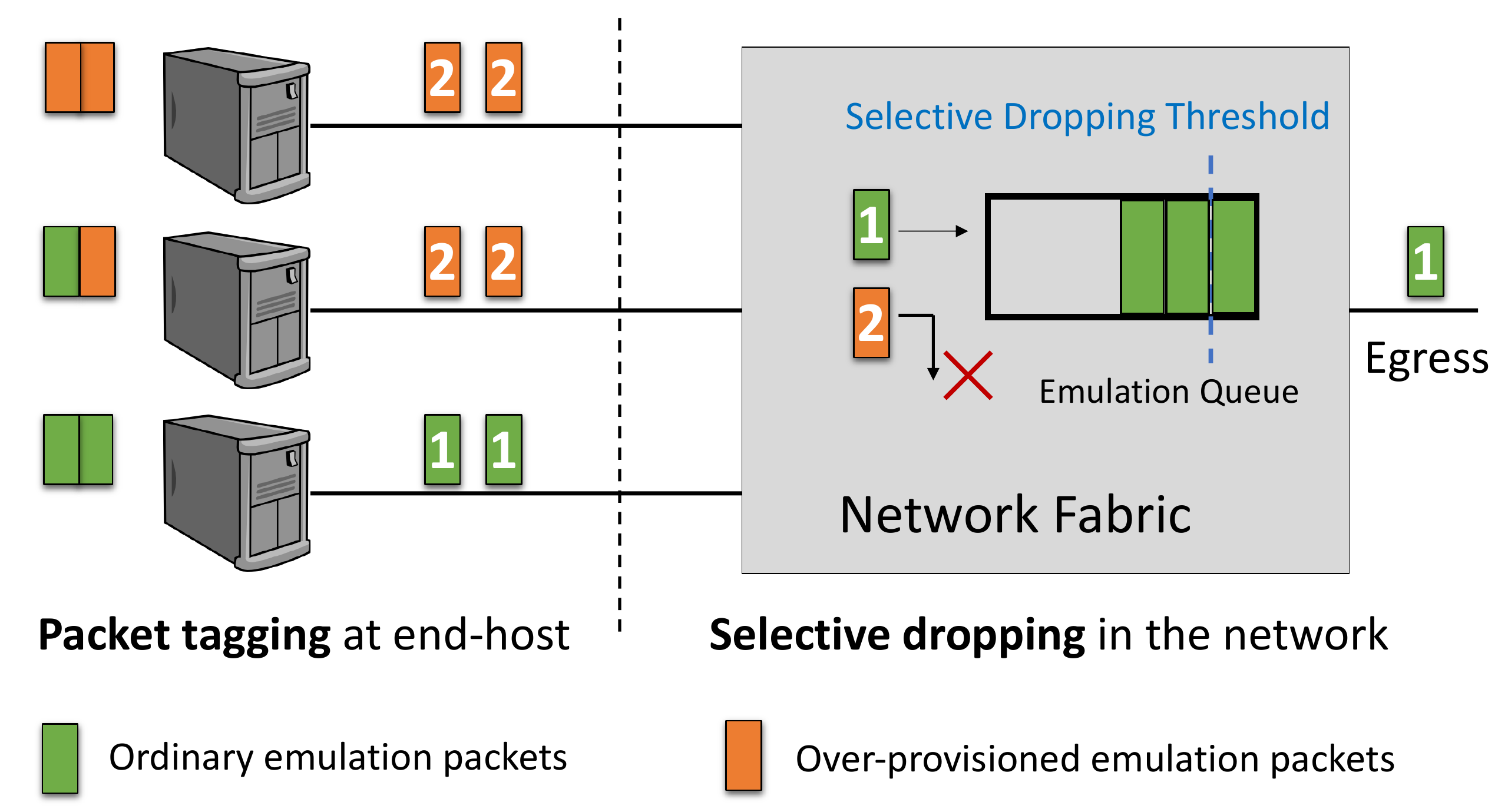}
\caption{Over-provisioning with selective dropping mechanism.} 
\label{fig:selective-dropping}
\end{figure}

Figure~\ref{fig:selective-dropping} illustrates the detailed credit over-provisioning and selective dropping mechanism of \sysB.

\textit{Packet tagging at the end-host.} The end-host maintains a per-flow estimation of the incoming credit packets ($C_{expected}$). Specifically, the sender end-host records the in-flight emulation packets ($E_{out}$). With the regulation of the emulation feedback control loop, the actual emulation packet loss rate is expected to be lower than $max\_target\_loss$ most of the time. Thus, the expected credits can be calculated following Equation~(\ref{equ:credit}):
\begin{equation}
\label{equ:credit}
C_{expected} >= E_{out} \times (1 - max\_target\_loss)
\end{equation}
When the expected credit amount $C_{expected}$ exceeds the remaining traffic demand\footnote{We assume traffic demand information is available. If not, we simply use the send buffer occupancy to calculate the remaining traffic demand~\cite{flow-size}.} of the flow, the sender end-host starts marking the emulation packets as over-provisioned, i.e., with high dropping priority label. Otherwise, emulation packets are marked as ordinary, i.e., with low dropping priority label.

\textit{Selective dropping on the network fabric.} Based on the dropping priority label in the emulation packets, network switches selectively drop the high dropping priority packets before low priority ones when network bandwidth is not enough. Commodity switching chips cannot push out packets that are already stored in the switch buffers. Thus, we can only selectively drop packets at the ingress queue. To this end, we adopt a feature of RED/ECN function, which is widely supported by commodity switches~\cite{wu2012tuning, judd2015morgan}. Specifically, when the switch queue length exceeds the ECN marking threshold, the switch will mark the arrival ECN-capable packets and drop the non-ECN-capable packets. Therefore, \sysB can repurpose this function to achieve selective dropping. Senders can simply set ordinary emulation packets as ECN-capable and over-provisioned emulation packets as non-ECN-capable, to enable selective dropping of the over-provisioned packets with high priority on the network fabric.
\section{Evaluation} 
\label{sec-flashpass:evaluation}

In this section, we present the detailed \sysB evaluation setup in~\S\ref{evaluation-setup}, and conduct extensive experiments to answer the following questions:
\begin{itemize}
\item \textit{Can \sysB achieve high throughput and low loss rate under static workload?} In static workload experiments (\S\ref{evaluation-static}), \sysB achieves near full throughput (9.14Gbps) with zero packet loss persistently across various settings under the shallow-buffered WAN. The throughput is up to 55.9\% higher than that of TCP Cubic.
\item \textit{Can \sysB reduce flow completion time (FCT) under realistic dynamic workload?} In realistic dynamic workload experiments (\S\ref{evaluation-dynamic}), \sysB with Aeolus enhancement achieves 28.5\%-32\% and 3.4\%-11.4\% smaller overall flow completion times compared to Cubic, and Aeolus-ehanced ExpressPass, respectively. 
\item \textit{How do different parameters and components of \sysB impact its network performance?} We show the impact of parameter settings of \sysB in \S\ref{dive-parameter}, and validate the effectiveness of the over-provisioning with selective dropping mechanism in \S\ref{dive-overprovisioning}.
\end{itemize}

\subsection{Evaluation Setup} \label{evaluation-setup}
\textbf{Schemes Compared:} We mainly compare the performance of \sysB with TCP Cubic~\cite{cubic} and ExpressPass~\cite{expresspass}. TCP Cubic is a loss-based reactive congestion control (RCC) protocol that is enabled by default in Linux. It is serving for the majority of the real-world wide-area traffic nowadays. ExpressPass is one of the seminal proactive congestion control (PCC) protocols. Based on our knowledge, it is the only existing PCC that can practically work on WAN. We use the default parameter settings in ExpressPass. We have evaluated both \sysB and ExpressPass with and without Aeolus~\cite{aeolus}. Aeolus is a building block for PCC solutions that improves their performance in the pre-credit phase (i.e., first RTT). 

\textbf{Experiment Configuration:} We build and run our NS2 simulation based on the open-sourced code from ExpressPass~\cite{expresspass-code}. In the static workload experiments (\S\ref{evaluation-static}), we mainly use a dumbbell network topology as shown in Figure~\ref{fig:topo-dumbbell}. In the dynamic workload experiments (\S\ref{evaluation-dynamic}), we mainly use a regional wide-area network as shown in Figure~\ref{fig:topo-wan}. By default, the switch data packet buffer is set to 20KB per port per Gbps according to the Table~\ref{tab:buffer}. The emulation/credit network queue buffer is set to 8 packets for both \sysB and ExpressPass. The selective dropping threshold is set to 2 packets for both Aeolus and over-provisioning mechanism of \sysB. The retransmission timeout is set to 200ms. 
The packet MTU is set to 1.5KB.

\begin{figure}[!t]
\centering
\includegraphics[width=0.9\textwidth]{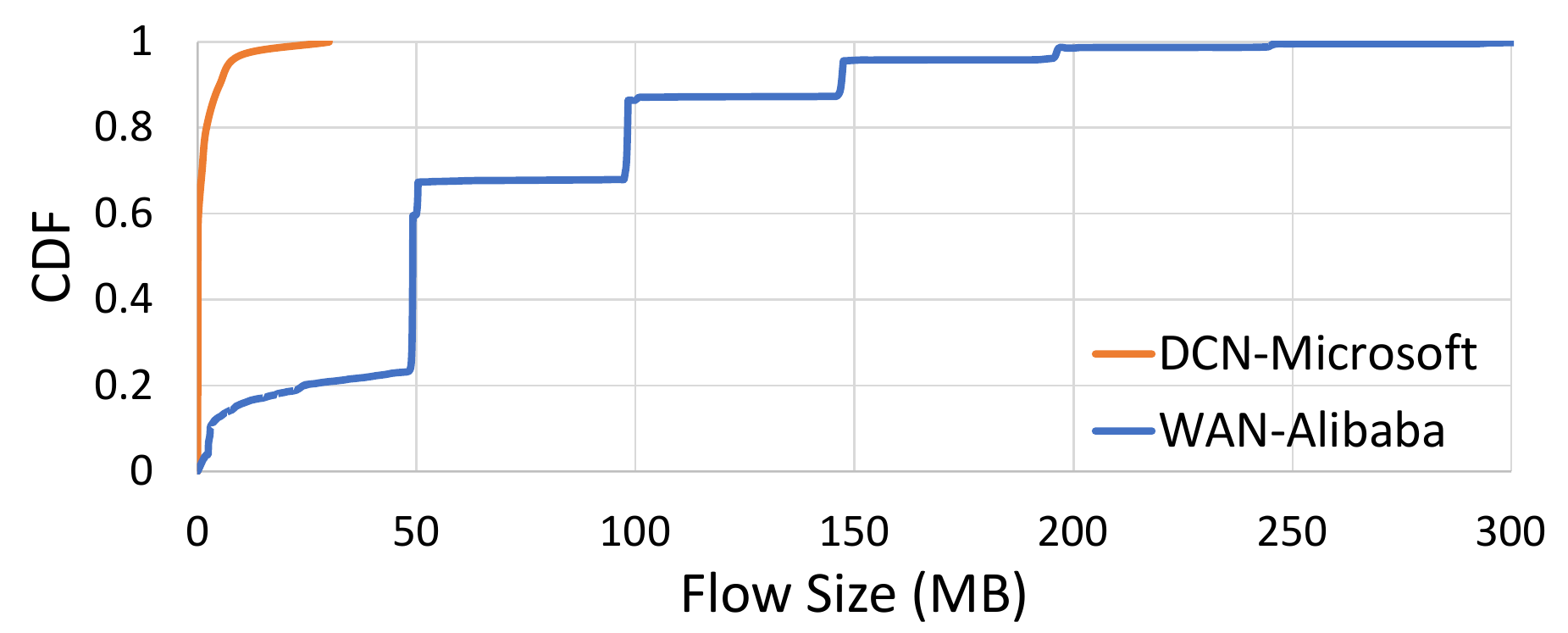}
\caption{Flow size distribution.} 
\label{fig:cdf-flow}
\end{figure}

\textbf{Realistic Workload:} We measured the flow size from a regional wide-area network of Alibaba. The data are collected on links between two data centers. The traffic running on those links are mainly from a data storage service. The flow size distribution is shown in Figure~\ref{fig:cdf-flow} and summarized in Table~\ref{tab:ali-wan}. Compared with the datacenter network workload~\cite{dctcp, vl2, inside-facebook}, wide-area workload has much larger flow size ($\sim$65MB) on average. Based on our classification, there are more than 84.2\% flows that are large-sized with more than 10MB traffic volume. 
The largest flows on WAN can get up to GBs (not shown in the figure), which again are much larger than those on DCN.
In our realistic workload experiments, workloads are generated based on traffic patterns measured from the production WAN. Flows arrive by the Poisson process. The source and destination is chosen uniformly random from different DCs.

\begin{table}[!t]
\centering
\begin{tabular}{|l|l|l|l|}
\hline
           & Small Flow & Large Flow        & Average \\ \hline
Flow Size (MB)  & 0-10       &  \textgreater{}10 & 65      \\ \hline
Flow Percentage & 15.8\%     & 84.2\%             & -       \\ \hline
\end{tabular}
\caption{Traffic characteristic of a production WAN.}
\label{tab:ali-wan}
\end{table}

\textbf{Performance Metrics:} In the static workload experiments (\S\ref{evaluation-static}), we mainly measure the throughput or network utilization, packet loss rate, and buffer occupancy. In the dynamic workload experiments (\S\ref{evaluation-dynamic}), we use flow completion time (FCT) as the major performance metric, which can directly reflect the data transfer speed of network applications.

\subsection{Evaluation Results} \label{evaluation-result}
We present our evaluation results in this subsection. We first show the results of the controlled static workload experiments in \S\ref{evaluation-static}, and then show the results under the realistic dynamic workload in \S\ref{evaluation-dynamic}.

\subsubsection{Static Workload Experiments} \label{evaluation-static}
\begin{figure}[!t]
\centering
\includegraphics[width=0.8\textwidth]{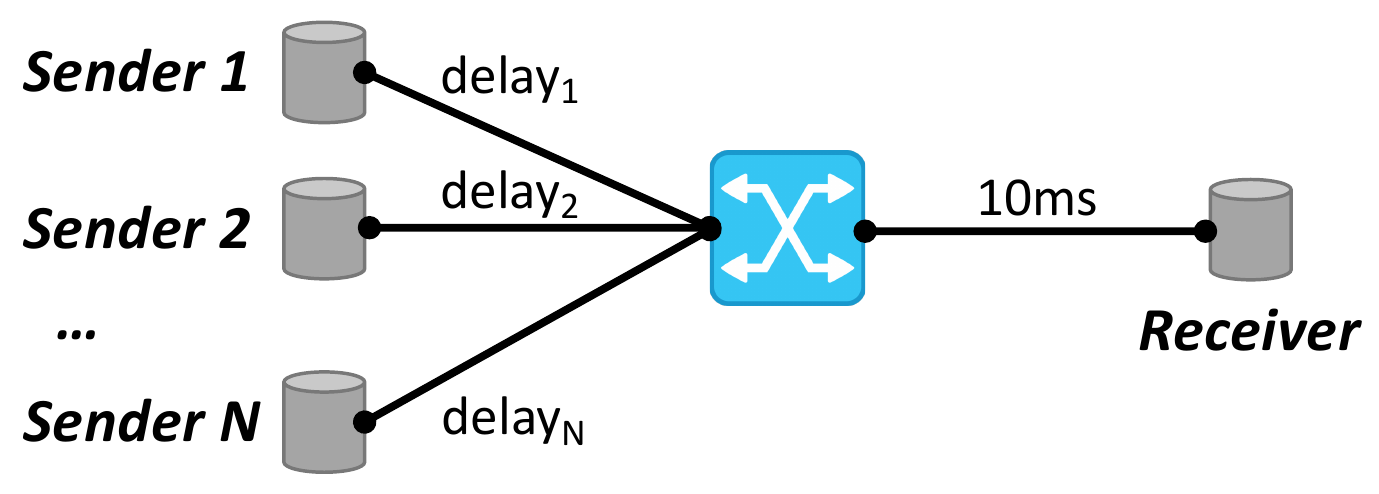}
\caption{Dumbbell network topology. 
There are N senders that simultaneously transfer bulk data to the same receiver. All links have 10Gbps capacity. The network delays vary for different senders to reach the bottleneck switch due to the distinct geographical distances on WAN.
} 
\label{fig:topo-dumbbell}
\end{figure}

In this part, we evaluate the performance of \sysB under static workload experiments. We mainly use the dumbbell network topology as shown in Figure~\ref{fig:topo-dumbbell}. All links have 10Gbps capacity.  There are N senders that simultaneously transfer data to the same receiver. The network delay from sender to the bottleneck switch is set to a uniformly random value between 0 and 20ms. We also test a case with identical delay of 10ms (i.e., 40ms RTT). The sender number is set to 20 or 200. Flows start randomly in the initial 0.2 second and run for 5 seconds. 

\begin{figure*}[!t]
\subfigure[$delay_{i}=Uniform(0, 20ms), i=1,2,...,N; \newline N=20$.]{
  \centering
  \includegraphics[width=0.45\linewidth]{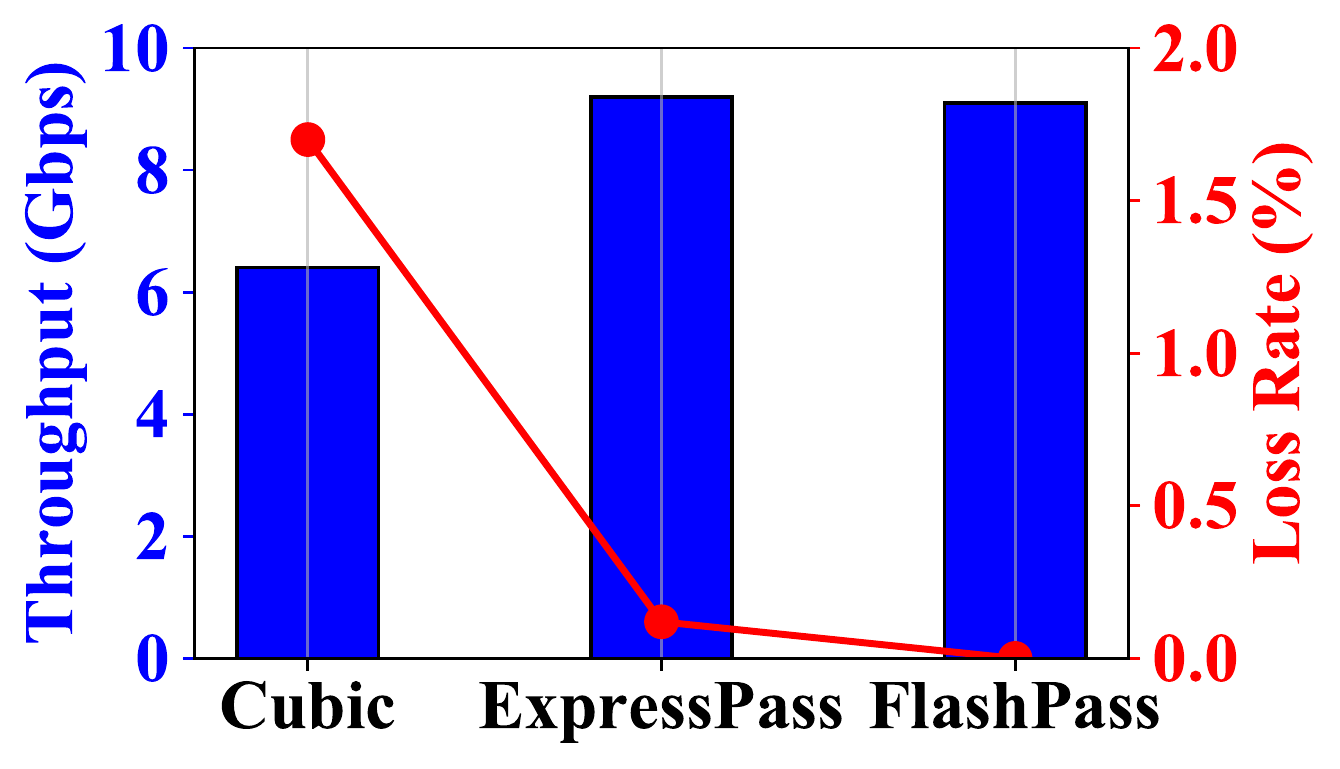}
  \label{fig:static_exp_n20_uniformdelay}
}
\subfigure[$delay_{i}=Uniform(0, 20ms), i=1,2,...,N; \newline N=200$.]{
  \centering
  \includegraphics[width=0.45\linewidth]{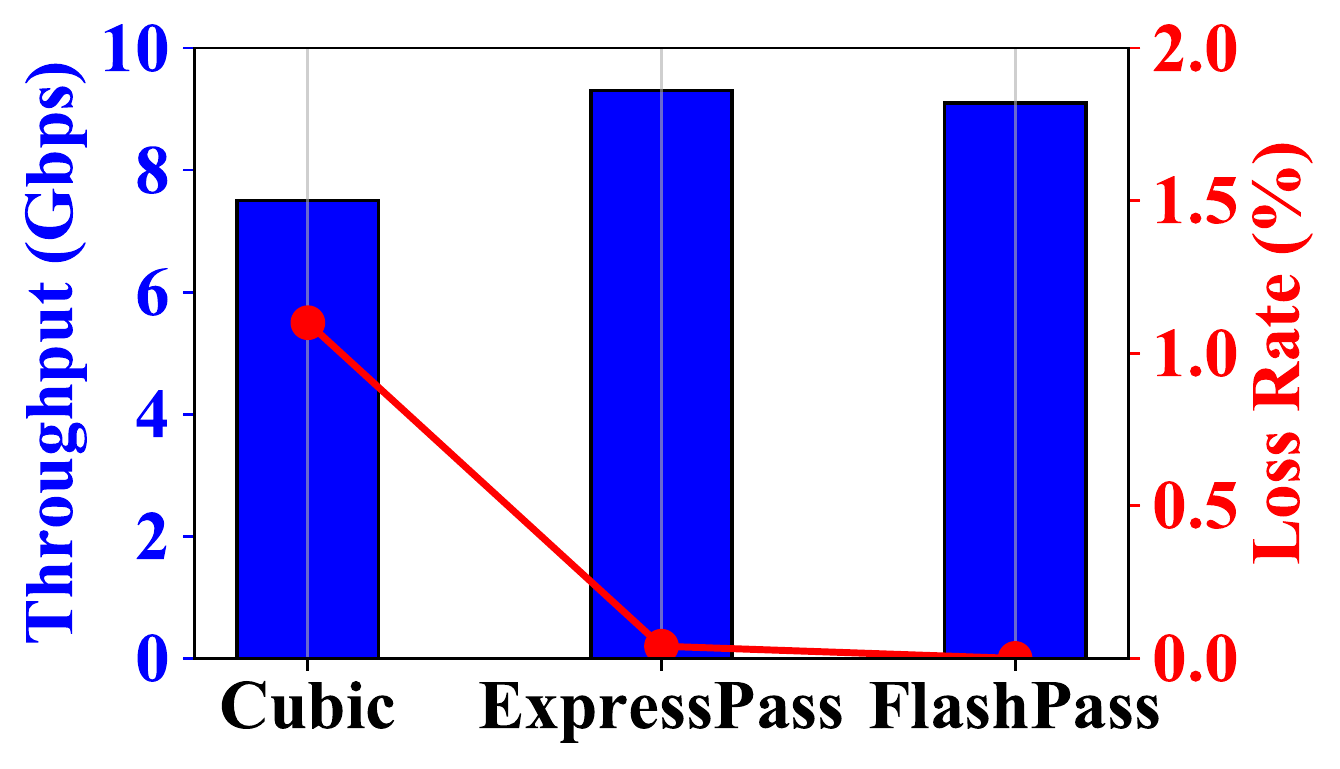}  
  \label{fig:static_exp_n200_uniformdelay}
}
\subfigure[$delay_{i}=10ms, i=1,2,...,N; N=20$.]{
  \centering
  \includegraphics[width=0.45\linewidth]{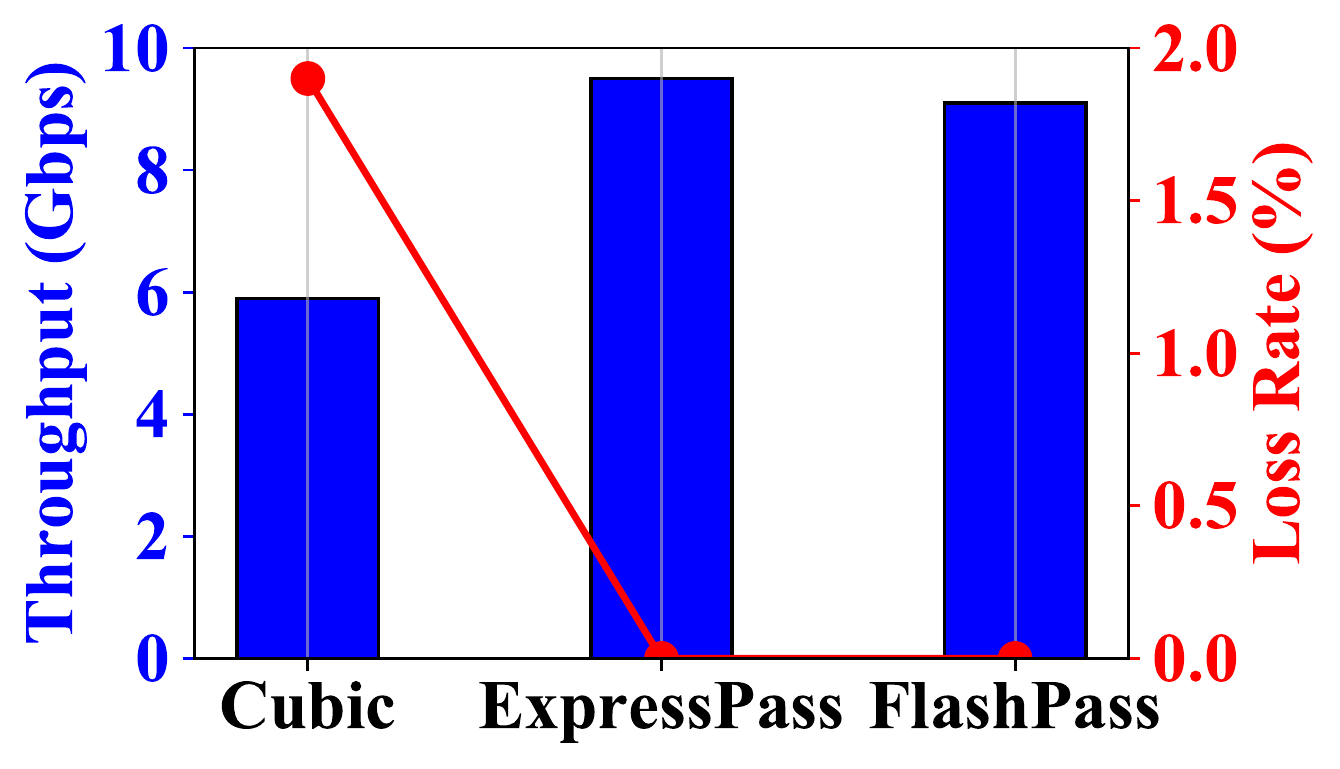}
  \label{fig:static_exp_n20_samedelay}
}
\centering
\caption{Static workload experiment results running on a dumbbell network topology shown in Figure~\ref{fig:topo-dumbbell}. The blue bar indicates the throughput (Gbps) performance on the left y-axis. The red line indicates the packet loss rate (\%) on the right y-axis.}
\label{fig:static_exp}
\end{figure*}

Figure~\ref{fig:static_exp} shows the experimental results. Specifically, Figure~\ref{fig:static_exp_n20_uniformdelay} shows the results under 20 long flows with different RTTs. The throughput is 6.4Gbps, 9.2Gbps, 9.1Gbps for TCP Cubic, ExpressPass, and \sysB. And the packet loss rate is 1.7\%, 0.12\%, and 0 for TCP Cubic, ExpressPass, and \sysB, respectively. In general, \sysB achieves the best performance compared to Cubic and ExpressPass. It is able to maintain near full throughput with zero packet loss throughout all experiments.

As a reactive protocol, TCP Cubic achieves the lowest throughput and worst loss rate throughout the experiments. Compared to \sysB, the throughput can get up to 35.9\% lower in the worst case (Figure~\ref{fig:static_exp_n20_samedelay}). This is because the ``try and backoff'' nature of RCC leads to inherently high data packet losses, and as well periodical under-utilization after rate ``backoff''. Besides, Cubic uses slow start to ramp up its sending rate at the initial phase, which can lead to high losses at the end of the slow start (significant losses are observed during time at $\sim$0.5 second in our experiments).

While ExpressPass achieves similar throughput with \sysB, it has observable packet losses in the scenario with different RTTs (i.e., Figure~\ref{fig:static_exp_n20_uniformdelay} and \ref{fig:static_exp_n200_uniformdelay}). While the packet losses only slightly reduce the throughput of large data transfer in this case, it has much larger negative impact on small flows as we will see in the dynamic workload experiments, especially when retransmission timeout is triggered. \sysB can effectively resolve the imperfect scheduling issue with its sender-driven emulation process. Notice that \sysB has a slightly lower throughput due to the emulation packet overhead. In the static experiments, there is one-way traffic only, which hides the reverse path credit overhead of ExpressPass. 

When comparing the case of 20 flows (Figure~\ref{fig:static_exp_n20_uniformdelay}) with that of 200 flows (Figure~\ref{fig:static_exp_n200_uniformdelay}), the throughput is better while the packet loss is worse for both Cubic and ExpressPass with smaller concurrent flows. This is because more flows can lead to better statistical multiplexing~\cite{buffer-sizing} and thus more stable throughput performance in general.
When comparing the case of different RTTs (Figure~\ref{fig:static_exp_n20_uniformdelay}) with that of identical RTTs (Figure~\ref{fig:static_exp_n20_samedelay}), we find that the packet losses of ExpressPass are indeed caused by the different RTTs on WAN. Identical RTTs also leads to lower throughput for Cubic because it increases the probability of synchronized ``sawtooth'' (i.e., larger variation in congestion window or sending rate).

\begin{figure}[!t]
\centering
\includegraphics[width=0.7\textwidth]{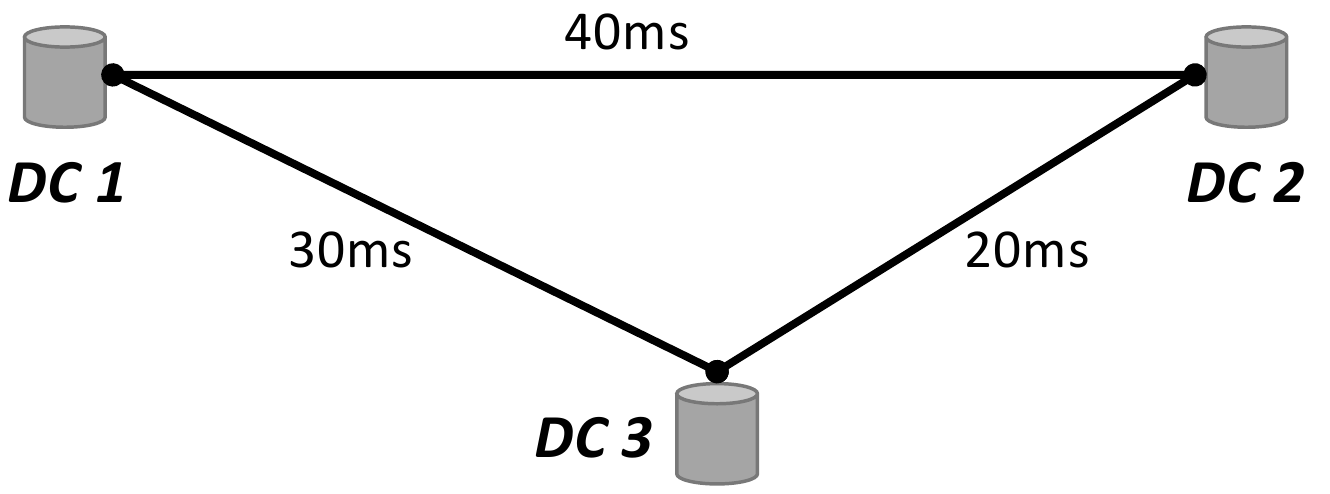}
\caption{Wide-area network (WAN) topology. 
Three datacenters (DCs) connect with each other by WAN links of 20ms, 30ms, and 40ms one-way delay, respectively. All wide-area links have 10Gbps capacity. There are 20 hosts in each DC connecting to the wide-area border router with 1Gbps links.
} 
\label{fig:topo-wan}
\end{figure}

\subsubsection{Dynamic Workload Experiments} \label{evaluation-dynamic}
In this part, we evaluate the performance of \sysB under realistic dynamic workload experiments. We mainly use the wide-area network (WAN) topology as shown in Figure~\ref{fig:topo-wan}. There are three datacenters (DCs) in the wide-area region, connecting with each other by WAN links of 20ms, 30ms, and 40ms one-way delay, respectively. All wide-area links have 10Gbps capacity. There are 20 hosts in each DC connecting directly to the wide-area border router each with a 1Gbps link. The DC link delay is set to 10 microseconds, which is negligible compared to that of WAN. The workloads are generated based on traffic patterns measured from the production WAN (Figure~\ref{fig:cdf-flow}). Flows arrive by the Poisson process. The source and destination is chosen uniformly random from different DCs. Therefore, the WAN links and DC links should have equal load on average. We vary the load from 0.4 to 0.8 of the full network capacity.

\begin{figure*}[!t]
\centering
\subfigure[Small Flow - Average.]{
  \centering
  \includegraphics[width=0.45\linewidth]{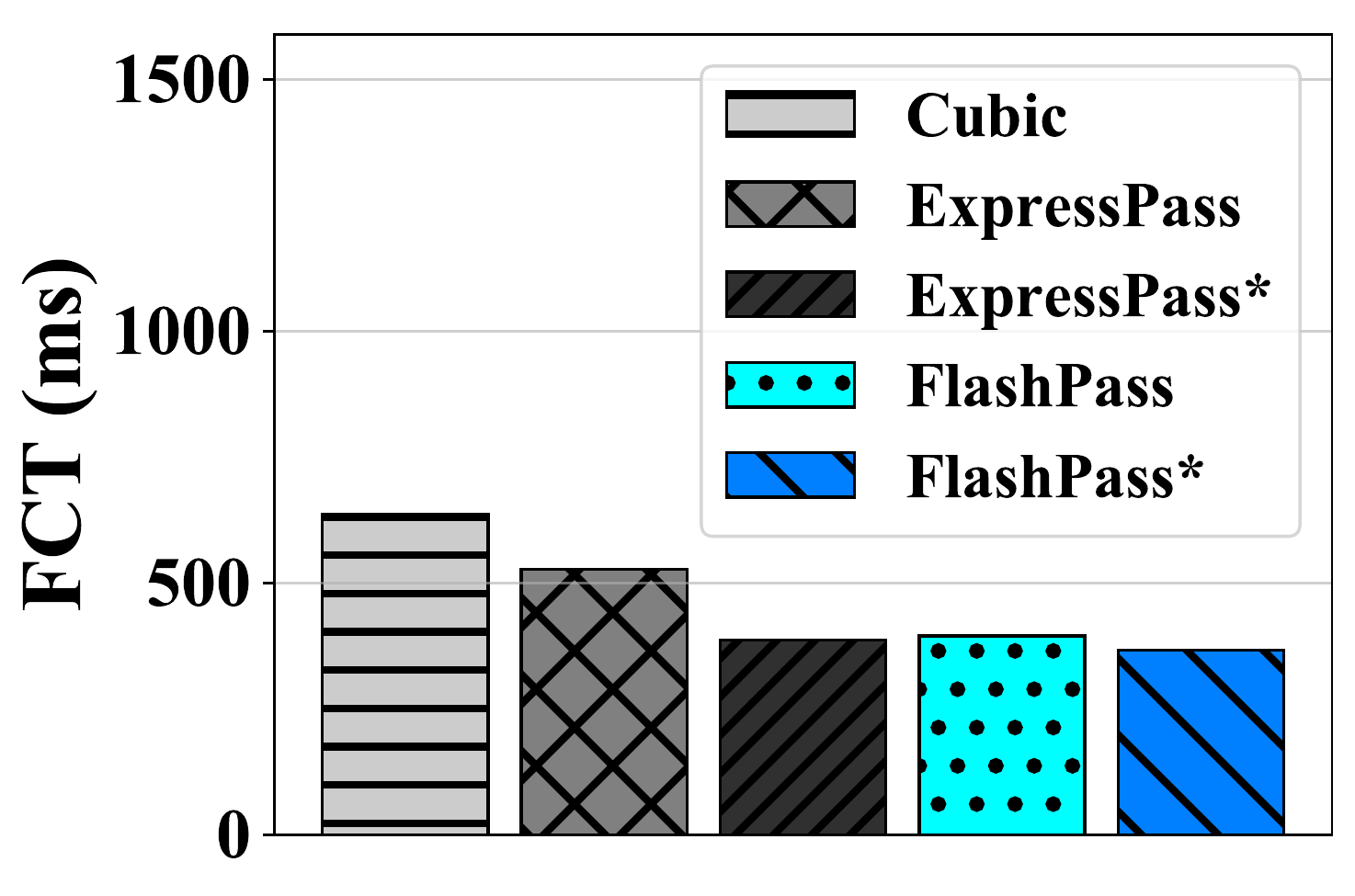}  
  \label{fig:Small_Flows_FCT_Avg_case1}
}
\subfigure[Small Flow - 99th Tail.]{
  \centering
  \includegraphics[width=0.45\linewidth]{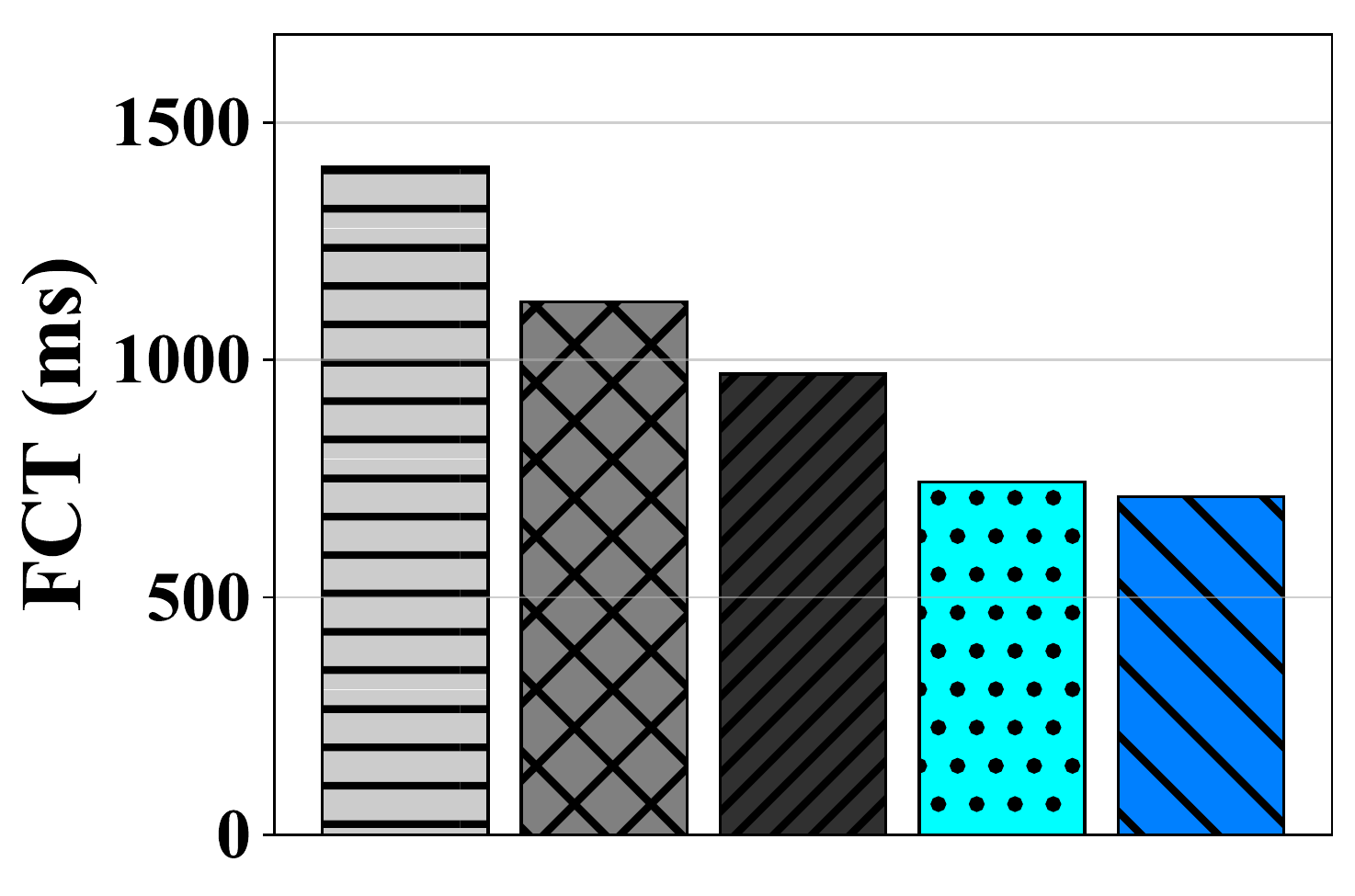}  
  \label{fig:Small_Flows_FCT_99th_case1}
}
\subfigure[Large Flow - Average.]{
  \centering
  \includegraphics[width=0.45\linewidth]{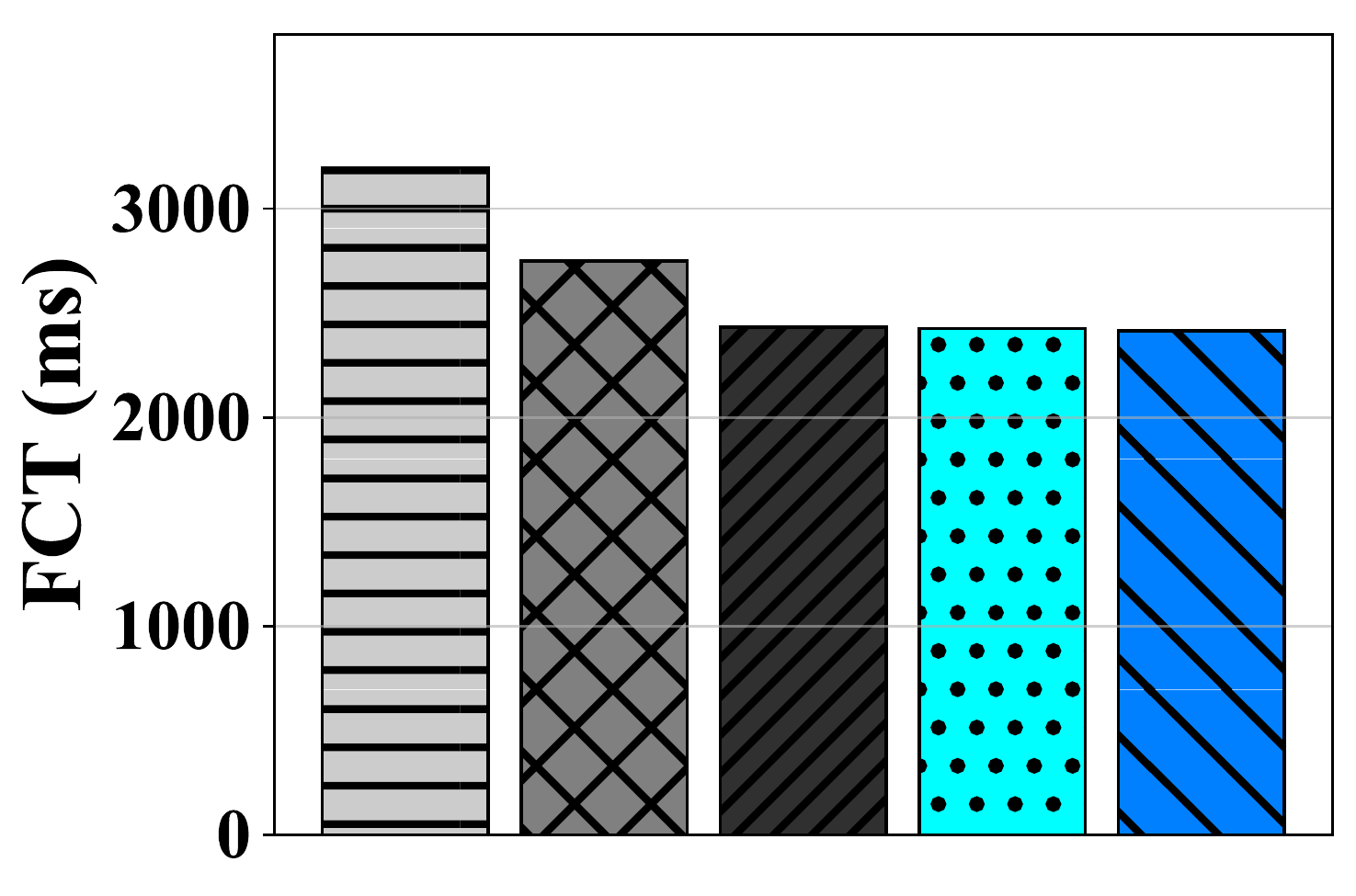}  
  \label{fig:Large_Flows_FCT_case1}
}
\subfigure[All Flow - Average.]{
  \centering
  \includegraphics[width=0.45\linewidth]{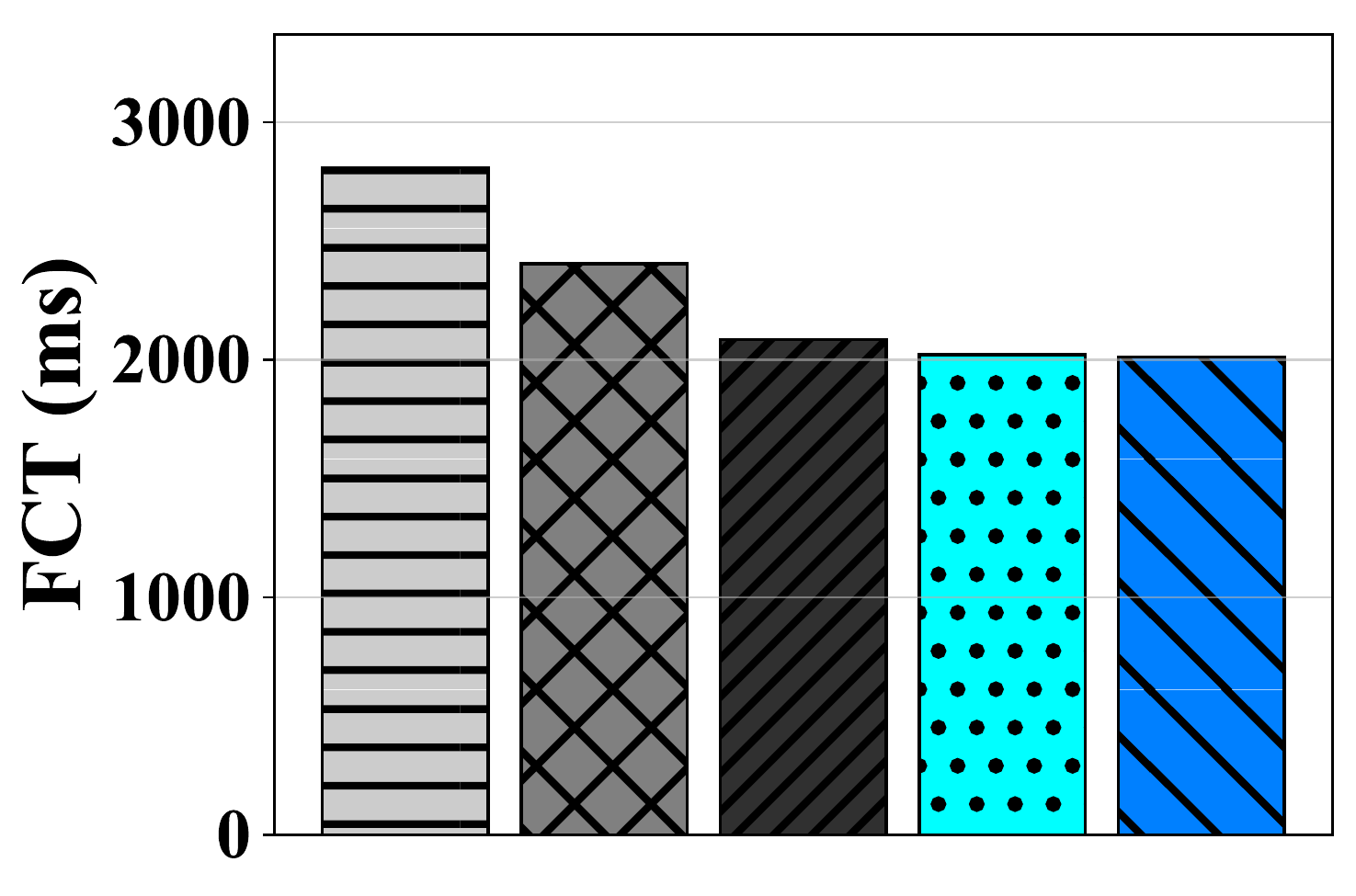}  
  \label{fig:Overall_Flows_FCT_case1}
}
\caption{Flow completion times (FCTs) of various transport protocols under realistic dynamic workload (average load = 0.4). ExpressPass* indicates the Aeolus-enhanced ExpressPass version. FlashPass* indicates the Aeolus-enhanced FlashPass version.}
\label{fig:dynamic_case1}
\end{figure*}

\begin{figure*}[!t]
\centering
\subfigure[Small Flow - Average.]{
  \centering
  \includegraphics[width=0.45\linewidth]{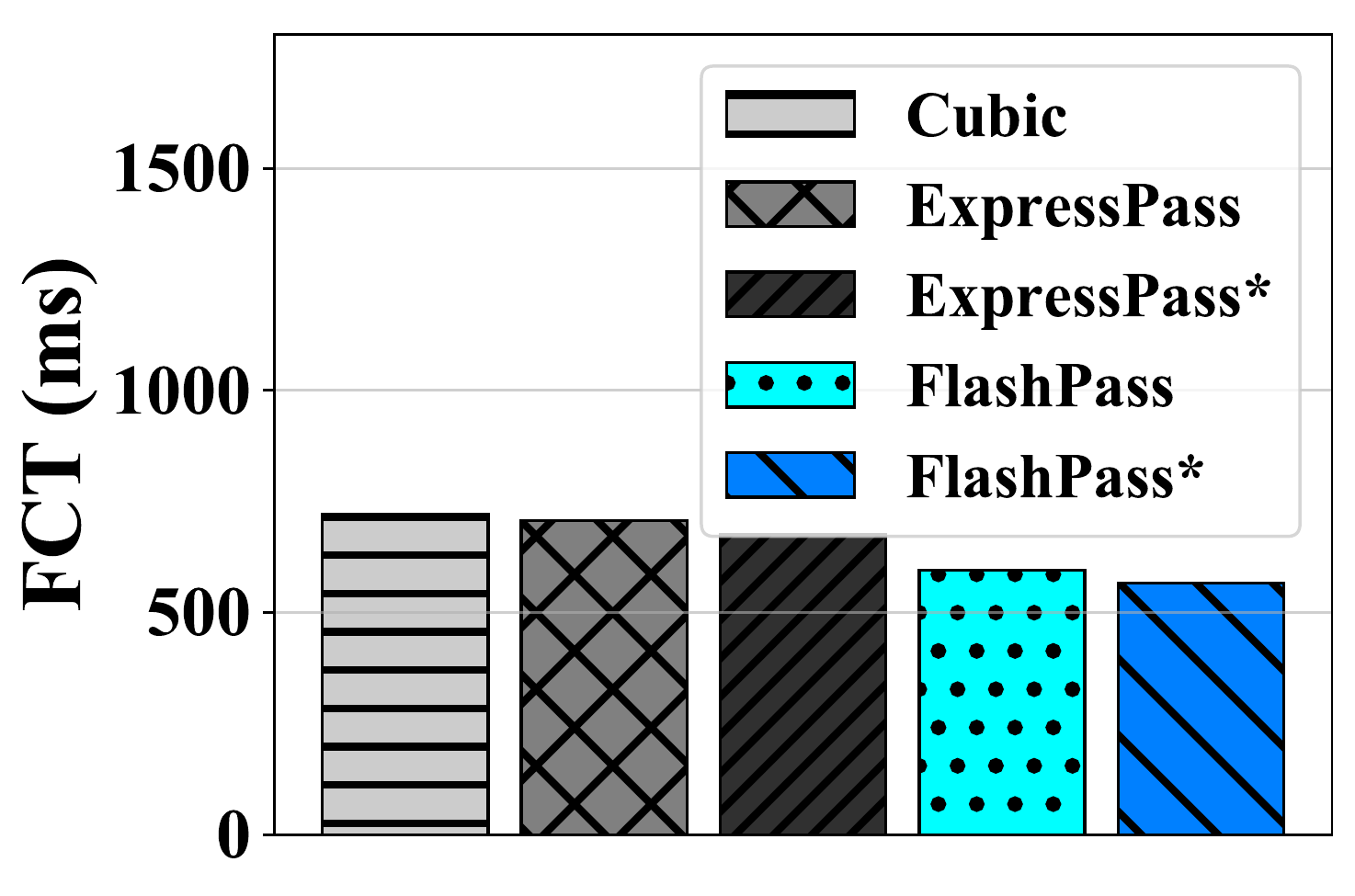}  
  \label{fig:Small_Flows_FCT_Avg_case2}
}
\subfigure[Small Flow - 99th Tail.]{
  \centering
  \includegraphics[width=0.45\linewidth]{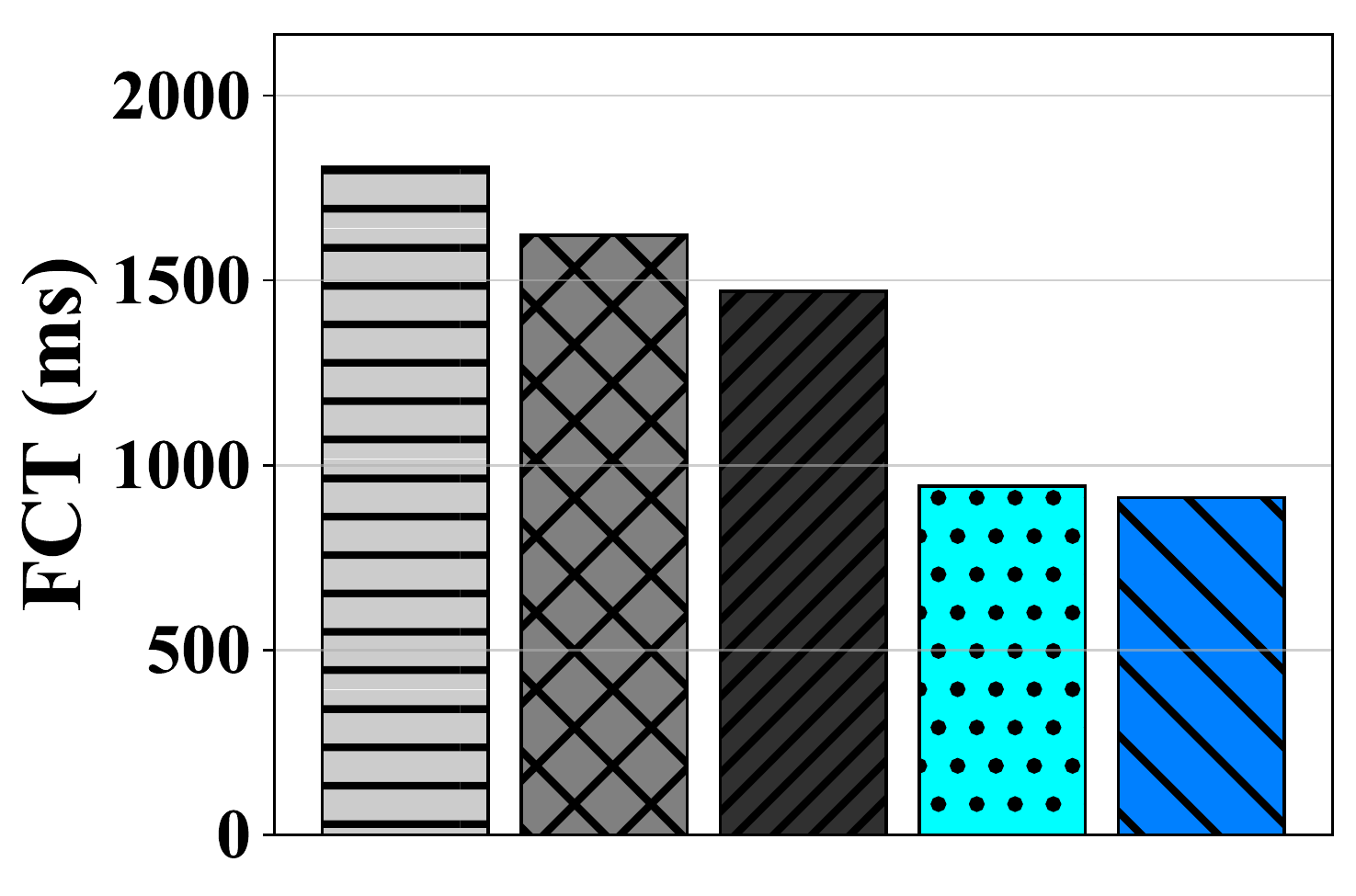}  
  \label{fig:Small_Flows_FCT_99th_case2}
}
\subfigure[Large Flow - Average.]{
  \centering
  \includegraphics[width=0.45\linewidth]{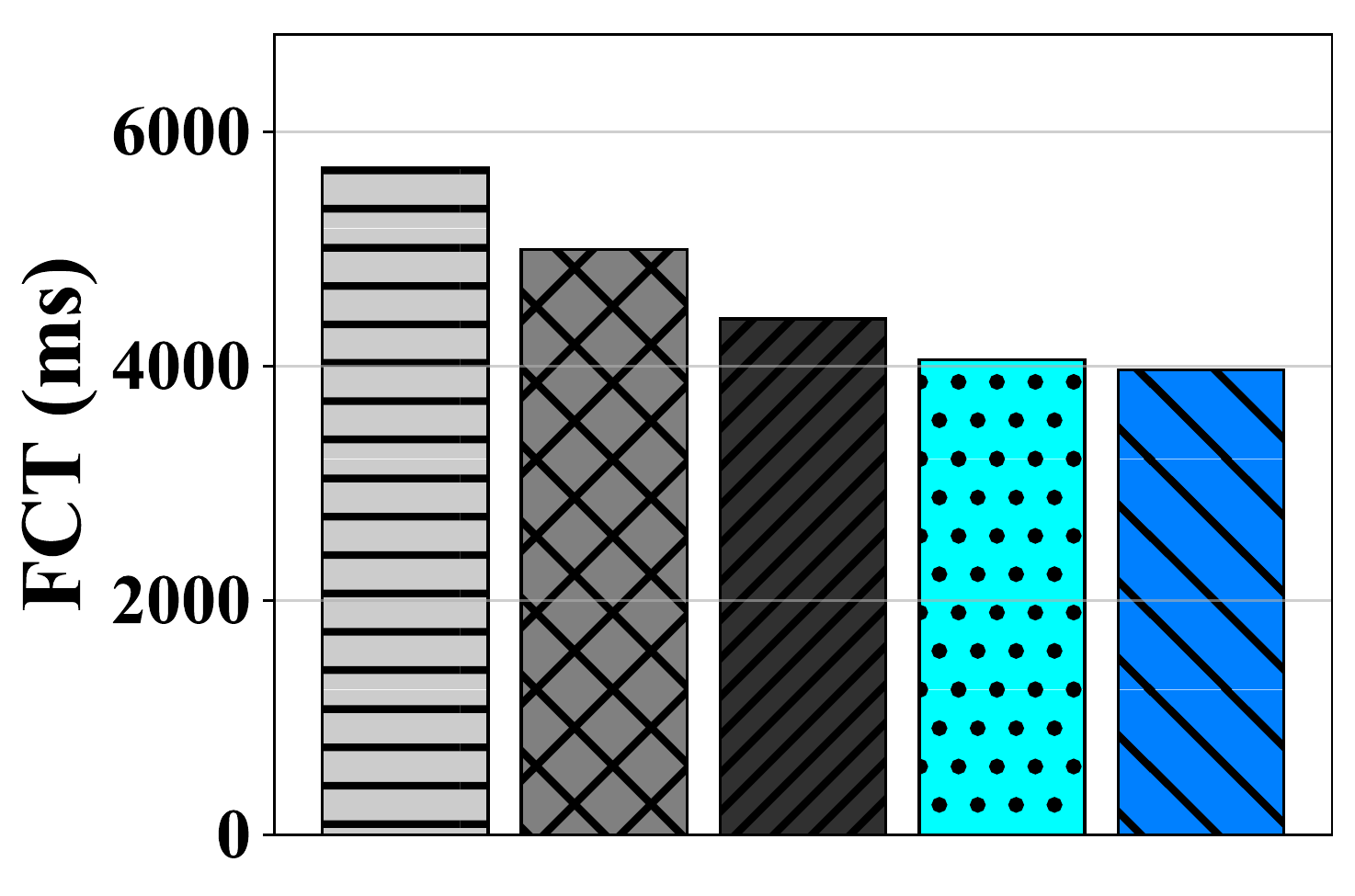}  
  \label{fig:Large_Flows_FCT_case2}
}
\subfigure[All Flow - Average.]{
  \centering
  \includegraphics[width=0.45\linewidth]{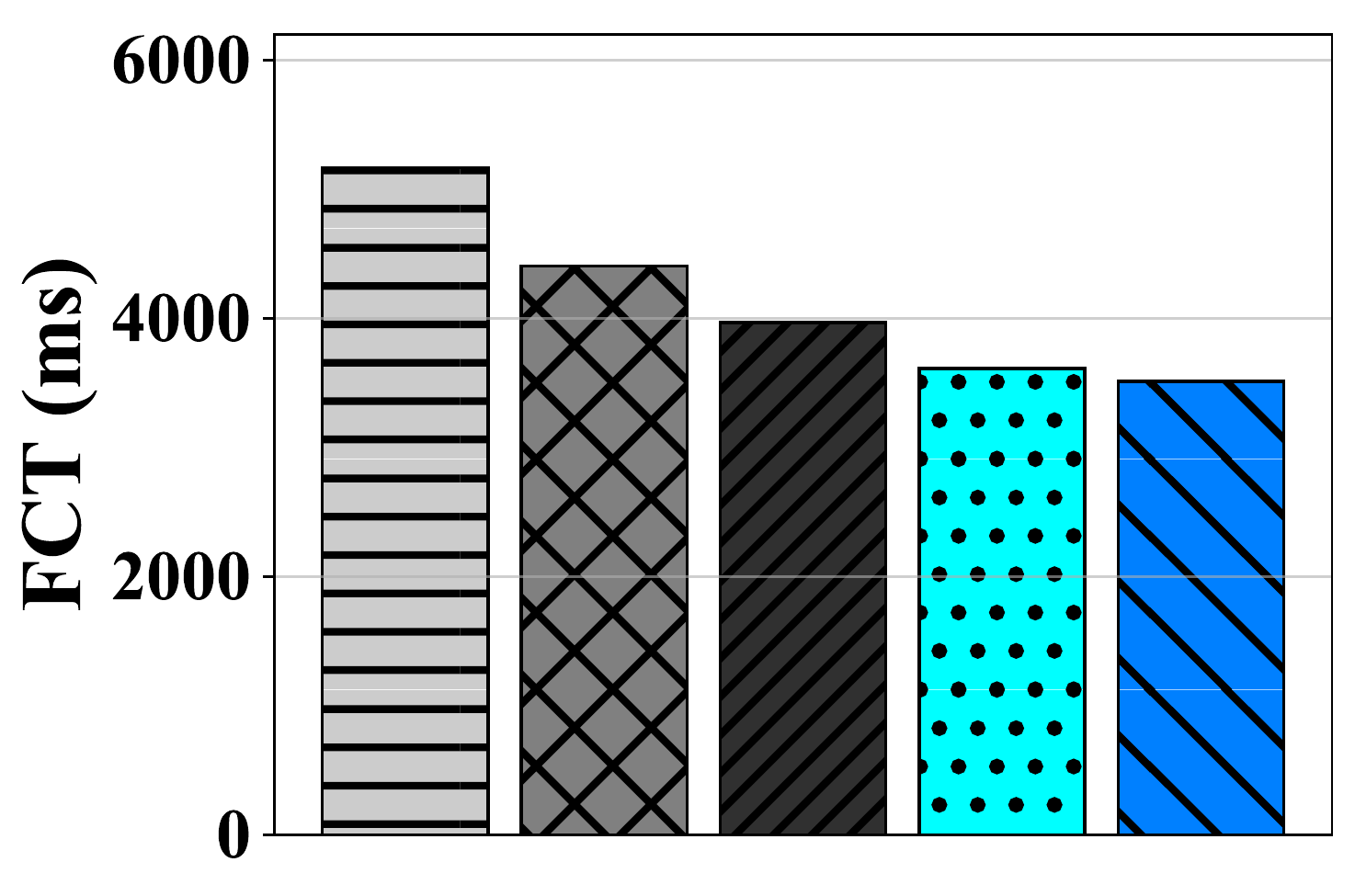}  
  \label{fig:Overall_Flows_FCT_case2}
}
\caption{Flow completion times (FCTs) of various transport protocols under realistic dynamic workload (average load = 0.8). ExpressPass* indicates the Aeolus-enhanced ExpressPass version. FlashPass* indicates the Aeolus-enhanced FlashPass version.}
\label{fig:dynamic_case2}
\end{figure*}

Figure~\ref{fig:dynamic_case1} and Figure~\ref{fig:dynamic_case2} show the experimental results. In general, \sysB performs the best for flow completion times (FCTs) of both small flows ($<$10MB) and large flows ($>$10MB) compared to Cubic and ExpressPass. Specifically, for the best version (i.e., enhanced by Aeolus), FlashPass* reduces the overall FCT by 28.5\%-32\%, and 3.4\%-11.4\% when comparing to Cubic and ExpressPass*, respectively.

For small flows, reactive congestion control like Cubic adds time delay for slow start and its inherent high packet loss also introduces high retransmission timeout overhead, leading to much larger completion times than those of proactive transports. While for ExpressPass, it probes for more bandwidth with much more aggressive credit control algorithm while still keeping low packet losses and thus low timeout. \sysB improves the emulation process and achieves zero losses, thus reducing the small flow completion times to a great extent. When enhanced by Aeolus with linerate start (i.e., ExpressPass* and FlashPass*), one more RTT is saved for credit packets to start data sending, resulting in even lower small flow FCT. This is especially important when the average network load is low. For example, FlashPass* reduces the small flow FCT by 5.1\%-16.4\% on average and 24.3\%-38.0\% at 99-th tail when comparing to ExpressPass*.

For large flows, due to shallow buffer and high packet losses, Cubic suffers from low throughput and thus achieves relatively high FCTs for large flows. While there are also some losses for ExpressPass data sending due to imperfect scheduling in time, the negative impact is very limited because it does not reset its sending rate when loss or timeout happens. However, it does waste some bandwidth due to the last RTT credit scheduling. Such a credit wastage introduces severe negative impact when the network load is high (e.g., when average load = 0.8). \sysB handles the last RTT credit scheduling issue with over-provisioning and selective dropping mechanism. Thus, the credit wastage is effectively avoided, leading to roughly 10.2\% reduction on large flow completion times when comparing FLashPass* with ExpressPass*.

\subsection{Evaluation Deep Dive} \label{evaluation-deep}

\subsubsection{How do parameters of \sysB affect its network performance} \label{dive-parameter}
\begin{figure}[!t]
\centering
\includegraphics[width=0.6\textwidth]{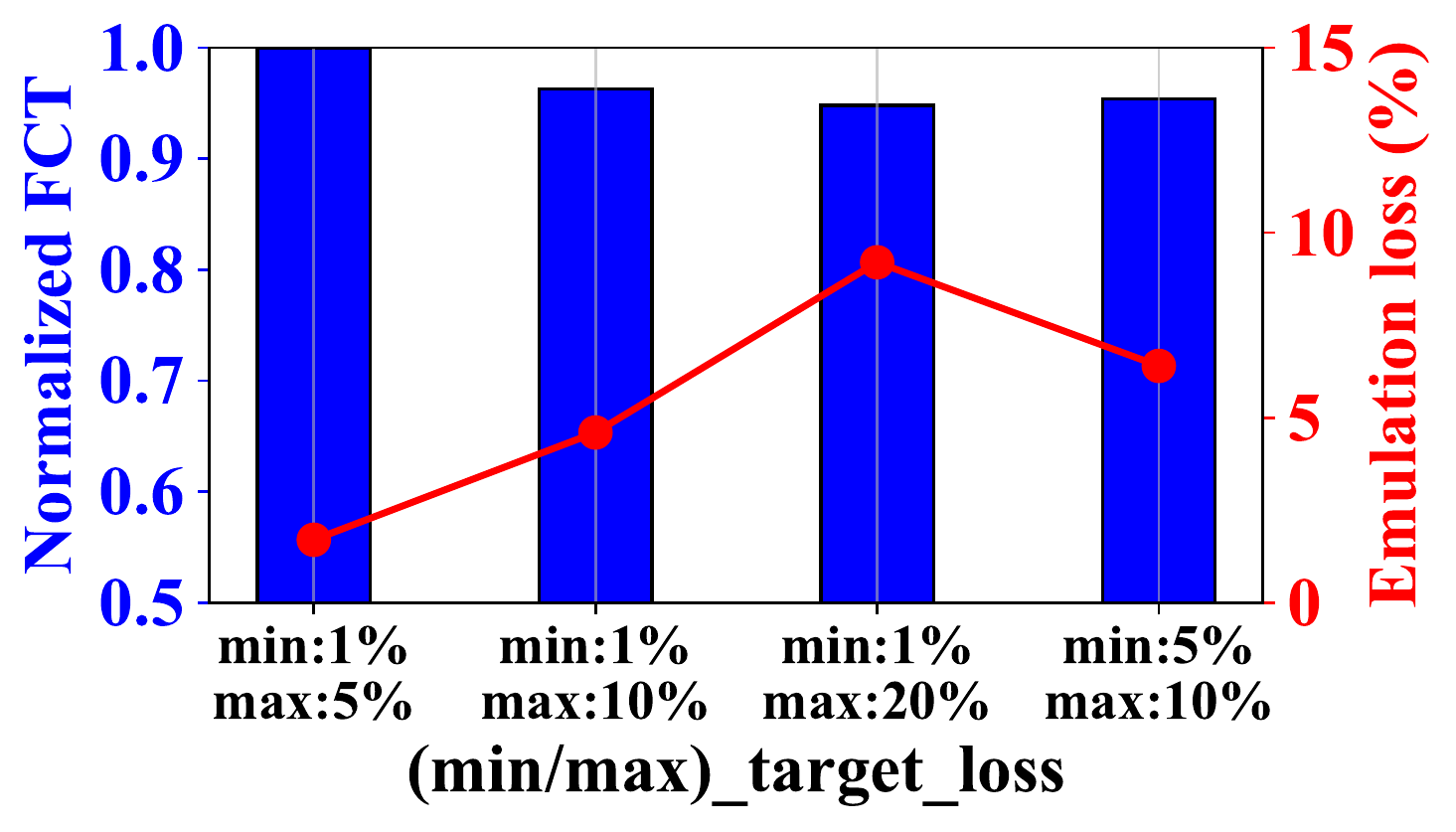}
\caption{FCT (blue bar on the left y-axis) and credit loss rate (red line on the right y-axis) of \sysB. The average FCT is normalized by the result of \sysB with $min\_target\_loss=1\%$ and $max\_target\_loss=5\%$.} 
\label{fig:target_loss}
\end{figure}

In this part, we evaluate the \sysB performance under various parameter settings. First of all, \sysB sets the initial credit rate to the maximum of the emulation network for new flows. This is enabled by the low loss penalty of credit packets as well as the negligible credit wastage with the help of selective dropping mechanism of \sysB. The linerate credit start helps to reduce short flow completion times, especially when the network is mostly idle.
Secondly, the \sysB emulation feedback control loop uses two parameters, i.e., $min\_target\_loss$ and $max\_target\_loss$, to control the target credit loss rate. We repeat the same experiments in \S\ref{evaluation-dynamic} and Figure~\ref{fig:target_loss} shows the results. In general, higher target loss rates lead to faster convergence under traffic dynamics and thus higher efficiency in utilizing network bandwidth, while lower ones have smaller emulation packet overhead. Based on the results, we recommend default parameter settings of $min\_target\_loss=1\%$ and $max\_target\_loss=10\%$.

\subsubsection{How effective is the over-provisioning with selective dropping mechanism of \sysB in avoiding bandwidth or credit wastage} \label{dive-overprovisioning}
\begin{figure}[!t]
\centering
\includegraphics[width=0.6\textwidth]{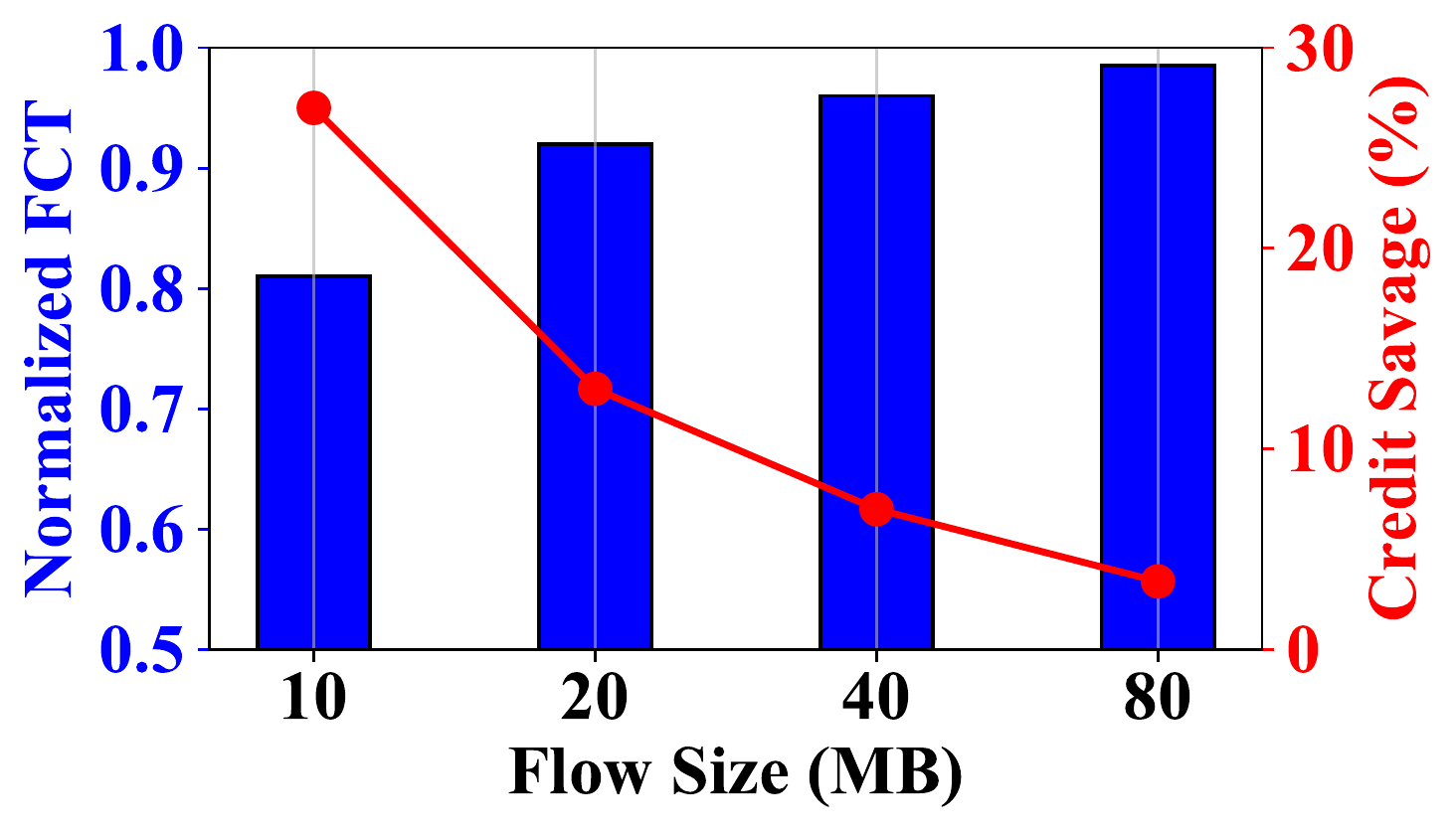}
\caption{FCT (blue bar on the left y-axis) and credit savage (red line on the right y-axis) of \sysB with over-provisioning and selective dropping mechanism over that without over-provisioning mechanism. The average FCT is normalized by the result of \sysB without over-provisioning.} 
\label{fig:overprovisioning}
\end{figure}

In this part, we compare the performance of \sysB with and without the over-provisioning with selective dropping mechanism. We again use the wide-area network topology shown in Figure~\ref{fig:topo-wan}. We generate synthetic workload~\cite{annulus}. Flow sizes are varied from 10MB to 80MB. 
The average network load is set to 0.8 of the full network capacity.
Figure~\ref{fig:overprovisioning} shows the experimental results. We find that the over-provisioning with selective dropping mechanism of \sysB helps to save credits in the last RTT by 3.4\%-27\% of the total traffic volume. This leads to a reduced flow completion time by up to 19\% comparing to \sysB without the over-provisioning and selective dropping mechanism. 
We also observe that the last RTT credit wastage has larger impact in cases with smaller flow sizes. For example, the credit wastage of the case with 10MB-sized flow is $\sim$7$\times$ more than that of the 80MB-sized flow case.

\section{Practical Deployment Analysis} 
\label{sec-flashpass:discussion}

For practical deployment on the enterprise WAN, there are some more requirements for \sysB than the pure end-to-end transport. Firstly, a separate emulation queue should be reserved and rate limited on network switches for emulation process. Secondly, to enable selective dropping (\S\ref{design2}), single-threshold ECN marking/dropping should be configured on the emulation queue. These requirements can be achieved with commodity switches, but need non-trivial configuration across the whole network. Thirdly, we also notice that PCC solutions cannot co-exist with the legacy TCP protocols. A straightforward workaround is to separate different traffic with multiple queues, but may bring about some overheads.

Lastly, an efficient implementation of proactive congestion control logic is required. Recent work~\cite{homa-implementation, homa-kernel} presents a Linux kernel implementation (based on Homa) that achieves magnitudes lower latency than TCP. Specifically, a variety of issues such as batching, load balancing, and realtime processing, have been well addressed in the work. There are some other efforts on realizing PCC protocols on user- space DPDK~\cite{aeolus}, or on congestion control plane (CCP)~\cite{ccp}. While these efforts have validated the feasibility of building an efficient PCC network stack, prototyping a fully functional FLASHPASS is our next step effort and is beyond the scope of this work. 
\section{Related Work} \label{sec-flashpass:related}

For wide-area cloud network optimization, there are three lines of work in general, each operating on a different granularity. Firstly, WAN traffic engineering~\cite{b4, swan, b4after} works on the datacenter level. It distributes network traffic to multiple site-to-site paths (usually hundreds of updates per day). Secondly, bandwidth allocation~\cite{bwe} applies to the tenant or task flow group level. It re-allocates the site-to-site bandwidths and split them among all competing flow groups. Thirdly, transport protocol regulates the per-flow sending rate in realtime. We focus on the transport design in this work.

\textit{Reactive Congestion Control (RCC):} 
The seminal work of TCP congestion control~\cite{reno} works in a  reactive manner. It detects congestion based on a delayed feedback signal (e.g., packet loss) from the network and reacts passively. Many variants have emerged since then, e.g., Cubic~\cite{cubic}, Compound~\cite{compound}, etc. Vegas~\cite{vegas} is the first delay-based protocol to avoid intrinsic high loss and queueing delay of loss-based transport. After that, many protocols~\cite{bbr, copa} are proposed to use delay signal (either explicitly or implicitly). However, these RCC protocols are insufficient to achieve desired high throughput and low loss under shallow-buffered WAN.

There are other RCC protocols designed in particular for DCN. DCTCP~\cite{dctcp} detects the network congestion with explicit congestion notification (ECN) signal. Following that, many ECN-based protocols~\cite{hull, d2tcp, l2dct, dcqcn} are proposed. More recently, HPCC~\cite{hpcc} leverages In-band Network Telemetry (INT) to detect rate mismatch, and adjust sending rates accordingly. These protocols require advanced congestion signals that are not available or well supported~\cite{gemini} on WAN.

\textit{Proactive Congestion Control (PCC):} 
To achieve low buffering and fast convergence, PCC protocols have been proposed to allocate bandwidth proactively. Centralized PCC~\cite{fastpass} regulates traffic rate with a centralized controller. Switch-assisted PCC~\cite{ndp} leverages switch assistance in bandwidth allocation. Both impose high requirements on network facilities (e.g., cutting payload~\cite{cp}) and hence are either unscalable or impractical. 
Some receiver-driven protocols like Homa~\cite{homa} employ simple credit scheduling on the receiver side. They assume single bottleneck link between the top-of-rack (ToR) switch and receiver, which does not hold on WAN. 
Other receiver-driven protocols like ExpressPass~\cite{expresspass} that leverage credit emulation on a separate queue for bandwidth allocation. However, they suffer from efficiency problems on WAN. In contrast, \sysB addresses these problems to fully unleash the power of PCC on shallow-buffered WAN.
\section{Final Remarks} \label{sec-flashpass:conclusion}
In this work, we reveal the trend of adopting shallow-buffered commodity switching chips on wide-area networks (WAN). We then investigate its impact on network performance, and find the insufficiency of the TCP-style reactive congestion control (RCC). To address it, we turn to the emerging proactive congestion control (PCC), and propose \sysB. \sysB is the first attempt to employ PCC on WAN. It is also the first sender-driven credit-scheduled PCC protocol. It leverages the sender-driven emulation process and over-provisioning with selective dropping mechanism to work practically and effectively for the shallow-buffered WAN. Extensive experiments are conducted and validate the superior performance of \sysB under realistic workload.

\chapter{Conclusion and Future Work} \label{sec-conclusion}

My PhD research centers around one question:
\emph{how to improve congestion control mechanisms for inter-datacenter network communication?}
Specifically, we find two key challenges that are underestimated before. Firstly, we observe that the heterogeneous characteristics lead to significant performance degradation in the inter-datacenter network communication. Secondly, the trending shallow-buffered inter-DCN imposes great buffer pressure on congestion control, resulting in low throughput or high packet losses or both.

\section{Summary}
Now we summarize the efforts we make to address each one of the above challenges.

To address the first challenge, we design \sys, a solution for cross-DC congestion control that integrates both ECN and delay signal. \sys uses the delay signal to bound the total in-flight traffic end-to-end, while ECN is used to control the per-hop queues inside a DCN. It further modulates ECN-triggered window reduction aggressiveness with RTT to achieve high throughput under limited buffer.
We implement \sys with Linux kernel and commodity switches.
Experiments show that \sys achieves low latency, high throughput, fair and stable convergence, and delivers lower FCTs compared to various transport protocols (\eg, Cubic, Vegas, DCTCP and BBR) in cross-DC networks.

TO address the second challenge, we seek help from the emerging proactive congestion control (PCC), and propose \sysB. \sysB is the first attempt to employ PCC on WAN. It is also the first sender-driven credit-scheduled PCC protocol. It leverages the sender-driven emulation process to achieve accurate bandwidth allocation, and employs over-provisioning with selective dropping mechanism to take advantage of the spare bandwidth without unnecessary wastage. Extensive experiments are conducted and validate the superior performance of \sysB under realistic workload.

\section{Future Directions}
We envision that the future inter-datacenter networks need to accommodate a continuously growing traffic demand, imposing even harsher pressure on the network communication. The opportunities are that the emerging technologies (\eg, RDMA, programmable switch) enable lots of network innovations to get adopted in practice. Given these trends, some new problems and opportunities will emerge.

\parab{Congestion Control based on Advanced Signal.}
With the prevalence of programmable switches, the wide-area inter-DC network may have support of advanced signals in the near future. For example, HPCC~\cite{hpcc} is the seminal work that leverages in-band telemetry (INT) for congestion control in DCN. However, there are still plenty of challenges to enable INT-based congestion control mechanisms to work for the inter-DC WAN. For example, protocols like HPCC add a prohibitive header overhead as the hop counts increase on the wide-area network, significantly degrading effective data throughput.

\parab{Congestion Control for Wide-area RDMA Network.}
As the RDMA technology getting deployed in DCN, bringing the benefits to the wide-area inter-DC network communication becomes a natural next step. However, unlike intra-DC network, guaranteeing lossless fabric on wide-area network is quite challenging because of the long delays and thus high buffer requirements when using PFC on WAN. Otherwise, with lossy fabric, how to handle packet losses efficiently becomes a new challenge because the RDMA NIC have only limited computation and memory capacity for packet processing.

\bibliographystyle{plain}
\bibliography{reference}

\newpage
\appendix
\chapter*{PhD's Publications}\label{publication}
\addcontentsline{toc}{chapter}{PhD's Publications}

\textbf{Conference \& Workshop Publications}
\begin{itemize}
\item
\textbf{Gaoxiong Zeng}, Li Chen, Bairen Yi, Kai Chen. ``Cutting Tail Latency in Commodity Datacenters with Cloudburst'', in \textbf{IEEE INFOCOM}, 2022.

\item
\textbf{Gaoxiong Zeng}, Jianxin Qiu, Yifei Yuan, Hongqiang Liu, Kai Chen. ``FlashPass: Proactive Congestion Control for Shallow-buffered WAN,'' in \textbf{IEEE ICNP}, 2021.

\item
Shuihai Hu, Wei Bai, \textbf{Gaoxiong Zeng}, Zilong Wang, Baochen Qiao, Kai Chen, Kun Tan, Yi Wang. ``Aeolus: A Building Block for Proactive Transport in Datacenters,'' in \textbf{ACM SIGCOMM}, 2020.

\item
\textbf{Gaoxiong Zeng}, Wei Bai, Ge Chen, Kai Chen, Dongsu Han, Yibo Zhu, Lei Cui. ``Congestion Control for Cross-Datacenter Networks,'' in \textbf{IEEE ICNP}, 2019.

\item
Jiacheng Xia, \textbf{Gaoxiong Zeng}, Junxue Zhang, Weiyan Wang, Wei Bai, Junchen Jiang, Kai Chen. ``Rethinking Transport Layer Design for Distributed Machine Learning,'' in \textbf{ACM APNet}, 2019.

\item
\textbf{Gaoxiong Zeng}, Wei Bai, Ge Chen, Kai Chen, Dongsu Han, Yibo Zhu. ``Combining ECN and RTT for Datacenter Transport,'' in \textbf{ACM APNet}, 2017.
\end{itemize}

\noindent
\textbf{Journal Publications}
\begin{itemize}
\item
Shuihai Hu, \textbf{Gaoxiong Zeng}, Wei Bai, Zilong Wang, Baochen Qiao, Kai Chen, Kun Tan, Yi Wang. ``Aeolus: A Building Block for Proactive Transport in Datacenter Networks,'' in IEEE/ACM Transactions on Networking (\textbf{ToN}), 2021.

\item
\textbf{Gaoxiong Zeng}, Shuihai Hu, Junxue Zhang, Kai Chen. ``Transport Protocols for Data Center Networks: A Survey,'' in ICT-CAS/CCF Journal of Computer Research and Development (\textbf{J-CRAD}), 2020.

\item
Wei Bai, Shuihai Hu, \textbf{Gaoxiong Zeng}, Kai Chen. ``Data Center Flow Scheduling,'' in Communications of the China Computer Federation (\textbf{CCCF}), 2019.
\end{itemize}

\noindent
\textbf{Miscellaneous}
\begin{itemize}
\item
\textbf{Gaoxiong Zeng}, Li Chen, Bairen Yi, Kai Chen. “Optimizing Tail Latency in Commodity Datacenters using Forward Error Correction”, in arXiv:2110.15157 [cs.NI], 2021.

\item
Ge Chen, \textbf{Gaoxiong Zeng}, Li Chen. ``P4COM: In-Network Computation with Programmable Switches,'' in arXiv:2107.13694 [cs.NI], 2021.

\item
Hao Wang, Jingrong Chen, Xinchen Wan, Han Tian, Jiacheng Xia, \textbf{Gaoxiong Zeng}, Weiyan Wang, Kai Chen, Wei Bai, Junchen Jiang. ``Domain-specific Communication Optimization for Distributed DNN Training,'' in arXiv:2008.08445 [cs.DC], 2020.

\item
Shuihai Hu, Kai Chen, \textbf{Gaoxiong Zeng}. ``Improved Path Compression for Explicit Path Control in Production Data Centers,'' in USENIX NSDI (Poster), 2016.
\end{itemize}


\end{document}